\documentclass{aa} 
\usepackage{xcolor}
\usepackage{support-caption}
\usepackage{subcaption}
\usepackage{graphicx}
\usepackage[flushleft]{threeparttable}

\newcommand{\oiidoub}{[\textrm{O}\textsc{ii}]\ensuremath{\lambda\lambda3727,3729}}

\newcommand{\oiiidoub}{[\textrm{O}~\textsc{iii}]\ensuremath{\lambda\lambda4959,5007}}
 
\newcommand{\ha}{\ifmmode {\rm H}\alpha \else H$\alpha$\fi}
\newcommand{\hb}{\ifmmode {\rm H}\beta \else H$\beta$\fi}
\newcommand{\lya}{\ifmmode {\rm Ly}\alpha \else Ly$\alpha$\fi}
\newcommand{\pg}{\ifmmode {\rm P}\gamma \else Pa$\gamma$\fi}
\newcommand{\lyb}{\ifmmode {\rm Ly}\beta \else Ly$\beta$\fi}
\newcommand{\lyg}{\ifmmode {\rm Ly}\gamma \else Ly$\gamma$\fi}
\newcommand{\ciii}{\textrm{C}\textsc{iii}]\ensuremath{\lambda1908}}

\newcommand{\siiinew}{[\textrm{S}\textsc{iii}]\ensuremath{\lambda9531}}

\newcommand{\siidoublet}{[\textrm{S}\textsc{ii}]\ensuremath{\lambda\lambda6716,6731}}

\newcommand{\nvdoub}{[\textrm{N}\textsc{V}]\ensuremath{\lambda\lambda1239,1243}}

\newcommand{\civmed}{\textrm{C}\textsc{iv}\ensuremath{\lambda 1550}}

\newcommand{\flyc}{\ifmmode \mathrm{f}_\mathrm{esc}\mathrm{(LyC)} \else $\mathrm{f}_\mathrm{esc}\mathrm{(LyC)}$\fi}

\def\kmsmpc{km s$^{-1}$ Mpc$^{-1}$}

\def\ergs{\ifmmode \mathrm{erg\hspace{1mm}s}^{-1} \else erg s$^{-1}$\fi}

\def\micron{\ifmmode \mu\mathrm{m} \else $\mu$m\fi}
\def\msun{\ifmmode \mathrm{M}_{\odot} \else M$_{\odot}$\fi}
\def\msunyr{\ifmmode \mathrm{M}_{\odot} \hspace{1mm}{\rm yr}^{-1} \else $\mathrm{M}_{\odot}$ yr$^{-1}$\fi}
\def\zsun{\ifmmode Z_{\odot} \else Z$_{\odot}$\fi}
\def\lsun{\ifmmode L_{\odot} \else L$_{\odot}$\fi}
\def\mstar{\ifmmode \mathrm{M}_{\star} \else M$_{\star}$\fi}
\newcommand{\hst}{HST}
\newcommand{\jwst}{JWST}

\usepackage{txfonts}
\newcommand{\orcid}[1]{\href{https://orcid.org/#1}{\includegraphics[width=10pt]{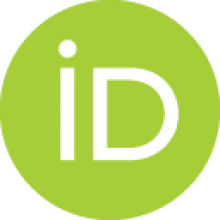}}}
\usepackage{hyperref}
\hypersetup{colorlinks, 
 linkcolor = blue,
 urlcolor = blue,
 citecolor = blue,
 anchorcolor = blue
}
\begin{document}


\title{Peering into cosmic reionization: the \lya\ visibility evolution from galaxies at z = 4.5 - 8.5 with JWST}



 \subtitle{}
 \author{L. Napolitano \orcid{0000-0002-8951-4408}
 \inst{1,2}
 \and L. Pentericci \orcid{0000-0001-8940-6768}
 \inst{1}
 \and P. Santini \orcid{0000-0002-9334-8705}
 \inst{1} 
 \and A. Calabrò \orcid{0000-0003-2536-1614}
 \inst{1}
 \and S. Mascia \orcid{0000-0002-9572-7813}
 \inst{1}
 \and M. Llerena \orcid{0000-0003-1354-4296}
 \inst{1} 
 \and M. Castellano \orcid{0000-0001-9875-8263}
 \inst{1}
 \and \\ M. Dickinson \orcid{0000-0001-5414-5131}
 \inst{9}
 \and S. L. Finkelstein \orcid{0000-0001-8519-1130}
 \inst{6}
 \and R. Amor\'in \orcid{0000-0001-5758-1000}
 \inst{12, 13} 
 \and P. Arrabal Haro \orcid{0000-0002-7959-8783}
 \inst{9}
 \and M. Bagley \orcid{0000-0002-9921-9218}
 \inst{6}
 \and R. Bhatawdekar \orcid{0000-0003-0883-2226}
 \inst{18}
 \and \\ N. J. Cleri \orcid{0000-0001-7151-009X}
 \inst{3,4}
 \and K. Davis 
 \inst{10}
 \and J. P. Gardner \orcid{0000-0003-2098-9568}
 \inst{19}
 \and E. Gawiser \orcid{0000-0003-1530-8713}
 \inst{11}
 \and M. Giavalisco \orcid{0000-0002-7831-8751}
 \inst{16}
 \and N. Hathi \orcid{0000-0001-6145-5090}
 \inst{5}
 \and W. Hu \orcid{0000-0003-3424-3230}
 \inst{3, 4}
 \and \\ I. Jung \orcid{0000-0003-1187-4240}
 \inst{5}
 \and J. S. Kartaltepe \orcid{0000-0001-9187-3605}
 \inst{14}
 \and A. M. Koekemoer \orcid{0000-0002-6610-2048}
 \inst{5}
 \and R. L. Larson \orcid{0000-0003-2366-8858}
 \inst{14}
 \and E. Merlin \orcid{0000-0001-6870-8900}
 \inst{1}
 \and B. Mobasher \orcid{0000-0001-5846-4404}
 \inst{17}
 \and \\ C. Papovich \orcid{0000-0001-7503-8482}
 \inst{3,4}
 \and H. Park \orcid{0000-0002-7464-7857}
 \inst{20}
 \and N. Pirzkal \orcid{0000-0003-3382-5941}
 \inst{15}
 \and J. R. Trump \orcid{0000-0002-1410-0470}
 \inst{10}
 \and S. M. Wilkins \orcid{0000-0003-3903-6935}
 \inst{7,8} 
 \and L.~Y.~A.~Yung \orcid{0000-0003-3466-035X}
 \inst{5}
 }
 \institute{\textit{INAF – Osservatorio Astronomico di Roma, via Frascati 33, 00078, Monteporzio Catone, Italy}\\ 
 \email{lorenzo.napolitano@inaf.it}
 \and 
 \textit{Dipartimento di Fisica, Università di Roma Sapienza, Città Universitaria di Roma - Sapienza, Piazzale Aldo Moro, 2, 00185, Roma, Italy} 
 \and 
 \textit{Department of Physics and Astronomy, Texas A\&M University, College Station, TX, 77843-4242 USA} 
 \and
 \textit{George P.\ and Cynthia Woods Mitchell Institute for Fundamental Physics and Astronomy, Texas A\&M University, College Station, TX, 77843-4242 USA} 
 \and
 \textit{Space Telescope Science Institute, 3700 San Martin Dr., Baltimore, MD 21218, USA} 
 \and
 \textit{Department of Astronomy, The University of Texas at Austin, Austin, TX, USA} 
 \and
 \textit{Astronomy Centre, University of Sussex, Falmer, Brighton BN1 9QH, UK} 
 \and
 \textit{Institute of Space Sciences and Astronomy, University of Malta, Msida MSD 2080, Malta} 
 \and
 \textit{NSF’s National Optical-Infrared Astronomy Research Laboratory, 950 N. Cherry Ave., Tucson, AZ 85719, USA} 
 \and
 \textit{Department of Physics, 196A Auditorium Road, Unit 3046, University of Connecticut, Storrs, CT 06269, USA} 
 \and
 \textit{Department of Physics and Astronomy, Rutgers University, Piscataway, NJ 08854, USA} 
 \and
 \textit{Departamento de Astronomía, Universidad de La Serena, Av. Juan Cisternas 1200 Norte, La Serena, Chile} 
 \and 
 \textit{ARAID Foundation. Centro de Estudios de F\'{\i}sica del Cosmos de Arag\'{o}n (CEFCA), Unidad Asociada al CSIC, Plaza San Juan 1, E--44001 Teruel, Spain} 
 \and
 \textit{Laboratory for Multiwavelength Astrophysics, School of Physics and Astronomy, Rochester Institute of Technology, 84 Lomb Memorial Drive, Rochester, NY, 14623, USA} 
 \and
 \textit{ESA/AURA Space Telescope Science Institute, USA} 
 \and 
 \textit{University of Massachusetts Amherst, 710 North Pleasant Street, Amherst, MA 01003-9305, USA} 
 \and 
 \textit{Department of Physics and Astronomy, University of California Riverside, 900 University Avenue, Riverside, CA 92521, USA } 
 \and
 \textit{European Space Agency (ESA), European Space Astronomy Centre (ESAC), Camino Bajo del Castillo s/n, 28692 Villanueva de la Cañada, Madrid, Spain} 
 \and
 \textit{Astrophysics Science Division, NASA Goddard Space Flight Center, 8800 Greenbelt Rd, Greenbelt, MD 20771, USA} 
 \and
 \textit{Lawrence Berkeley National Laboratory, CA 94720, USA} 
}

\date{Accepted XXX. Received YYY; in original form ZZZ}
 
\abstract{The resonant scattering interaction between \lya\ photons and neutral hydrogen implies that a partially neutral intergalactic medium can significantly impact the detectability of \lya\ emission in galaxies. The redshift evolution of the \lya\ equivalent width distribution of galaxies thus offers a key observational probe of the degree of ionization during the Epoch of Reionization (EoR). Previous in-depth investigations at $z$ $\geq$ 7 were limited by ground-based instrument capabilities. We present an extensive study of \lya\ emission from galaxies at 4 < $z$ < 8.5, observed as part of CEERS and JADES surveys in the JWST NIRSpec/PRISM configuration. The sample consists of 235 galaxies, among which we identify 65 as \lya\ emitters. 
We first measure \lya\ escape fractions from \lya\ to Balmer line flux ratios, and explore correlations with the inferred galaxies' physical properties, which are similar to those found at lower redshift. We also investigate the possible connection between the escape of \lya\ photons and  the inferred escape fractions of LyC photons obtained from indirect indicators. We then analyze the redshift evolution of the \lya\ emitter fraction, finding lower average values at $z$ = 5 and 6 compared to previous ground-based observations. At $z$ = 7 we find a very large difference in \lya\ visibility between the EGS and GOODS-South fields,
possibly due to the presence of early reionized regions in the EGS. Such large variance is also expected in  the Cosmic Dawn II radiation-hydrodynamical simulation. Our findings suggest a scenario 
in which the ending phase of the EoR is characterized by $\sim$ 1 pMpc ionized bubbles around a high fraction of moderately bright galaxies. Finally, we characterize such two ionized regions found in the EGS at $z$ = 7.18 and $z$ = 7.49 by estimating the radius of the ionized bubble that each of the spectroscopically-confirmed members could have created.}

 



 \maketitle

 


\section{Introduction} \label{sec:intro}
Cosmic reionization is a crucial event in the early history of the Universe, a transition through which the intergalactic medium (IGM) changes from being largely neutral to nearly completely ionized, and therefore transparent to Ultra-Violet (UV) photons. While observations established that reionization occurred \citep[e.g.,][]{Fan2006, Banados2018, Planck2020, Becker2021}, the timeline's characterization of the Epoch of Reionization (EoR) is still highly debated \citep[e.g.,][]{Gaikwad2020, Wang2020}. 
The Thomson electron scattering optical depth to cosmic microwave background (CMB) photons measurement \citep{Planck2020} only provides an integral constraint of the EoR, suggesting $z \sim$ 7.7 as the midpoint of reionization. On the other hand, the transmitted flux in quasars (QSOs) spectra \citep[e.g.,][]{Becker_2015, Yang_2020} gives information about the ending phases of reionization, showing it is mostly complete by $z \sim$ 6, with neutral islands remaining down to $z \sim$ 5.2 -- 5.7 \citep[e.g.,][]{becker15b,bosman22}. The paucity of known QSOs at $z >$ 7 and the high Ly$\alpha$ absorption saturation for low volume-averaged neutral fractions ($X_{\mathrm{HI}}$) make it challenging to peer deeper through the EoR with QSOs. \\
 Compared to QSOs, galaxies offers a complementary way to  trace the fraction of neutral hydrogen $X_{\mathrm{HI}}$ across cosmic epochs.  Large scale surveys are needed  since many studies have shown that reionization is a spatially patchy process and it is subject to field-to-field variations \citep[e.g.,][]{Castellano2016, Jung2019, jung2020, Leonova2022}. This picture is also supported by the latest large-volume radiation-hydrodynamics simulations of the EoR \citep[e.g.,][]{Taha_2018, Ocvirk_2021, Ucci_2021}. \\
Currently, faint star-forming galaxies (SFGs) are considered the main candidate sources that provided most of the the Lyman continuum (LyC; $\lambda < 912$ \AA) ionizing radiation needed to complete cosmic reionization \citep[e.g.,][]{Robertson2013, Bouwens2015b, Finkelstein2015, Livermore2017, Bhatawdekar2019, Finkelstein2019, Yung2020a, Yung2020b}, while Active Galactic Nuclei (AGNs) had a minor role in the process \citep[e.g.,][]{Giallongo2015, Hassan2018, Matsuoka2018, Parsa2018, Finkelstein2019, Kulkarni2019, Yung2021, Jiang2022, Matthee2023}. In particular, \lya\ emitting galaxies (LAEs) \citep[e.g.,][]{Hu1996, Steidel1996} may provide our current strongest probe to study the EoR, since this line is commonly observed in high-redshift star-forming galaxies \citep[e.g.,][]{Stark2010} and is highly sensitive to the IGM neutral content. The resonant scattering interaction between \lya\ photons and neutral hydrogen causes a partially neutral IGM to heavily impact the detectability of \lya\ photons  \citep[see][for a review]{Ouchi_2020}.
In recent years, this effect was explored by comparing  the \lya\ luminosity function (LF) with the UV-continuum LF \citep[e.g.,][]{Ota2008, Ouchi2010, Zheng2017, Itoh2018, Hu2019, Konno2018}. The redshift evolution of the \lya\ LF is determined by galaxy evolution and \lya\ opacity of the IGM, while the UV-continuum LF is  governed only by galaxy evolution. It was found that \lya\ LF rapidly drops from $z \sim 6.5$ to $7.5$, while the UV-continuum LF experiences a milder decrease, thus suggesting  an increase of $X_{\mathrm{HI}}$
A complementary approach is to measure $X_{\lya}$, that is the fraction of \lya\ emitting galaxies from all of the UV continuum-selected galaxies at a given redshift. Ground based systematic efforts have been conducted by many authors for more than a decade \citep[e.g.,][]{Fontana2010, Stark2010, Pentericci2011, Schenker2012, Treu2012, Caruana2014, Schenker2014, Tilvi2014, Arrabal_Haro2018, Caruana2018,  Pentericci_2018b, Mason2019, jung2020, Yoshioka2022}, and the results show a drop of $X_{\lya}$ above $z \sim 6$, again probing solid evidences that at higher redshifts the Universe was partially neutral. 
However, there is still significant scatter associated with this measurement and the precise evolution of $X_{\lya}$ is still debated. This is primarily caused by large statistical uncertainties associated with relatively small datasets at high redshift. Biases may be brought in by the different methods used for selecting target samples (color or photometric redshift selection), the diverse sample cuts in the UV absolute magnitude ($M_{\mathrm{UV}}$), the choice of the instrument configuration (Integral Field Unit, slit spectroscopy, or Narrow-Band surveys) to identify LAEs amongst SFGs, and the limitations of using ground-based telescopes. High redshift \lya\ observations from the ground were necessarily restricted to the brightest sources \citep[e.g.,][]{Larson_2018, Harikane_2019ApJ, Taylor_2021} as at $z>7$ Ly$\alpha$ moves into the near infrared (near-IR), where sky background and atmospheric telluric lines significantly limit spectroscopic sensitivity. 
This makes the non-detection of \lya\ challenging to interpret, as it hinders the determination of the galaxy's actual redshift. Probing the full EoR with Ly$\alpha$ constraints has thus not been fully achieved.\\ 
In this context, the advent of of the James Webb Space Telescope \citep[\jwst ,][]{Gardner2006, Gardner2023} has led to significant progress in systematically discovering galaxies at very early cosmic epochs. Early Release Science programs \citep[e.g.,][]{TreuGlass2022, Bagley2023_NGDEEP, Finkelstein2023} found many high-redshift galaxy candidates at $z$ > 9 \citep[e.g.,][]{Castellano2022b, Finkelstein2022b, Naidu2022b, Adams2023, Atek2023, Bouwens2023, Casey2023, Harikane2023}. Moreover \jwst/Near InfraRed Spectrograph \citep[NIRSpec,][]{Jakobsen2022}  demonstrated to be successful at identifying emission lines of high-redshift galaxies  \citep[e.g.,][]{Bunker2023B, Jones2023, Jung2023, Roy2023, Saxena2023, Tang2023}. 
Unlike ground-based telescopes, JWST can accurately measure spectroscopic redshifts using other optical or UV rest frame emission lines, \textit{regardless} of whether \lya\ emission is present.\\
In this paper our aim is to construct a sample of high-z galaxies with robust spectroscopic redshift and completeness estimates for \lya\ rest frame equivalent width, probing a wide range of redshifts throughout the EoR. We analyze data that are part of the Cosmic Evolution Early Release Science (CEERS) survey \citep{Finkelstein2023}, to select 
\lya\ emitters and study their physical properties and the evolution of the line visibility.
The paper is organized as follows. We discuss our parent sample construction in Sect.~\ref{sec:Data_and_sample_selection} and the methodology used in Sect.~\ref{sec:Method}. We present the derived $f_{\mathrm{esc}}^{\lya}$ and \lya\ fraction $X_{\lya}$ measurements and discuss the correlations found within our data, in Sect.~\ref{sec:fescLyA} and Sect.~\ref{sec:XlyA} respectively. We summarize our findings in Sect.~\ref{sec:Conclusion}.\\
In the following, we adopt the $\Lambda$CDM concordance cosmological model ($H_0 = 70$ \kmsmpc, $\Omega_M = 0.3$, and $\Omega_{\Lambda} = 0.7$). We report all magnitudes in the AB system \citep{Oke1983} and EWs to rest-frame values.

\section{Data} \label{sec:Data_and_sample_selection}
\subsection{CEERS data}\label{sec:CEERS}
In this work we employ the publicly released \jwst/NIRCam and NIRSpec data from the Cosmic Evolution Early Release Science survey (CEERS; ERS 1345, PI: S. Finkelstein). CEERS targets the CANDELS Extended Growth Strip (EGS) field \citep[][]{Davis2007, Grogin2011, Koekemoer2011}, by observing this region in 12 pointings using the \jwst/NIRSpec, NIRCam, and MIRI instruments. All the NIRCam pointings are uniformly covered in the broad band filters F115W, F150W, F200W, F277W, F356W, and F444W, along with F410M medium-band filter. Arrabal Haro et al. (in prep.) \citep[see also][]{ArrabalHaro2023} will present the CEERS NIRSpec spectra, Finkelstein et al. \citep[in prep., see also][]{Finkelstein2022, Finkelstein2022b} will discuss target selection. For the present work we note that only a handful of objects with previously identified redshifts were inserted in the MSA. Details on NIRCam imaging data reduction procedure are contained in \cite{Bagley2023}. 
Photometric redshifts were obtained from version v0.51.2 of the CEERS Photometric Catalog (Finkelstein et al. in prep.) with \textsc{EAZY} \citep{Brammer2008}, following the methodology described in \cite{Finkelstein2023} by including new templates from \cite{Larson2023B}, which improve the photometric redshifts' accuracy for high-z galaxies. \\
For this work, we only consider data obtained in the NIRSpec PRISM/CLEAR configuration, that provides continuous wavelength coverage in the 0.6-5.3 $\mu$m wavelength range. The spectral resolution $R=\lambda / \Delta\lambda$ of the instrument is $\sim$ 30 - 300. Each pointing was observed for a total of 3107 s, divided into three exposures of 14 groups each, utilizing the NRSIRS2 readout mode. A three-point nod pattern was employed for each observation, to facilitate background subtraction. As detailed also in \cite{ArrabalHaro2023} and \cite{Arrabal_Haro2023Nature}, for data processing and reduction we make use of the STScI Calibration Pipeline\footnote{\url{https://jwst-pipeline.readthedocs.io/en/latest/index.html}} version 1.8.5 and the Calibration Reference Data System (CRDS) mapping 1029, with the pipeline modules separated into three modules. In brief, the \textsc{calwebb\_detector1} module addresses detector 1/f noise, subtracts dark current and bias, and generates count-rate maps (CRMs) from the uncalibrated images. The \textsc{calwebb\_spec2} module creates two-dimensional (2D) cutouts of the slitlets, corrects for flat-fielding, performs background subtraction using the three-nod pattern, executes photometric and wavelength calibrations, and resamples the 2D spectra to correct distortions of the spectral trace. The \textsc{calwebb\_spec3} module combines images from the three nods, utilizing customized extraction apertures to extract the one-dimensional (1D) spectra. Finally, both 2D and 1D spectra are simultaneously examined with the \textsc{Mosviz} visualization tool \citep{Developers2023} to mask potential remaining hot pixels and artifacts in the spectra. After masking image artifacts, data from three consecutive exposure sequences are combined to produce the final 2D and 1D spectral products. Arrabal Haro et al. (in prep.) will present a more detailed description of the CEERS NIRSpec data reduction. \\
We do not consider PRISM NIRSpec data observed in pointings 9 and 10, since due to a short circuit issue, they are contaminated and lack secure flux calibrations.
To investigate potential residual issues in the absolute flux calibration caused by slit losses (although \textsc{calwebb\_spec2} step of the pipeline already employs a slit path loss correction) or other inaccuracies in flux calibration files, as a first step we check the consistency with the broad band photometry, by integrating the spectra across the NIRCam filter bandpasses. We then compare this synthetic photometry  with the measured NIRCam photometry. From this procedure we obtain  the correction factors for the spectra in each NIRCam filter. For the high redshift sample we consider in this work, the multiplicative flux correction factors have an average value of $\sim 1.4$, in agreement to the values found by \cite{Arrabal_Haro2023Nature}. Most importantly, we find that these corrections remain constant across wavelength. Given that in this paper we will deal with EWs and line flux ratios, we consider the spectra derived by the standard pipeline, without applying any further corrections. The only exception is the $M_{\mathrm{UV}}$ calculation (see Sect.~\ref{sec:Muv}).\\
In our analysis we also consider \hst\ photometry, in the F606W, F814W, F125W, F140W, and F160W filters, which are presented in the official EGS photometric catalog \citep[see][]{Stefanon2017}. 
In the following section, we provide a brief summary of our sample selection criteria. 
\subsection{CEERS Parent sample selection}\label{sec:parent_sample}
\begin{figure*}[!h]
    \centering
    \begin{subfigure}{\textwidth}
        \includegraphics[width=\linewidth]{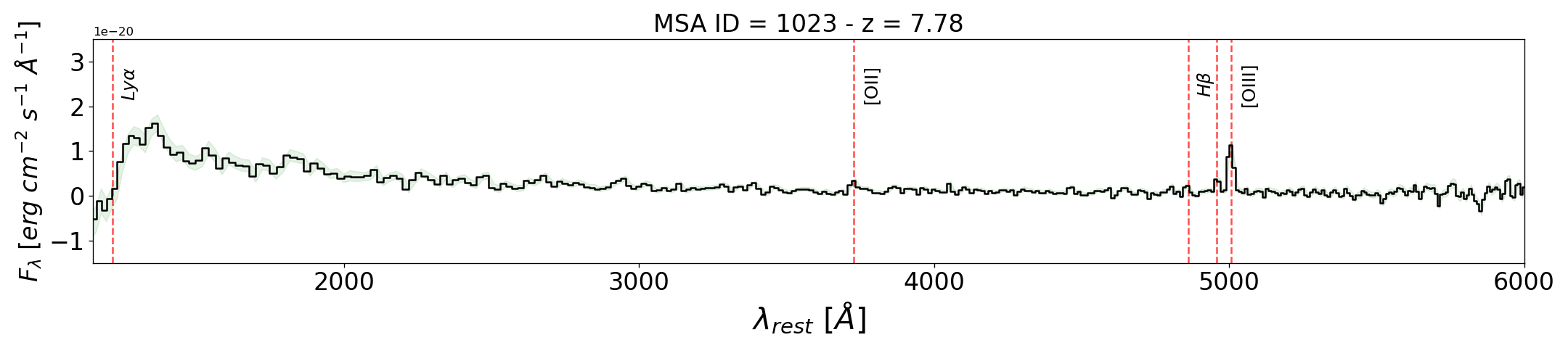}
    \end{subfigure}
    
    \medskip
    
    \begin{subfigure}{\textwidth}
        \includegraphics[width=\linewidth]{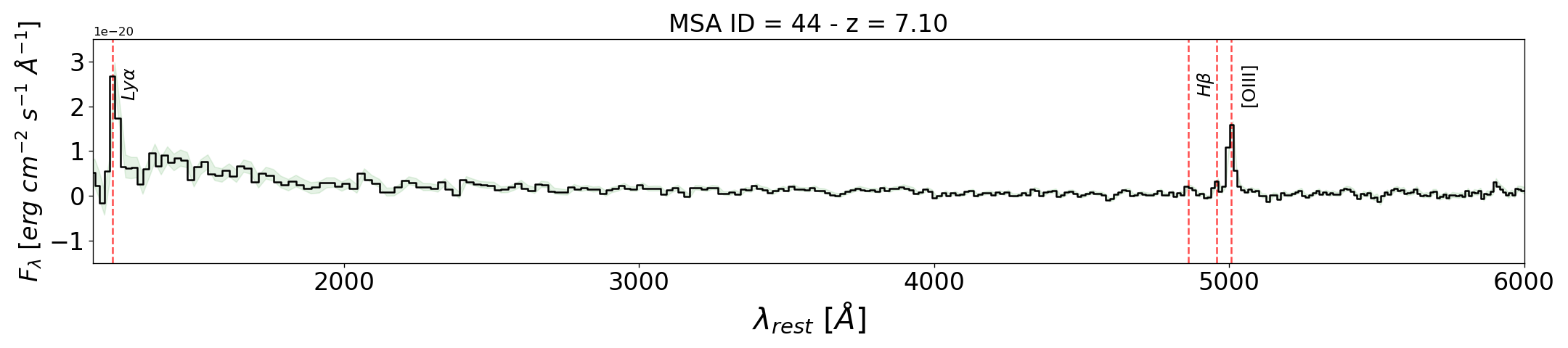}
    \end{subfigure}

    \medskip
    
    \begin{subfigure}{\textwidth}
        \includegraphics[width=\linewidth]{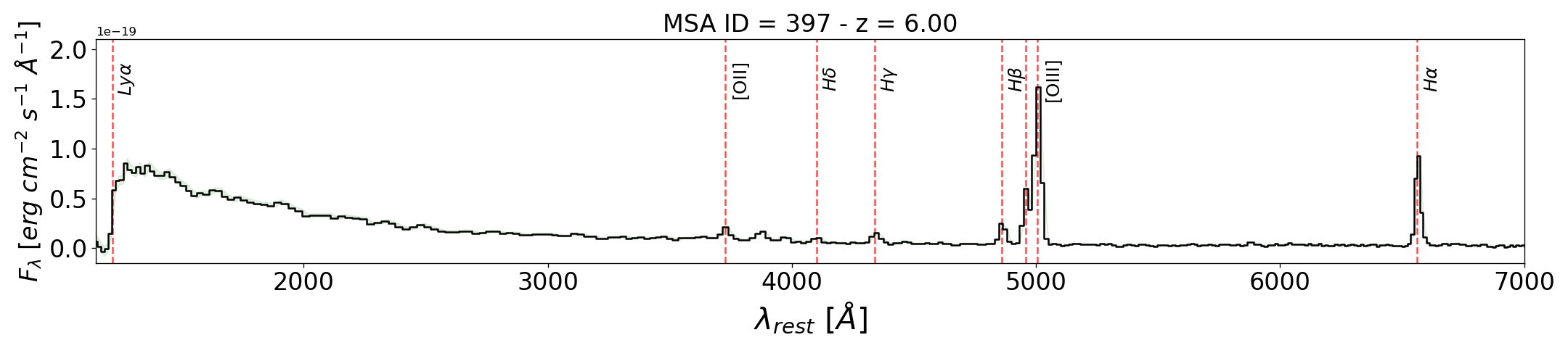}
    \end{subfigure}

    \medskip
    
    \begin{subfigure}{\textwidth}
        \includegraphics[width=\linewidth]{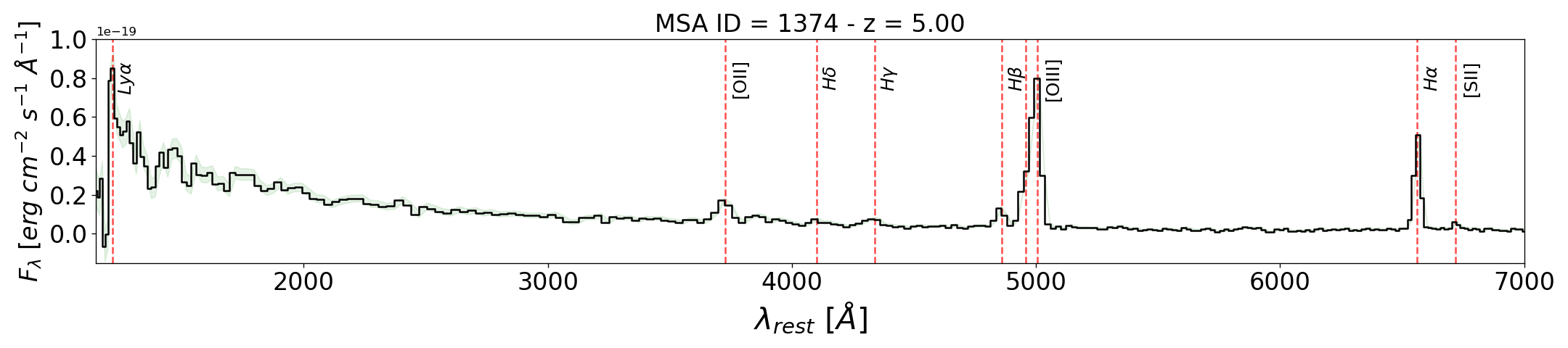}
    \end{subfigure}

    \medskip
    
    \begin{subfigure}{\textwidth}
        \includegraphics[width=\linewidth]{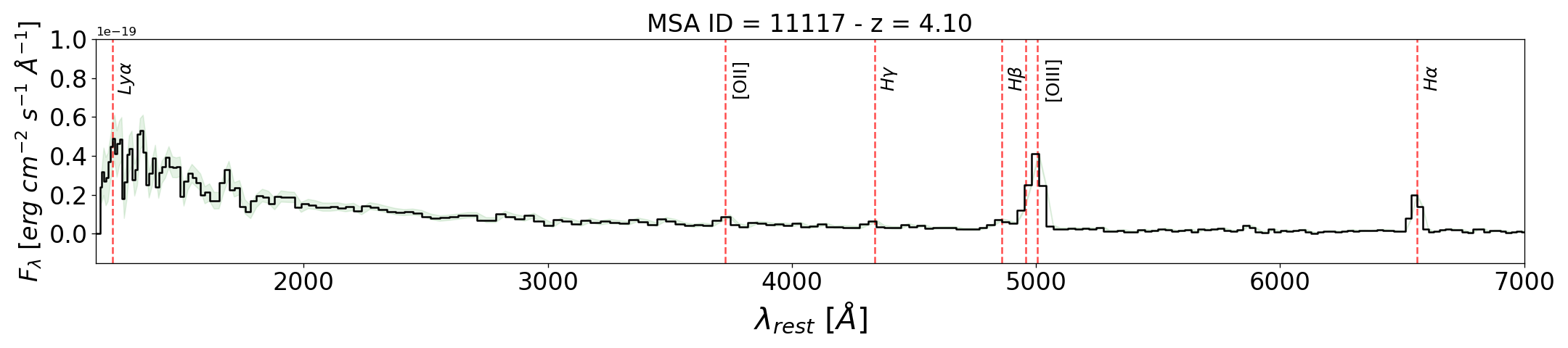}
    \end{subfigure}
    
    \caption{1D spectra examples from the parent sample. Back solid line and green shaded region represent the flux and associated error respectively. For each galaxy, we report the rest-frame emission line features used for the spectroscopic redshift identification. The position of the \lya\ line is also highlighted. MSA ID = 44 and 1374 are \lya\ emitters with S/N > 3, while MSA ID = 1023, 397, and 11117 have no \lya\ in emission.}
    \label{fig:z_vetting}
\end{figure*}

\begin{figure}[h!]
\centering
\includegraphics[trim={0.1cm 0.1cm 0.1cm 0.1cm},clip,width=\linewidth]{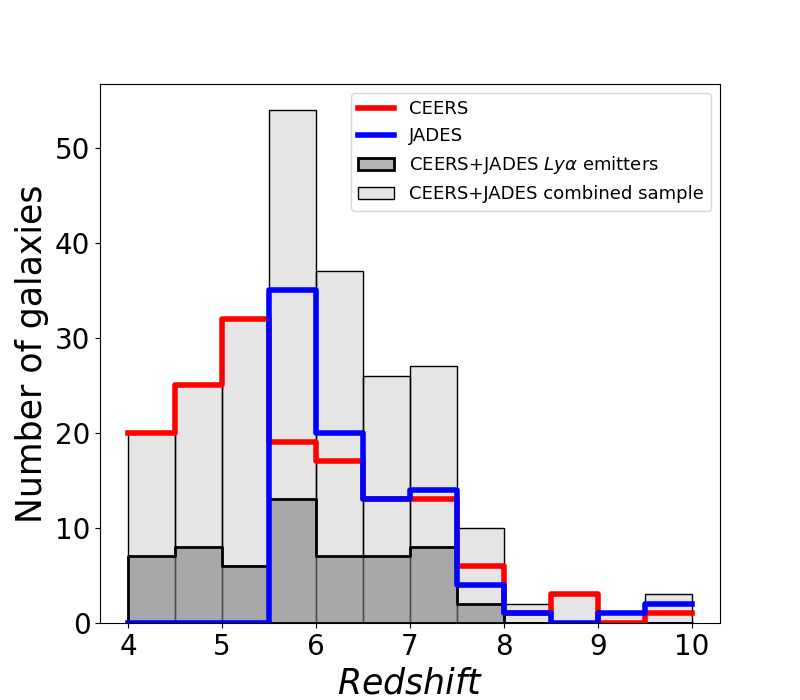}
\caption{Redshift distribution of the 150 sources (AGNs included) identified in CEERS and of the 91 galaxies presented in \cite{Jones2023} from the JADES survey. \lya\ emitting galaxies with S/N > 3 from the combined sample are reported in dark grey.} \label{fig:RedshiftHisto}
\end{figure}

\begin{figure*}[ht!]
\begin{minipage}{\textwidth}
\centering
\includegraphics[width=\linewidth]{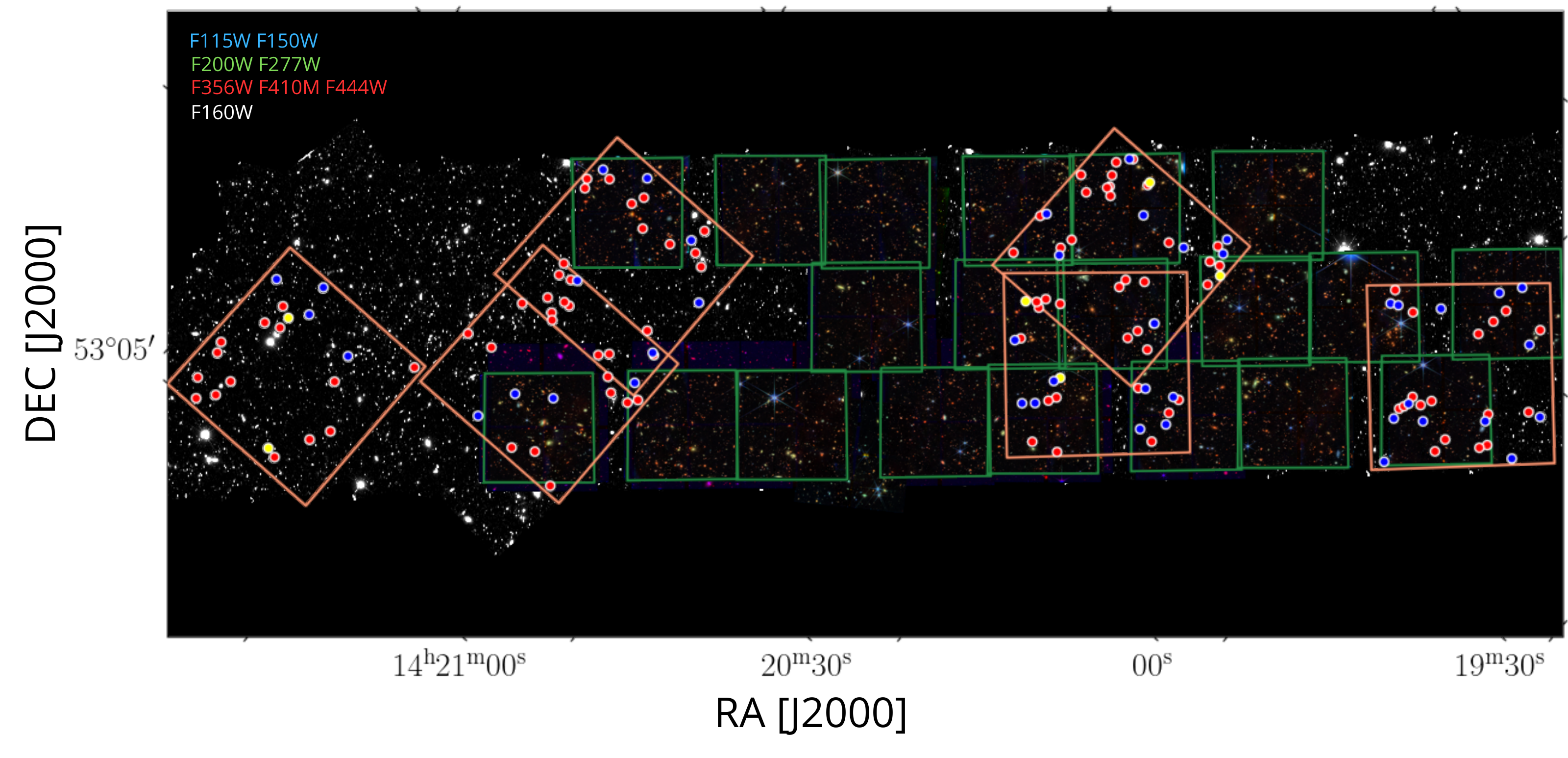}
\end{minipage}
\caption{Spatial distribution of the 150 sources (AGNs included) spectroscopially identified in CEERS. Blue (red) dots are galaxies that do (do not) show \lya\ emission with S/N > 3. The AGNs are represented by the yellow dots. 
Green and orange squares represent the pointings for NIRCam and NIRSpec, respectively. The \hst\ and \jwst\ images of the field are shown in the background. 
} \label{fig:pointing}
\end{figure*}

For our purposes, we needed to assemble the largest possible sample of high redshift sources with secure spectroscopic redshifts, whose spectra contain information about the \lya\ line within the observed wavelength range. The range covered by the PRISM/CLEAR configuration  sets a lower limit of $z \sim$ 4 on the range where we could probe \lya\ emission. 
We therefore selected all the sources with a photometric redshift higher than 3 (to allow for even large photometric redshift uncertainties), and visually examined the 2D and 1D spectra simultaneously in order to derive a spectroscopic redshift. This was done by searching for the relatively bright optical lines (e.g., \oiiidoub, \oiidoub, Balmer lines, \siidoublet, and \siiinew) and the Ly-break feature, as a whole set. We report some examples in Fig.~\ref{fig:z_vetting}, covering the redshift range probed. We determined spectroscopic redshifts for281 galaxies (82\% success rate), of which 150 are at $z >4.1$ and include \lya\ line spectral information in the observed wavelength range. 
In the following, we will refer to the latter as our parent sample. We acknowledge that the visual redshift identification process we employed may have limited precision.
The spectroscopic redshift identified were checked independently by many team members and no dubious cases were reported. In the case of galaxies in our parent sample with previously published spectroscopic redshifts, we cross-checked our estimates to ensure their consistency with values reported in the literature. Specifically there are 27 galaxies in common with \cite{Chen2023}, 10 with \cite{Davis2023}, 7 with \cite{Harikane2023b}, 1 with \cite{Jung2023}, 2 with \cite{Kocevski2023}, 45 with \cite{Mascia2023_CEERS}, 118 with \cite{Nakajima2023}, 15 with \cite{Nakane2023}, and 10 with \cite{Tang2023}. In all cases, our spectroscopic redshifts are consistent with those reported in the literature, up to the second decimal place. Note that some CEERS lyman-break galaxies (LBGs) might not not be included in our sample because of the initial photometric redshift selection adopted. Arrabal Haro et al. (in prep.) will present the complete catalog of spectroscopically confirmed sources in CEERS.\\
In Fig.~\ref{fig:RedshiftHisto} and in Fig.~\ref{fig:pointing} we present the redshift and spatial distribution respectively of all the CEERS galaxies in the parent sample (150 sources). In Table~\ref{tab:summary_data}, we report our spectroscopic redshift measurements for the \lya\ emitting galaxies subset (50 galaxies with a positive \lya\ flux detection, see Sect.~\ref{sec:LyAmodel}), together with spectroscopic and physical properties, derived as detailed in the following sections.

\subsection{CEERS AGN identification} \label{sec:AGN}
\jwst\ has identified an unexpected number of AGNs at high redshift \citep[e.g.,][]{Harikane2023b, Maiolino2023, Matthee2023}. From our follow up analysis we need to exclude possible AGNs from our parent sample, since the mechanisms that allow \lya\ to escape from an AGN are different than in galaxies and the presence of even a few such objects might bias the measurement of the line visibility statistics. In addition, since we will also derive the physical properties of our sources through standard SED fitting, a treatment of AGN would require ad hoc templates which we do not include in our tool \citep[\textsc{zphot}][see Sect.~\ref{sec:SED_fitting} ]{fontana00}. \\ 
To exclude AGNs from the parent sample, we first visually examined all spectra to search for any high ionization emission lines (such as \civmed\ ) in the NIRSpec PRISM/CLEAR configuration. Whenever available we also inspected the NIRSpec medium-resolution ($R \approx 1000$) grating (G140M/F100LP, G235M/F170LP and G395M/F290LP) observation for the same sources, to search for typical broad optical emission lines, which would be difficult to identify using the PRISM, because of the low resolution. We are limited in this procedure, since only 55 sources in the parent sample have a grating observation as well. We identified 5 AGNs following the described procedure, i.e., source MSA ID = 2782 which shows broad H$\alpha$ \citep{Kocevski2023} and bright \civmed\ ; MSA ID = 746 which has broad H$\alpha$ \citep{Kocevski2023, Harikane2023b}; MSA ID = 1244 which has broad H$\alpha$ \citep{Harikane2023b} and bright \ciii; MSA ID = 80457 a new "candidate" AGN which reveals broad H$\alpha$; MSA ID = 82294 which shows bright \civmed.
We also added MSA ID = 1665 (not flagged from our visual inspection) to the AGN list, given the high Akaike Information Criterion (AIC) score \citep[see][for more details]{Harikane2023b}. \\
We finally checked the X-ray emission from the sources in our parent sample using the AEGIS-X Deep (AEGIS-XD) survey's catalog \citep{Nandra2015}. Due to the relatively shallow flux limit probed by these observations, we do not find any X-ray sources that match with our parent sample. 
In conclusion, from the 150 galaxies in the parent sample we find 6 AGNs in total that are excluded from any further analysis.
%
\subsection{JADES data} \label{sec:otherJWST}
We include in our study  all data presented in \cite{Jones2023} which were observed with the same NIRSpec PRISM/CLEAR configuration as the CEERS sources. These data come from the \jwst\ Advance Deep Extragalactic Survey \citep[JADES;][]{Bunker2020, Bunker2023, Eisenstein2023} observing the GOODS \citep[The Great Observatories Origins Deep Survey;][]{Giavalisco2004} north and south fields.\\
\cite{Jones2023} present IDs, coordinates, spectroscopic redshifts, and spectroscopic properties of GOODS-South sources. In total they report 15 galaxies with a detected \lya\ emission in the NIRSpec PRISM/CLEAR configuration and 76 non emitters. \cite{Saxena2023} provides the UV-$\beta$ slopes, and the \lya\ escape fraction measurements for 11 of the above emitters, while \cite{Rieke2023} presents catalogs with photometry and half-light radii information. 
To exclude possible AGNs from this sample, we checked the flags given by \cite{Luo2017} for the GOODS-South field. No matches were found, so we proceed under the assumption that the sample, which we will hence refer to as the JADES sample throughout this study, does not contain any AGNs.\\
In Fig.~\ref{fig:RedshiftHisto} we present the redshift distribution of the entire  sample from CEERS (144 galaxies) and JADES (91 galaxies) which consists of 235 galaxies with homogeneous data from \jwst.

\section{Methods} \label{sec:Method}

\subsection{\lya\ emission line measurements} \label{sec:LyAmodel}
\begin{figure*}[ht!]
\begin{minipage}{0.33\textwidth}
\centering
\includegraphics[width=\linewidth]{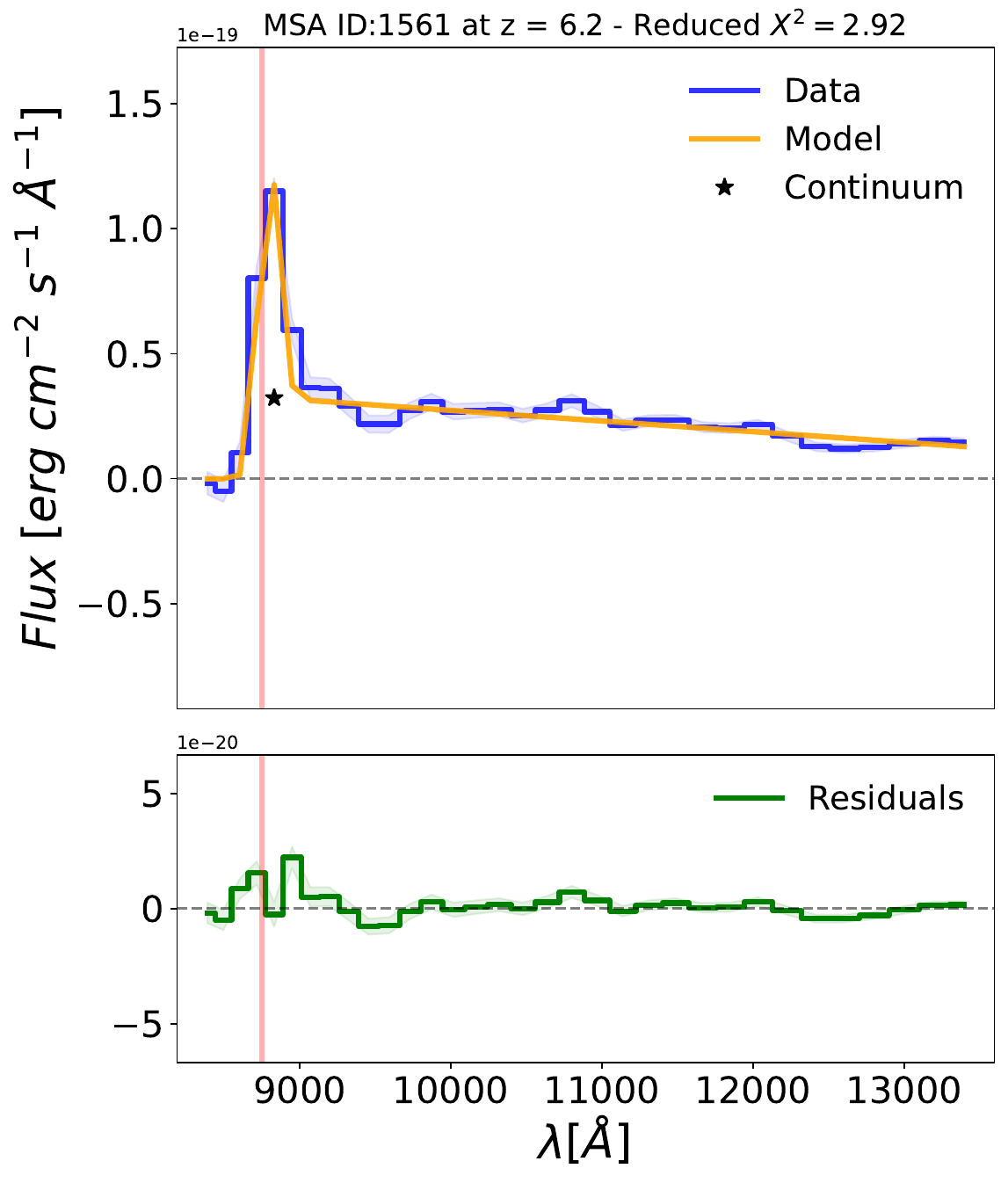}
\end{minipage}
\begin{minipage}{0.33\textwidth}
\centering
\includegraphics[width=\linewidth]{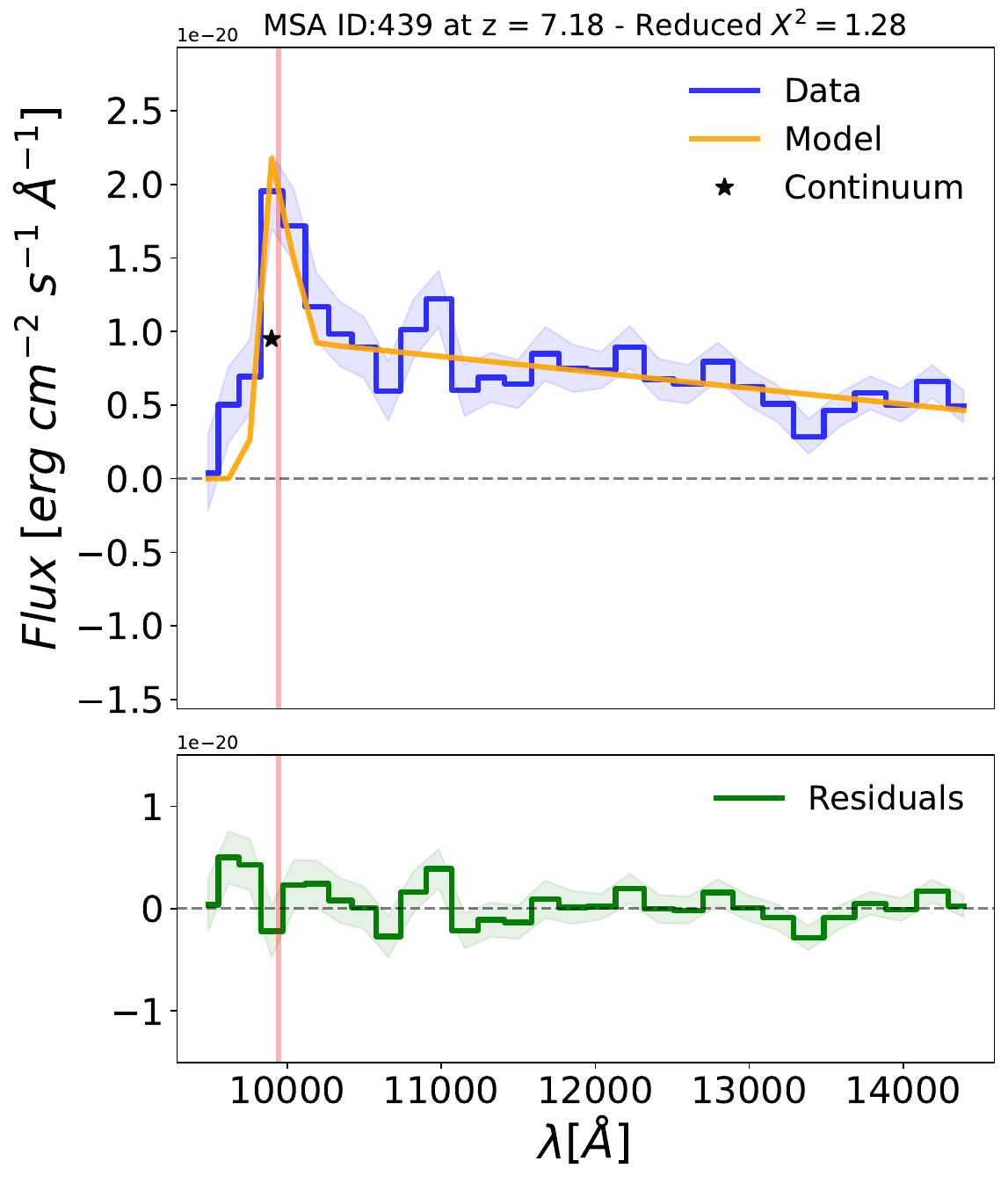}
\end{minipage}
\begin{minipage}{0.33\textwidth}
\centering
\includegraphics[width=\linewidth]{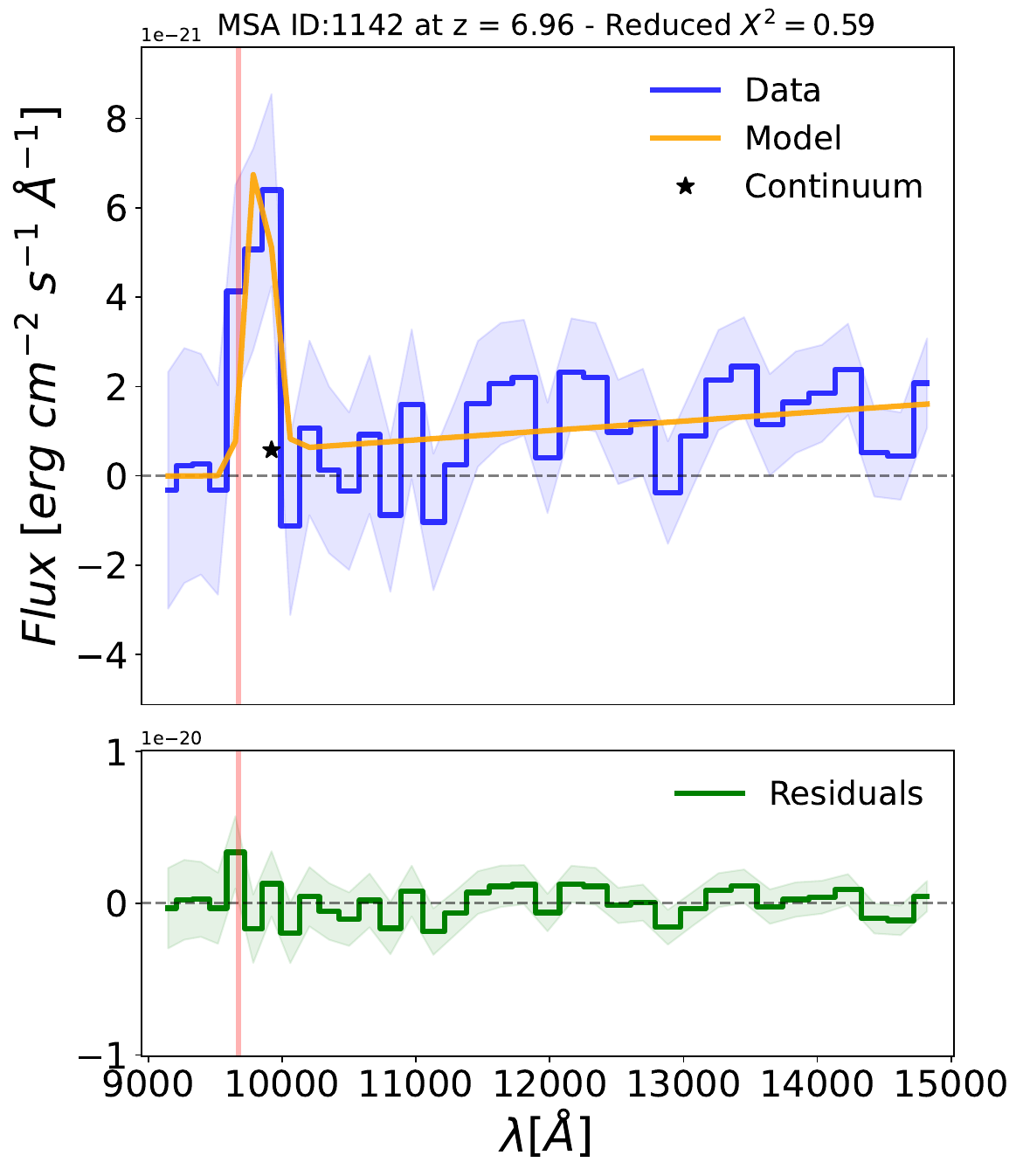}
\end{minipage}
\caption{Few examples of fitted \lya\ emitters in the sample for different values of the emission line S/N. Left: MSA ID = 1561 with S/N = 23; Center: MSA ID = 439 with S/N = 7; Right: MSA ID = 1142 S/N = 3. In the upper panels, the blue solid line and shaded area denote the flux and error measurement as a function of the observed wavelength. The best fit-model is represented by an orange solid line, with the related $\chi^2_{\mathrm{red}}$ reported at the top of each panel. The fitted continuum value at the \lya\ line peak is represented by the star symbol. The vertical red line indicates the \lya\ expected wavelength at the systemic redshift of the source. \text The lower panels show the residuals from the fit. In the appendix (Sect.~\ref{sec:app_figures}) we present the rest of the  \lya\ emitting galaxies.}
\label{fig:EW_fit}
\end{figure*}

\begin{figure}
    \centering
    \includegraphics[width=\linewidth]{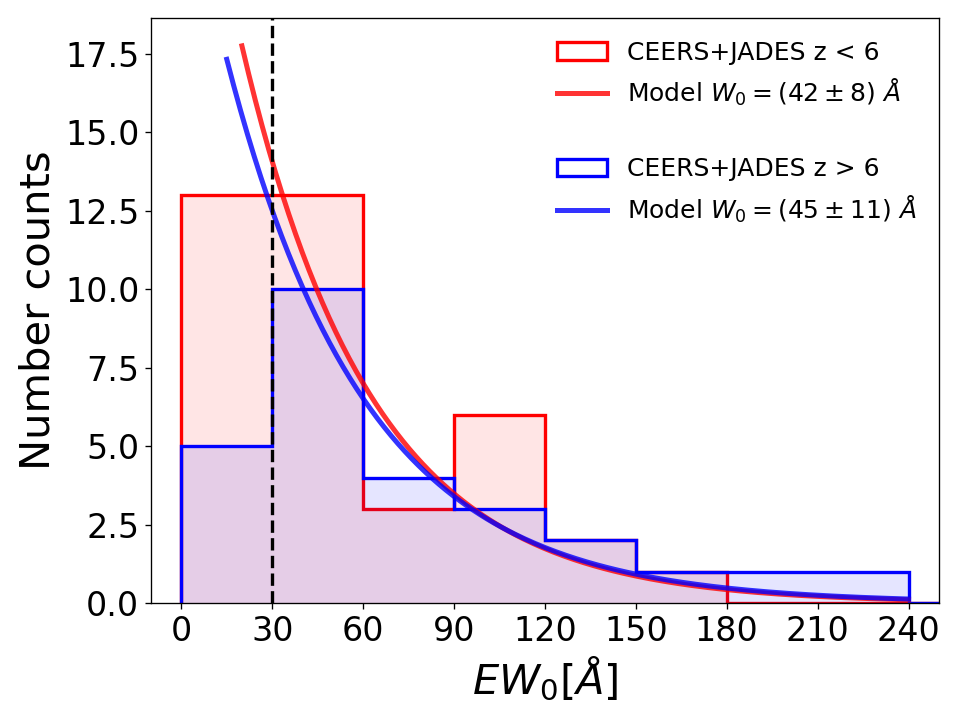}
    \caption{EW$_0$ distribution of the \lya\ emitting galaxies in the combined CEERS + JADES sample. The red (blue) histogram shows the population at $z$ < 6 ($z$ > 6). The red (blue) solid line shows the best fit exponential declining distribution $P(EW) \propto e^{-EW_0/W_0}$ using \textsc{emcee} \citep[][]{Foreman_Mackey2013}. The two populations are sampled in the range -21 < $M_{\mathrm{UV}}$ < -17.}
    \label{fig:ewhisto_zsubsample}
\end{figure}

The method employed to extract the \lya\ line information from our parent sample is similar to the one adopted by  \cite{Jones2023}. In this section, we give a brief description of the main points followed in our analysis. \\ 
For each source we considered the spectroscopic redshift identified (see Sect.~\ref{sec:parent_sample}) to fit the emission region of the spectrum where \lya\ is in the observed frame ($\sim$ 4 -- 5 pixels). The low resolution of the PRISM requires fitting models which account for both the \lya\ emission line and the adjacent continua. The red continuum ($\lambda > 1216$ \AA) was derived by directly fitting a linear function to the data, weighed for the inverse of their flux error from the spectrum. The wavelength range considered for the linear fit spans from 1900 \AA\ rest-frame to the closest pixel red-ward of the \lya\ emission ($\sim$ 3 pixels from the peak), to avoid the possible presence of the \ciii\ emission line. 
The blue continuum ($\lambda < 1216$ \AA) is expected to vary depending on the neutral hydrogen absorption and it is dependent on the redshift of the source. Thus, it was obtained by averaging the flux blue-ward of the emission line. 
We then defined a modified step model, whose flux is equal to the constant blue continuum value where $\lambda < (1+z) \times 1216$ \AA\, and to the red continuum values obtained by the linear fit for $\lambda \geq (1+z) \times 1216$\AA. \\
The low resolution of the PRISM configuration prevents us from characterizing an asymmetric \lya\ profile. We thus decided to fit each line emission with a library of Gaussian profile models, as detailed below. Because of the scattering nature of the \lya\ line, the peak of the emission is often slightly shifted from the systemic value, up to few hundred km/s \citep[e.g.,][]{Steidel2010, Verhamme2018, Marchi2019}, so we fixed the mean of the Gaussian models to the observed emission peak rather than the systemic redshift. The other two parameters of the Gaussian profiles were constrained to be within the following ranges: the FWHM $\in [100, 1500]$ km/s and the peak amplitude $\in [0.01, 10] \times 10^{-18}$ erg s$^{-1}$ cm$^{-2}$ \AA$^{-1}$. \\ 
We then added the modified step model to the library of Gaussians and convolved the result with a Gaussian kernel, whose standard deviation is defined as $\sigma_R (\lambda) [\AA] = 1216(1+z) / 2.355 R(\lambda)$. The convolution is applied to mimic the resolution of the instrument $R(\lambda)$ at the peak of the \lya\ emission in the observed frame. $R(\lambda)$ is provided by the \jwst\ documentation\footnote{\url{https://jwst-docs.stsci.edu/jwst-near-infrared-spectrograph/nirspec-instrumentation/nirspec-dispersers-and-filters}} with the assumption of a source that illuminates the slit uniformly. The resulting models were also re-sampled at the finite set of wavelengths observed in the real spectra. 
As a final step, we employed \textsc{emcee} \citep{Foreman_Mackey2013} to perform a Markov Chain Monte Carlo (MCMC) analysis for each target spectrum, identifying the best-fitting model free parameters. The best model parameters and the integrated \lya\ flux from the intrinsic emission line Gaussian Model are determined through the posterior distributions resulting from the MCMC fitting routine. Uncertainties are calculated based on the 68-th percentile highest posterior density intervals. We derive the signal-to-noise ratio (S/N) from the integrated \lya\ flux.
The rest-frame \lya\ equivalent width (EW$_0$) is computed based on the integrated \lya\ flux, taking into account both the continuum flux determined at the \lya\ line's position from the red continuum fit and the spectroscopic redshift. In Fig.~\ref{fig:EW_fit} we provide few examples of results obtained from the described fitting procedure for a range of emission line S/N. 
\\
The measurement of rest frame \lya\ equivalent width is challenging and careful modeling is needed to recover it, due to the low resolution at the blue end of the NIRSpec/PRISM instrument. To further prove this point, for each source we extracted again the EW$_0$ by directly integrating the continuum-subtracted line profile over the same emission region of the spectrum ($\sim$ 4 -- 5 pixels centered on the emission peak). We consistently obtain lower values, that are on average smaller by 30\% than the reference results from our modeling. This effect was also noted by \cite{Chen2023}, who report potential underestimation from direct integration by as much as 30\%--50\%. This is indeed expected in the case in which a fraction of the \lya\ photons fall on pixels dominated by the continuum and break.\\
We expect the shift of the \lya\ emission line compared to the systemic redshift to be typically below 1 pixel ($\Delta v < 2500$~km~s$^{-1}$ given the PRISM dispersion at these wavelengths). This is indeed true in all cases except for four galaxies which show shifts up to 2 pixels (MSA ID = 1420, 2089, 2168, and 80445). The identified lines could be the unresolved \nvdoub\ instead of \lya\ (although it would be the only sign of AGN emission) or a spurious detection. However, we are aware that there could be residual issues with the wavelength calibration of the spectra, due to the highly variable spectral resolution of the PRISM which have also been observed in other spectra, with discrepancies between red and blue regions. For this reason we decide to keep the four objects in the \lya\ list.\\
In total we find 50 \lya\ emitting galaxies (43 with a S/N > 3 and 7 tentative emitters with 2 < S/N < 3), whose properties are reported in Table~\ref{tab:summary_data}, whereas 94 galaxies have no \lya\ emission. 
In the appendix (see Sect.~\ref{sec:app_figures}) we show all the \lya\ line profiles fitted (including the tentative ones) and in Fig.~\ref{fig:ewhisto_zsubsample} we show the distribution of the measured EW$_0$ for CEERS, together with the published JADES data \citep{Jones2023}.
For all sources (including those where \lya\ is not detected) we also derive  a limit on the lowest EW$_0$ that could be measured given the  spectroscopic redshift ($z$), the red continuum value next to \lya\ ($F_{\lambda}^{\mathrm{cont\ }}$), and the flux error at the observed \lya\ peak of the spectrum ($E(\lambda ^{\lya})$), adopting equation (2) from \cite{Jones2023}
\begin{equation} \label{eq:EW1sigma}
    EW_{\mathrm{0,lim}} = \frac{\sqrt{2 \pi} \ E(\lambda ^{\lya}) \sigma_{R} (\lambda)}{(1+z)F_{\lambda}^{\mathrm{cont\ }}}
\end{equation}
where the numerator stands for the integrated flux of a Gaussian, whose amplitude and standard deviation equal to the flux error at the observed \lya\ peak of the spectrum $E(\lambda ^{Ly \alpha})$ and to the Gaussian kernel $\sigma_{R}$ that accounts for the instrument resolution, respectively.
We find EW$_{\mathrm{0,lim}}$ values from few \AA\ to 155 \AA\, with a median of 11 \AA\ . \\
As a final check, we inspected the grating spectra of all galaxies when available. Of the 29 galaxies at $z$ > 6.98 (where \lya\ emission is in the detectable range of G140M/F100LP), 6 have medium resolution observations. One (MSA ID = 20) is a tentative emitter according to the PRISM measurement, but does not show \lya\ in the grating spectrum. The other five are non-emitters according to the PRISM observations: of these only one (MSA ID = 1027) shows a faint \lya\ line in the medium resolution spectrum, whose EW$_0$ \citep[][]{Larson2022, Tang2023} is below the PRISM detection limit.

\subsection{$M_{\mathrm{UV}}$ and $\beta$ } \label{sec:Muv}
\begin{figure}[t]
\centering
\includegraphics[width=\linewidth]{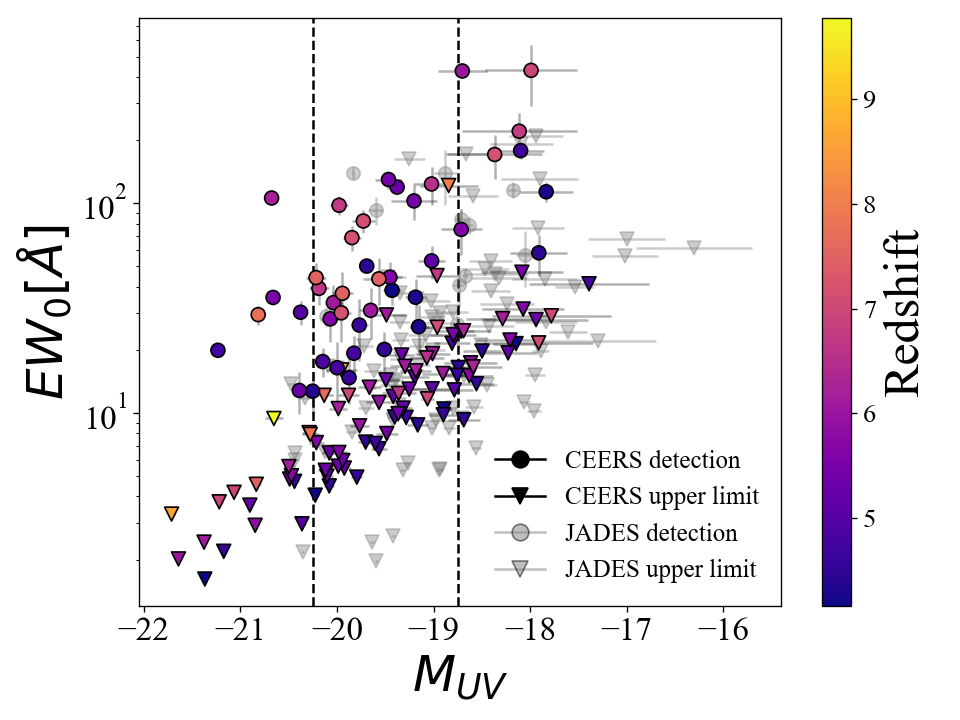}
\caption{The \lya\ EW$_0$ vs $M_{\mathrm{UV}}$ for our sample. Circles denote measured EW$_0$ with S/N>3, while triangles represent galaxies with just $EW_{\mathrm{0,lim}}$ upper limits. CEERS data are color coded by redshift, while JADES galaxies are reported in grey for comparison. The black dashed lines ($M_{\mathrm{UV}}=$ -20.25 and -18.75) divide the sample from the bright and faint ends.} \label{fig:EWvsMuv}
\end{figure}

We calculate the UV absolute magnitudes ($M_{\mathrm{UV}}$) directly from the observed PRISM spectra, after correcting them for the $\times$1.4 average factor already discussed in Sect.~\ref{sec:CEERS} to match the photometry. We used this average factor for the whole sample, since for 45 galaxies in the sample we do not have individual NIRCam photometry. 
We measured the median flux density and error within the rest-frame 1400--1500 \AA\ range to compute the UV absolute magnitudes (to be consistent with the values reported in \cite{Jones2023} for the JADES data). In Fig.~\ref{fig:EWvsMuv} we report the relation between the measured EW$_0$ and $M_{\mathrm{UV}}$. As expected for the fainter galaxies, we are only able to measure the \lya\ emission for higher EW$_0$ due to the flux limited nature of the spectroscopic observations. The data from  \cite{Jones2023} are reported in grey for a direct comparison. \\
The UV-$\beta$ slopes were measured as in \cite{Mascia2023_CEERS} and \cite{Calabro2021} by employing all the photometric bands whose bandwidth ranges fall between 1216 -- 3000 \AA\ rest-frame. The former limit is set to exclude the Lyman-break. We then fitted the available photometric bands amongst \hst\ F814W, F125W, F140W, F160W and \jwst-NIRCam F115W, F150W or F200W data (see Sect.~\ref{sec:Data_and_sample_selection}), depending on the exact redshift of the sources and the accessibility of data. Notably, 16 out of the 56 emitters in the CEERS parent sample lack \jwst-NIRCam data, therefore we are limited to consider the \hst\ bands for fitting these sources.
We fitted a single power-law of the form f($\lambda$) $\propto \lambda^\beta$ \citep{Calzetti1994, Meurer1999}. Typically we employed between 2 and 4 bands. For each galaxy, the measured $\beta$ and its uncertainty are obtained as the mean and standard deviation of a $n = 1000$ Monte Carlo approach, through which fluxes in each band are extracted according to their error.\\
The $M_{\mathrm{UV}}$ and UV-$\beta$ slopes for our \lya\ emitting galaxies are reported in Table~\ref{tab:summary_data}. For the JADES subset, \cite{Jones2023} provide $M_{\mathrm{UV}}$ for all galaxies, while \cite{Saxena2023} provide UV-$\beta$ slopes for 11 out of 15 emitters.

\begin{table*}
\caption{Physical and spectroscopic properties of the \lya\ emitting galaxies in the CEERS sample. Galaxies associated to a detected \lya\ emission with S/N>3 are listed first. Tentative emitters with S/N<3 are reported at the bottom of the table.}\label{tab:summary_data}
\tiny{
$$ %
\begin{array}{lccccccccc}
\hline \hline
\noalign{\smallskip}
\text{MSA ID} & \text{RA} \ [\text{deg}] & \text{DEC} \ [\text{deg}] & z_{\mathrm{spec}} &\mathrm{EW}_0(\lya) \ [\AA] & f_{\mathrm{esc}}^{\lya} & f_{\mathrm{esc}}^{\mathrm{LyC, pred}} & M_{\mathrm{UV}} \ [\text{mag}] & \beta & r_{e} \ [\text{kpc}]\\

\noalign{\smallskip}
 \hline
 \noalign{\smallskip}
686 & 215.150862 & 52.989562 & 7.75 & 29.5 \pm 3.1 & 0.164 \pm 0.018 & 0.59 [ 0.17 , 1.00 ] & -20.82 \pm 0.05 & -3.7 \pm 0.9 & 0.25 \pm 0.06 \\
80445^* & 214.843115 & 52.747886 & 7.51 & 44 \pm 11 & 0.38 \pm 0.14 & 0.08 [ 0.03 , 0.12 ] & -19.57 \pm 0.16 & -1.98 \pm 0.04 & 0.74 \pm 0.10 \\
80372^* & 214.927798 & 52.850003 & 7.49 & 37 \pm 9 & 0.040 \pm 0.010 & > 0.21 & -19.94 \pm 0.11 & -1.87 \pm 0.05 & <0.13 \\
80432^* & 214.812056 & 52.746747 & 7.48 & 44 \pm 8 & 0.110 \pm 0.034 & 0.34 [ 0.13 , 0.56 ] & -20.22 \pm 0.10 & -2.04 \pm 0.03 & 0.15 \pm 0.02 \\
80374^* & 214.898074 & 52.824895 & 7.18 & 171 \pm 41 & > 0.39 & 0.23 [ 0.08 , 0.38 ] & -18.37 \pm 0.49 & -2.25 \pm 0.01 & 0.26 \pm 0.12 \\
439^* & 214.825364 & 52.863065 & 7.18 & 69 \pm 9 & 0.42 \pm 0.15 & 0.16 [ 0.06 , 0.25 ] & -19.85 \pm 0.10 & -2.60 \pm 0.15 & 0.15 \pm 0.03 \\
498^* & 214.813045 & 52.834249 & 7.18 & 30 \pm 8 & 0.146 \pm 0.041 & 0.02 [ 0.01 , 0.02 ] & -19.96 \pm 0.10 & -2.50 \pm 0.07 & 0.30 \pm 0.02 \\
44^* & 215.001115 & 53.011269 & 7.10 & 82 \pm 11 & 0.47 \pm 0.07 & > 0.39 & -19.73 \pm 0.12 & -2.58 \pm 0.12 & < 0.13\\
1142 & 215.060716 & 52.958708 & 6.96 & 430 \pm 140 & 0.36 \pm 0.12 & 0.09 [ 0.03 , 0.22 ] & -17.99 \pm 0.48 & -1.56 \pm 0.44 & 0.46 \pm 0.12 \\
80925^* & 214.948680 & 52.853273 & 6.76 & 221 \pm 49 & 0.143 \pm 0.032 & 0.05 [ 0.02 , 0.08 ] & -18.1 \pm 0.6 & -1.97 \pm 0.05 & 0.65 \pm 0.12 \\
81049^* & 214.789822 & 52.730789 & 6.74 & 98 \pm 10 & 0.196 \pm 0.021 & 0.22 [ 0.08 , 0.36 ] & -19.98 \pm 0.09 & -2.08 \pm 0.04 & 0.32 \pm 0.05 \\
1414^* & 215.128029 & 52.984936 & 6.68 & 39 \pm 7 & 0.047 \pm 0.008 & 0.03 [ 0.02 , 0.05 ] & -20.19 \pm 0.10 & -1.88 \pm 0.03 & 0.235 \pm 0.018 \\
80596^* & 214.771865 & 52.778189 & 6.54 & 124 \pm 26 & 0.50 \pm 0.10 & 0.08 [ 0.03 , 0.13 ] & -19.02 \pm 0.20 & -2.00 \pm 0.05 & 0.74 \pm 0.19 \\
1561 & 215.166097 & 53.070755 & 6.20 & 106.0 \pm 4.6 & 0.478 \pm 0.026 & 0.38 [ 0.14 , 0.66 ] & -20.68 \pm 0.06 & -3.5 \pm 0.7 & 0.48 \pm 0.07 \\
355^* & 214.806482 & 52.878827 & 6.11 & 34 \pm 7 & 0.084 \pm 0.018 & 0.05 [ 0.02 , 0.09 ] & -20.04 \pm 0.08 & -2.05 \pm 0.09 & 0.46 \pm 0.06 \\
603^* & 214.867247 & 52.836737 & 6.06 & 31 \pm 8 & 0.094 \pm 0.026 & 0.01 [ 0.01 , 0.02 ] & -19.65 \pm 0.12 & -2.17 \pm 0.10 & 2.05 \pm 0.06 \\
476^* & 214.805561 & 52.836345 & 6.01 & 428 \pm 33 & 0.64 \pm 0.08 & > 0.07 & -18.70 \pm 0.25 & -2.05 \pm 0.09 & < 0.15\\
80916^* & 214.891630 & 52.815943 & 5.67 & 44 \pm 8 & 0.098 \pm 0.018 & > 0.33 & -19.45 \pm 0.12 & -2.06 \pm 0.04 & < 0.15\\
323^* & 214.872556 & 52.875949 & 5.67 & 22 \pm 7 & 0.018 \pm 0.006 & > 0.50 & -19.16 \pm 0.20 & -2.37 \pm 0.08 & < 0.15 \\ 
80944^* & 214.917041 & 52.817472 & 5.66 & 75 \pm 19 & 0.59 \pm 0.20 & 0.14 [ 0.05 , 0.22 ] & -18.71 \pm 0.21 & -2.41 \pm 0.09 & 0.30 \pm 0.08 \\
2168 & 215.152602 & 53.057062 & 5.66 & 28 \pm 6 & 0.073 \pm 0.019 & 0.02 [ 0.01 , 0.02 ] & -20.07 \pm 0.09 & -1.58 \pm 0.41 & 2.7 \pm 0.6 \\
1334 & 214.768356 & 52.717641 & 5.50 & 35.6 \pm 2.8 & 0.0236 \pm 0.0019 & > 0.19 & -20.662 \pm 0.041 & -1.4 \pm 0.6 & < 0.14 \\
80573^* & 214.773924 & 52.780599 & 5.44 & 130 \pm 12 & 0.137 \pm 0.022 & 0.32 [ 0.12 , 0.54 ] & -19.47 \pm 0.14 & -2.34 \pm 0.09 & 0.240 \pm 0.023 \\
81026^* & 214.809841 & 52.754218 & 5.43 & 103 \pm 20 & 0.169 \pm 0.034 & 0.23 [ 0.07 , 0.38 ] & -19.20 \pm 0.24 & -2.50 \pm 0.38 & 0.52 \pm 0.11 \\
2123 & 214.824580 & 52.845726 & 5.28 & 12.8 \pm 2.9 & 0.041 \pm 0.009 & 0.03 [ 0.01 , 0.06 ] & -20.39 \pm 0.05 & -1.79 \pm 0.11 & 1.07 \pm 0.12 \\
82069^* & 214.730322 & 52.754972 & 5.25 & 120 \pm 5 & 0.82 \pm 0.19 & > 0.12 & -19.38 \pm 0.07 & -2.24 \pm 0.04 & < 0.16 \\
82171^* & 214.741550 & 52.736014 & 5.15 & 53 \pm 9 & 0.61 \pm 0.14 & 0.16 [ 0.06 , 0.27 ] & -19.02 \pm 0.14 & -1.88 \pm 0.28 & 0.195 \pm 0.045 \\
1374^* & 214.943911 & 52.850042 & 5.01 & 30.2 \pm 3.8 & 0.066 \pm 0.008 & 0.05 [ 0.02 , 0.09 ] & -20.38 \pm 0.06 & -2.02 \pm 0.02 & 0.610 \pm 0.018 \\
2140 & 214.796009 & 52.715878 & 4.89 & 17.6 \pm 2.8 & 0.134 \pm 0.021 & > 0.12 & -20.15 \pm 0.05 & -2.3 \pm 0.7 & < 0.12 \\
2000^* & 214.859629 & 52.888130 & 4.81 & 16 \pm 5 & > 0.066 & 0.11 [ 0.04 , 0.19 ]  & -20.00 \pm 0.06 & -2.10 \pm 0.04 & 0.625 \pm 0.043 \\
1565 & 215.057502 & 52.993715 & 4.79 & 19.3 \pm 4.0 & 0.27 \pm 0.06 & 0.03 [ 0.02 , 0.04 ] & -19.83 \pm 0.07 & -2.60 \pm 0.18 & 1.46 \pm 0.55 \\
1449^* & 215.080005 & 52.956786 & 4.76 & 19.9 \pm 1.3 & 0.0644 \pm 0.0043 & 0.12 [ 0.04 , 0.20 ] & -21.233 \pm 0.023 & -2.10 \pm 0.01 & 0.477 \pm 0.011 \\
82372^* & 214.924614 & 52.868844 & 4.73 & 178 \pm 16 & - & > 0.07 & -18.10 \pm 0.24 & -2.04 \pm 0.08 & < 0.17\\
3584^*  & 214.988752 & 52.998044 & 4.64 & 14.7 \pm 3.7 & 0.021 \pm 0.005 & 0.20 [ 0.08 , 0.32 ] & -19.88 \pm 0.06 & -2.14 \pm 0.02 & 0.283 \pm 0.014 \\
2089 & 214.999175 & 52.973301 & 4.64 & 26 \pm 9 & 0.0119 \pm 0.0039 & 0.02 [ 0.01 , 0.03 ] & -19.77 \pm 0.09 & -1.56 \pm 0.19 & 1.9 \pm 0.5 \\
1767 & 215.172758 & 53.035788 & 4.55 & 20.1 \pm 4.4 & 0.058 \pm 0.013 & 0.04 [ 0.01 , 0.07 ] & -19.51 \pm 0.08 & -2.55 \pm 0.21 & 2.4 \pm 1.2 \\
1400^* & 215.116105 & 52.974184 & 4.49 & 50.2 \pm 3.3 & 0.419 \pm 0.027 & 0.25 [ 0.09 , 0.44 ] & -19.692 \pm 0.038 & -2.21 \pm 0.05 & 0.253 \pm 0.014 \\
14777 & 215.022828 & 52.957766 & 4.47 & 58 \pm 13 & 0.55 \pm 0.17 & 0.13 [ 0.03 , 0.23 ] & -17.91 \pm 0.30 & -2.64 \pm 0.44 & 0.8 \pm 0.5 \\
1651 & 215.169217 & 53.054766 & 4.39 & 12.7 \pm 3.0 & 0.059 \pm 0.014 & 0.22 [ 0.06 , 0.39 ] & -20.253 \pm 0.042 & -2.37 \pm 0.19 & 0.62 \pm 0.27 \\
82043^* & 214.719986 & 52.750255 & 4.32 & 36 \pm 8 & 0.087 \pm 0.019 & 0.45 [ 0.19 , 0.74 ] & -19.19 \pm 0.11 & -2.28 \pm 0.16 & 0.239 \pm 0.031 \\
83779^* & 214.821417 & 52.754838 & 4.30 & 26 \pm 7 & 0.091 \pm 0.025 & 0.10 [ 0.03 , 0.16 ] & -19.15 \pm 0.10 & -2.23 \pm 0.16 & 1.07 \pm 0.15 \\
83502^* & 214.905847 & 52.811906 & 4.25 & 114 \pm 12 & 0.419 \pm 0.048 & 0.10 [ 0.04 , 0.15 ] & -17.83 \pm 0.28 & -1.81 \pm 0.26 & 0.277 \pm 0.048 \\
12221 & 214.758001 & 52.766495 & 4.17 & 39 \pm 6 & > 0.46 & 0.25 [ 0.09 , 0.43 ] & -19.43 \pm 0.10 & -2.15 \pm 0.45 & 0.36 \pm 0.14 \\
 \noalign{\smallskip}
\hline
\noalign{\smallskip}
\multicolumn{10}{c}{\text{\textbf{Tentative emitters with 2 < S/N < 3}}} \\
\noalign{\smallskip}
\hline 
\noalign{\smallskip}
20^* & 214.830685 & 52.887771 & 7.77 & 510 \pm 180 & 0.092 \pm 0.034 & 0.03 [ 0.01 , 0.06 ] & -17.2 \pm 1.3 & -1.25 \pm 0.31 & 0.200 \pm 0.030 \\
80239^* & 214.896054 & 52.869853 & 7.49 & 64 \pm 31 & 0.12 \pm 0.06 & 0.04 [ 0.02 , 0.07 ]& -19.18 \pm 0.20 & -1.35 \pm 0.13 & 0.61 \pm 0.15 \\
829^* & 214.861594 & 52.876159 & 7.16 & 41 \pm 18 & >0.21 & 0.08 [ 0.03 , 0.13 ] & -19.56 \pm 0.16 & -2.05 \pm 0.20 & 0.35 \pm 0.07 \\
535^* & 214.859175 & 52.853587 & 7.13 & 33 \pm 12 & >0.21 & 0.05 [ 0.02 , 0.08 ] & -19.73 \pm 0.13 & -2.08 \pm 0.04 & 0.46 \pm 0.08 \\
83764^* & 214.815305 & 52.755600 & 5.42 & 57 \pm 21 & 0.19 \pm 0.07 & 0.09 [ 0.03 , 0.14 ] & -18.72 \pm 0.33 & -2.22 \pm 0.14 & 0.3 \pm 0.1 \\
1420 & 215.092864 & 52.960698 & 5.29 & 46 \pm 16 & 0.0040 \pm 0.0014 & 0.02 [ 0.01 , 0.02 ] & -19.17 \pm 0.17 & -1.8 \pm 0.6 & 2.9 \pm 1.1 \\
80072^* & 214.890850 & 52.813941 & 5.28 & 107 \pm 49 & 0.19 \pm 0.09 & 0.26 [ 0.08 , 0.46 ] & -17.51 \pm 0.68 & -2.2 \pm 0.8 & 0.31 \pm 0.12 \\

\noalign{\smallskip}
\hline
\hline
\end{array}
$$
\begin{tablenotes}
 \item \small $^*$: NIRCam photometry available.
 \end{tablenotes}}
\end{table*}

\subsection{Measurements of physical properties} \label{sec:SED_fitting}
Physical properties were derived following the method described in \cite{Santini2022CANDELS}, fixing the redshift of each source to the spectroscopic value. We measured the stellar mass (Mass), star formation rates (SFR) and dust reddening ($E(B-V)$) 
by fitting synthetic stellar templates with the SED fitting code \textsc{zphot} \citep{fontana00}. For the fit we used the seven-band NIRCam photometry of the sources combined to the \hst\ photometry, if both available. Otherwise for the 16 emitters that do not have NIRCam photometry (see Sect.~\ref{sec:Data_and_sample_selection}) only \hst\ photometry was used. In this case, the derived stellar masses have higher uncertainties and are slightly biased to higher values (we defer a more exhaustive discussion of this issue to to Calabrò et al. in preparation).  
We fitted the observed photometry (see Sect.~\ref{sec:Data_and_sample_selection}), adopting \cite{Bruzual2003} models, the \cite{Chabrier2003} IMF and assuming delayed star formation histories (SFH($t$) $\propto (t^2/\tau) \cdot \exp(-t/\tau)$), with $\tau$ ranging from 100 Myr to 7 Gyr. The age could vary between 10 Myr and the age of the Universe at each galaxy redshift, while metallicity assumed values of 0.02, 0.2, 1, or 2.5 times solar metallicity. For the dust extinction, we used the \cite{Calzetti2000} law with $E(B-V)$ ranging from 0 to 1.1. Nebular emission was included following the prescriptions of \cite{Castellano2014} and \cite{Schaerer2009} and assuming a null LyC escape fraction.\\
We used the same method to measure stellar mass, SFR, and $E(B-V)$ from the 15 emitters from the \cite{Jones2023} sample. 
We employed the published \cite{Rieke2023} photometric catalog for this purpose. Total fluxes in all bands were obtained multiplying the reported fluxes and uncertainties in fixed circular apertures of 0.15$''$ (corresponding to $\sim$2 FWHM in F444W), computed on PSF-matched images, with a scaling factor given by the ratio of the total (Kron) flux and the aperture flux in the detection band \citep[see e.g.][]{Merlin2022}.

\subsection{Optical line flux measurements and dust correction} \label{sec:balmer}
We measured the total flux of each detected H$\alpha$, H$\beta$, \oiiidoub\ and \oiidoub\ with a single Gaussian fit. Given the resolution of the instrument the \oiidoub\ is unresolved and we treat it as a single feature. For this part of our analysis we followed the same procedure as described in \cite{Mascia2023_CEERS}, using \textsc{Mpfit} code\footnote{\url{http://purl.com/net/mpfit}} \citep{markwardt2009}. For H$\beta$ we also derived rest-frame equivalent width values.\\
Limited by the wavelength coverage of the NIRSpec PRISM configuration, a direct dust correction estimate is not available for each target by considering only the H$\alpha/$H$\beta$ Balmer ratio. For this reason, we adopted dust attenuation based on the Balmer decrement when H$\alpha/$H$\beta$ is observed and the gas reddening values provided by the SED fitting otherwise (see Sect.~\ref{sec:SED_fitting}). 
As reported in \cite{Osterbrock2006}, we consider the intrinsic H$\alpha/$H$\beta$ ratios to be 2.86 by assuming case B recombination with a density $n_e$ = 100 cm$^{-3}$ and temperature $T_e$ = 10,000 K. 
Then we corrected H$\alpha$ and H$\beta$ fluxes by dust extinction using the reddening curve values provided by \cite{Calzetti2000}. \\
We also employed the O32 line ratios in \cite{Mascia2023_CEERS}. We considered the O32 values provided by \cite{Saxena2023} for 11 out of 15 emitters in JADES.


\subsection{UV half-light radius measurements} \label{sec:re}

We measured the half-light radius $r_e$ of each galaxy in the rest-frame UV using the same procedure adopted by \cite{Mascia2023_CEERS}, with the python software \textsc{Galight}\footnote{\url{ https://github.com/dartoon/galight}} \citep{Ding2020}. The latter adopts a forward-modeling technique to fit a model to the observed luminosity profile of a source. We assume that galaxies are well represented by a Sérsic profile, constraining the axial ratio $q$ to the range $0.1$--$1$ and the Sérsic index $n$ to $1$, which has been shown to be the best suitable choice for high-z star forming galaxies \citep[e.g.,][]{Hayes2014, Morishita2018, Yang2022b}, and also for the CEERS sources, as detailed in \cite{Mascia2023_CEERS}. We remark also that LAEs tend to be more compact in their UV emission than the general population of LBGs, as recently shown by \cite{Napolitano2023} and \cite{Ning2023}. We visually checked from residuals that the luminosity profiles were well fitted by the Sérsic function.
The fit was performed in the F150W (F115W) NIRCam images for all sources at $z>5.5$ ($z<5.5$), to ensure the best  homogeneity in the rest-frame range. For the 14 sources that do not have NIRCam photometry, HST-WFC3 observations in the F160W (F125W) were used for the two redshift ranges. For unresolved sources we place an upper limit. The values are reported in Table~\ref{tab:summary_data}.
More information about both the procedure and the assumptions' justification can be found in \cite{Mascia2023_CEERS}, where we also briefly describe the simulations implemented to determine the minimum measurable radius in the various bands.\\
For the JADES subset, since all galaxies are at $z$ > 5.5 (see Fig.~\ref{fig:RedshiftHisto}) we consider the half-light radii obtained from F150W photometry \citep[see][for further details]{Rieke2023}. Only 10 out of the 15 emitters found by \cite{Jones2023}, have a half-light radius measurements in the published catalog.

\section{The properties of \lya\ emitters } \label{sec:fescLyA}
In this section we further analyze the derived properties of the combined sample of emitters, from both CEERS (50 galaxies) and JADES (15 galaxies).

\subsection{$f_{\mathrm{esc}}^{\lya}$ measurement} \label{sec:fescLyAmeasure}
We estimate $f_{\mathrm{esc}}^{\lya}$ as the ratio between the observed \lya\ emission flux measured in Sect.~\ref{sec:LyAmodel} and the expected intrinsic \lya\ emission flux calculated from the detected Balmer emission lines. For the intrinsic \lya\ emission flux, we adopt Case B recombination assuming density $n_e$ = 100 cm$^{-3}$ and temperature $T_e$ = 10,000 K. As reported in \cite{Osterbrock2006}, we consider the intrinsic \lya/H$\alpha$ and H$\alpha/$H$\beta$ ratios to be 8.2 and 2.86 respectively. Modifying these assumptions within the typical range for star-forming regions (5,000 K < $T_e$ < 30,000 K and 10 cm$^{-3}$ < $n_e$ < 500 cm$^{-3}$) results only in a few percent change in the above ratios \citep[e.g.,][]{Chen2023, Sandles2023}.\\
To derive $f_{\mathrm{esc}}^{\lya}$ for our 50 \lya\ emitting galaxies (see Table \ref{tab:summary_data}), we only consider fluxes measured from PRISM/CLEAR configuration spectra (see Sect.~\ref{sec:LyAmodel} and Sect.~\ref{sec:balmer}). We use the measured H$\alpha$ flux whenever it is observed in the spectral range (30 galaxies), otherwise we use the H$\beta$ flux (14 galaxies). For 5 galaxies with H$\beta$ S/N < 3, we can only derive lower limits on $f_{\mathrm{esc}}^{\lya}$.
For 1 emitter (MSA ID = 82372) we could not measure the $f_{\mathrm{esc}}^{\lya}$ or an upper limit, due to lack of both H$\alpha$ and H$\beta$ in the spectra covered by our observations. In Table~\ref{tab:summary_data} we report the $f_{\mathrm{esc}}^{\lya}$ value with the uncertainty derived by propagating the error of \lya\ and Balmer fluxes. The median value for the sample is 0.13, but the values span all the way to $\sim 0.8$. To attain such a high $f_{\mathrm{esc}}^{\lya}$, it might be necessary for the nearest predominantly neutral IGM patch to be situated at a distance of at least 1–2 physical Mpc (pMpc). We further investigate this in Sect.~\ref{sec:bubble}.

\subsection{Correlations between $f_{\mathrm{esc}}^{\lya}$ and physical properties} \label{sec:fesc_corr}
\begin{figure*}
    \centering
    \begin{subfigure}{0.48\textwidth}
        \includegraphics[width=\linewidth]{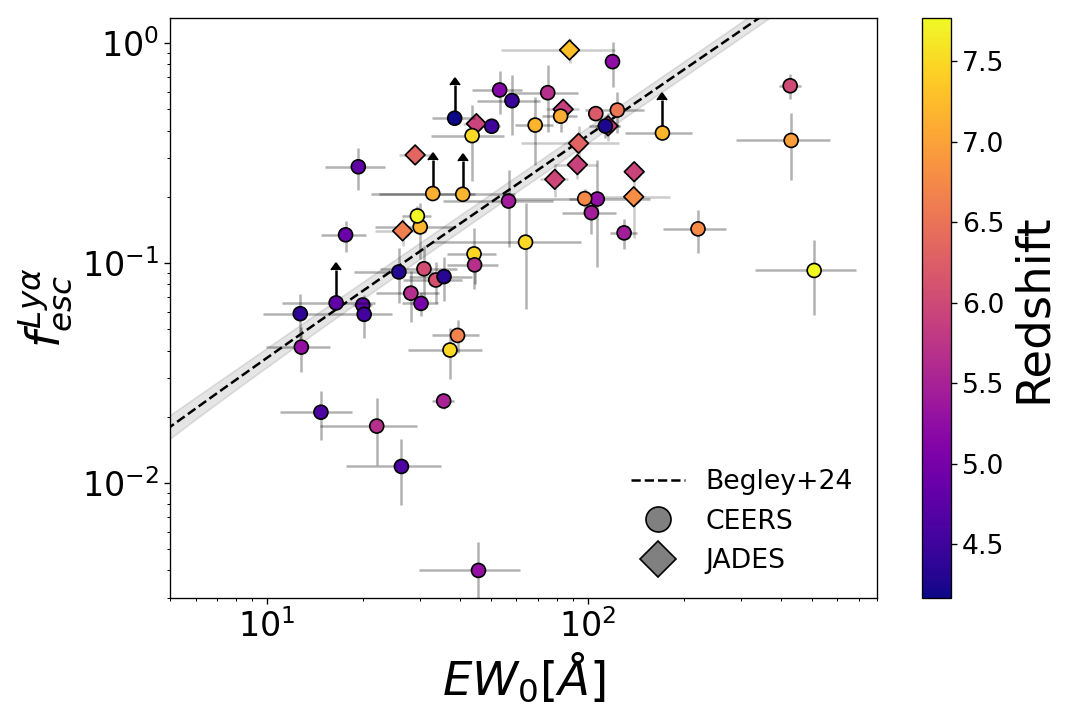}
    \end{subfigure}
    \begin{subfigure}{0.48\textwidth}
        \includegraphics[width=\linewidth]{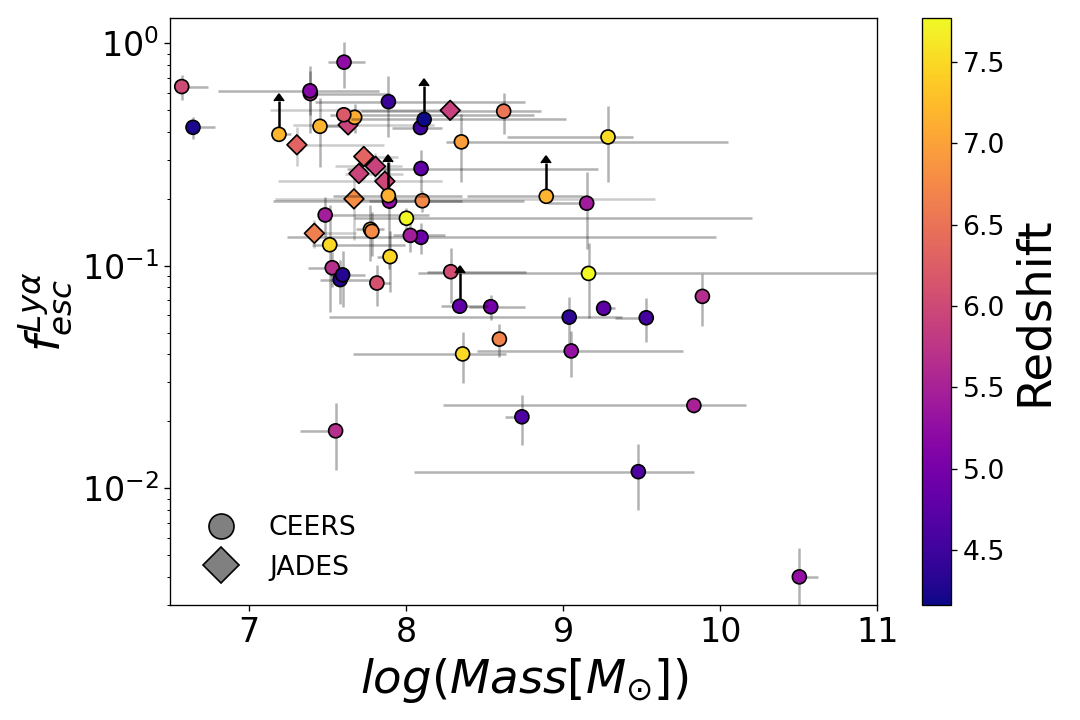}
    \end{subfigure}
    
    \medskip
    
    \begin{subfigure}{0.48\textwidth}
        \includegraphics[width=\linewidth]{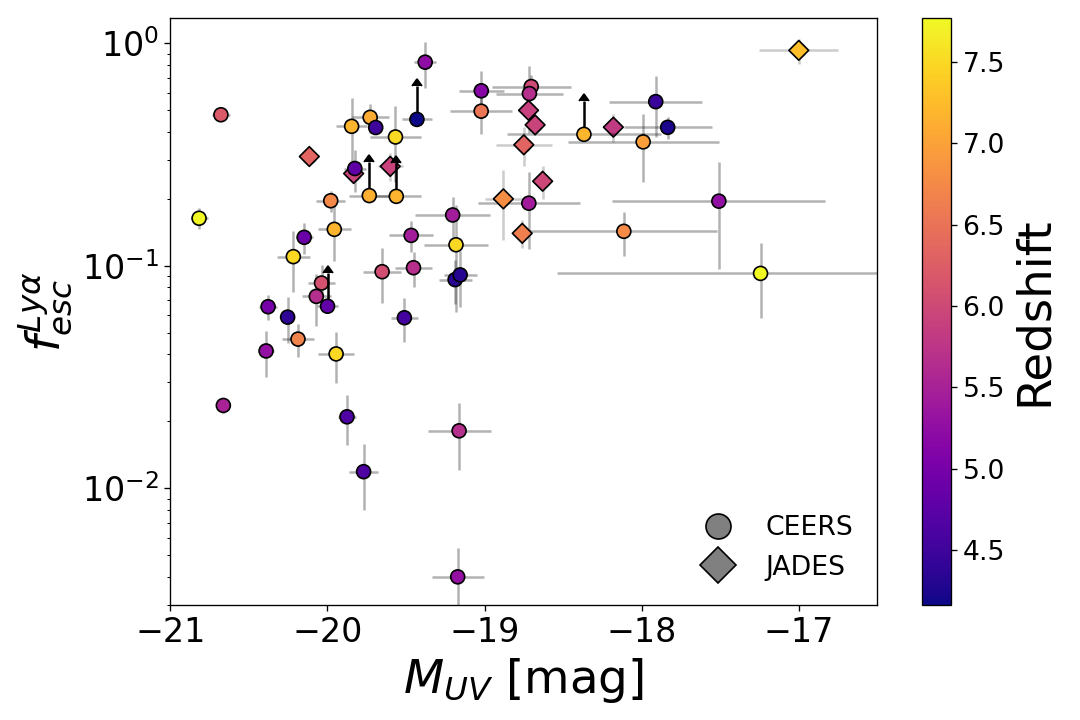}
    \end{subfigure}
    \begin{subfigure}{0.48\textwidth}
        \includegraphics[width=\linewidth]{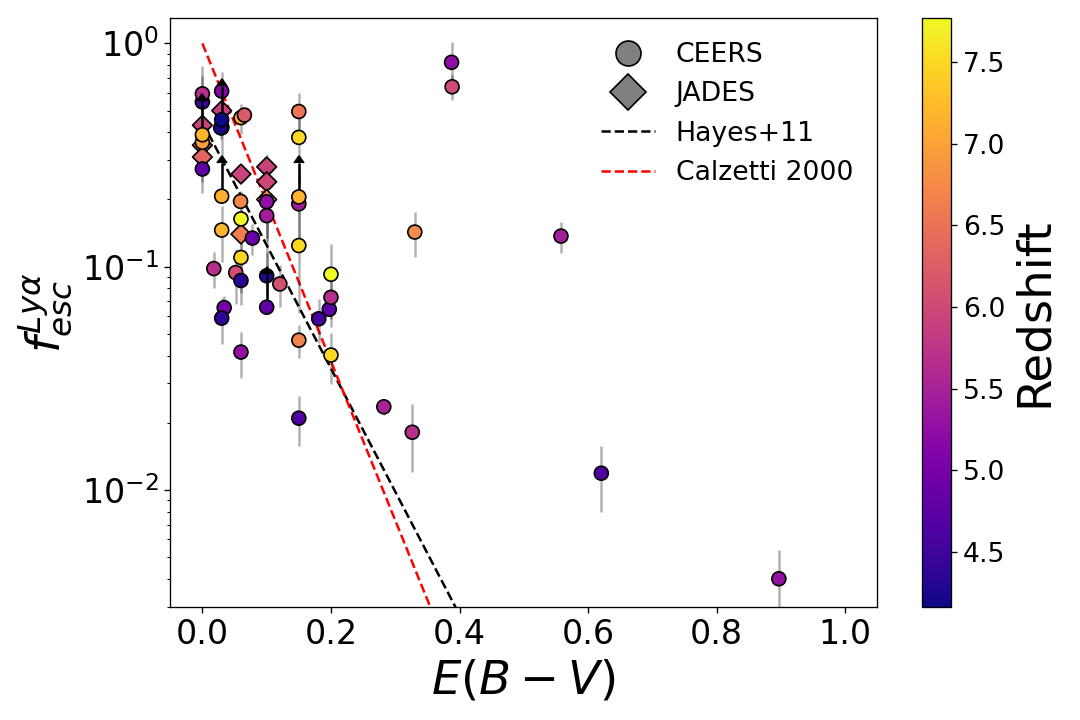}
    \end{subfigure}

    \medskip

    \begin{subfigure}{0.48\textwidth}
        \includegraphics[width=\linewidth]{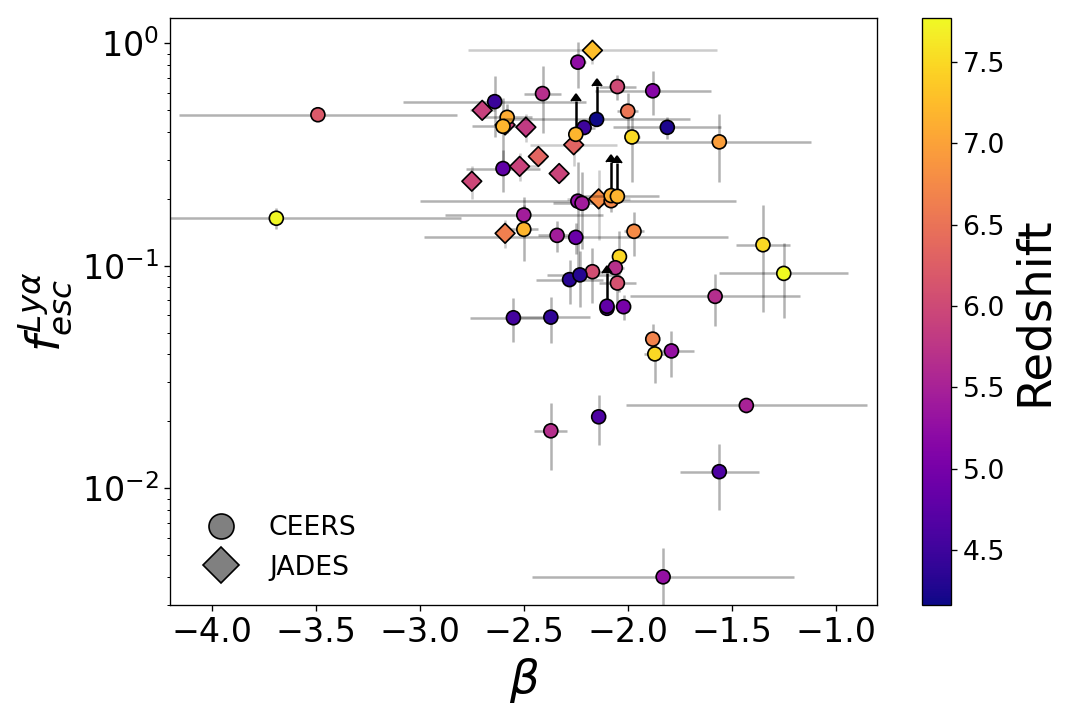}
    \end{subfigure}
    \begin{subfigure}{0.48\textwidth}
        \includegraphics[width=\linewidth]{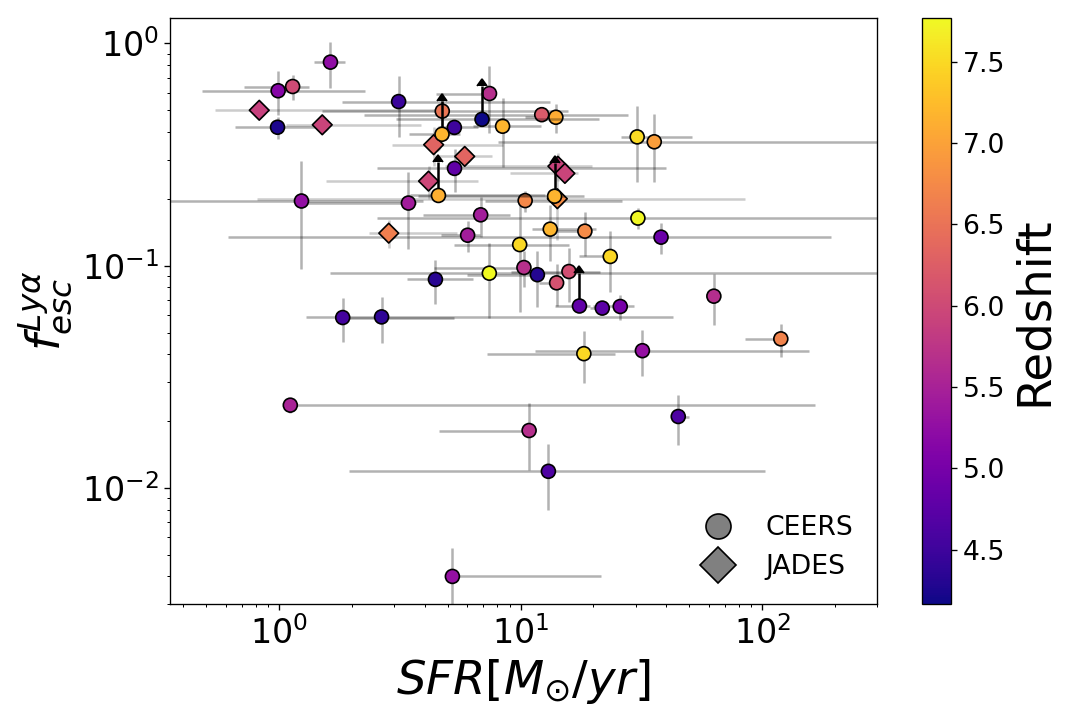}
    \end{subfigure}
    
    \caption{The \lya\ escape fraction as a function of rest frame \lya\ equivalent width, Stellar Mass, UV absolute magnitude, reddening, UV $\beta$ slope, and star formation rate. CEERS and JADES data are represented by circles and diamonds respectively. We report lower limits with black arrows, error bars are in grey instead. In the top left panel we show the fit relation found by \cite{Begley2024} in VANDELS $z \sim 4-5$ data. In the middle right panel we show the best fitting relations found by \cite{Hayes2011}, which exhibit slight deviation from the trend proposed by \cite{Calzetti2000}.}
    \label{fig:fescVSall}
\end{figure*}

In this section we further investigate the dependencies between the \lya\ escape fraction and the physical properties of the emitters.\\
In Fig.~\ref{fig:fescVSall} we present the correlations we found between $f_{\mathrm{esc}}^{\lya}$ and EW$_0$, the stellar mass, UV absolute magnitude, reddening, UV slope $\beta$, and SFR of all the CEERS and JADES emitters derived in Sect.~\ref{sec:Method}.
To quantify the existence of  correlations, we ran a Spearman rank test between $f_{\mathrm{esc}}^{\lya}$ and the derived properties. Whenever lower limits are present we use these values multiplied by $\sqrt{2}$.
We consider a correlation to be present whenever the p-value is $p(r_s)< 0.01$ (see Table \ref{tab:Spearman}). 
As expected the strongest correlations are with the EW$_0$ and $M_{\mathrm{UV}}$, and the strongest anti-correlations are with the Mass and the E(B-V).
As noted by \cite{Saxena2023}, since by definition $f_{\mathrm{esc}}^{\lya}$ is obtained from the observed ratio of \lya\ to H$\alpha$ (or H$\beta$) fluxes, the observed strong correlation with EW$_0$ implies that the H$\alpha$ (or H$\beta$) flux does not scale with \lya. Also \cite{Roy2023} and \cite{Begley2024} report the same trend from intermediate redshift emitters, respectively from GLASS and VANDELS data. The interpretation of the correlation between $f_{\mathrm{esc}}^{\lya}$ and $M_{\mathrm{UV}}$ is two-fold. On one hand, as already discussed in Sect.~\ref{sec:Muv} for the faint population this is produced by the flux limited nature of spectroscopic observations. However, the fact that we do not observe high $f_{\mathrm{esc}}^{\lya}$ from the UV-bright population is a real effect, and suggests an increasing neutral gas fraction and dust in more luminous galaxies. This result, also in agreement with \cite{Saxena2023}, is supported by the anti-correlation we find with stellar mass. UV bright massive systems are likely to have an increasing neutral hydrogen content in their interstellar media (ISM), which due to the resonant scattering nature of \lya\ \citep[for a review, see][]{Dijkstra2017} attenuates its emission along the line of sight, because the probability of the \lya\ photons to be absorbed by dust also increases \citep{Verhamme2015, GurungLopez2022}. The key role of dust in the process is also highlighted by the typical steep UV $\beta$ slopes, low reddening and low star formation rates values we find for the galaxies with the highest values of $f_{\mathrm{esc}}^{\lya}$. 
We report in Fig.~\ref{fig:fescVSall} the best-fitting relation identified by \cite{Hayes2011} between $f_{\mathrm{esc}}^{\lya}$ and reddening. This trend differs slightly from the dust attenuation model proposed by \cite{Calzetti2000}, as it enforces $f_{\mathrm{esc}}^{\lya}$ < 1 for a zero value of E(B-V).
Finally we find only a marginal anti-correlation between the $f_{\mathrm{esc}}^{\lya}$ and $r_e$, which is consistent with the compact of \lya\ emitters as already described in \cite{Napolitano2023}. However the p-values are higher than the threshold and cannot be considered conclusive.
All the above trends were known to exist at lower redshift: our results indicate that the nature of \lya\ emitting galaxies and the mechanisms that favour the line visibility do not change much with redshift and seem to be still primarily  associated to the galaxies  physical properties.
If confirmed with larger samples of \lya\ emitters at $z$ > 7, this could imply that the role of the IGM in suppressing \lya\ visibility is more likely an on/off effect, with line of sights that are almost free for the \lya\ photons to escape unattenuated and others where the \lya\ is completely absorbed (similar to the simple number evolution scenario suggested by \cite{Tilvi2014}). The inhomegeneity of reionization will be discussed further in Sect.~\ref{sec:bubble}.    

\begin{table}
\caption{Spearman correlation coefficients with the \lya\ escape fraction for the \lya\ emitting galaxies sample. Features are ranked by increasing p-values.}\label{tab:Spearman}
\begin{tabular}{lccc}
\hline \hline
\noalign{\smallskip}
    \textbf{Feature} &
    \textbf{Coefficient} &
    \textbf{p-value} &
    \textbf{Null hypothesis rejected} \\ \hline
\noalign{\smallskip}
 \hline
 \noalign{\smallskip}
EW$_0$ & 0.54 & <$10^{-3}$ & Yes \\ 
\noalign{\smallskip}
\hline
\noalign{\smallskip}
Mass & -0.45 & <$10^{-3}$ & Yes \\ 
\noalign{\smallskip}
\hline
\noalign{\smallskip}
M$_{\mathrm{UV}}$  & 0.40 & <$10^{-3}$ & Yes \\ 
\noalign{\smallskip}
\hline
\noalign{\smallskip}
E(B-V)  & -0.36 & $3 \times 10^{-3}$ & Yes \\ 
\noalign{\smallskip}
\hline
\noalign{\smallskip}
$\beta$  & -0.33 & $7 \times 10^{-3}$ & Yes \\ 
\noalign{\smallskip}
\hline
\noalign{\smallskip}
SFR  & -0.34 & $7 \times 10^{-3}$ & Yes \\
\noalign{\smallskip}
\hline
\noalign{\smallskip}
$r_e$  & -0.27 & $3 \times 10^{-2}$ & No \\
\hline
\end{tabular}
\end{table}


\subsection{Lyman Continuum escape in \lya\ emitters} \label{sec:fesc_lyCmeasure}
\begin{figure*}[ht!]
\begin{minipage}{0.5\textwidth}
\centering
\includegraphics[width=\linewidth]{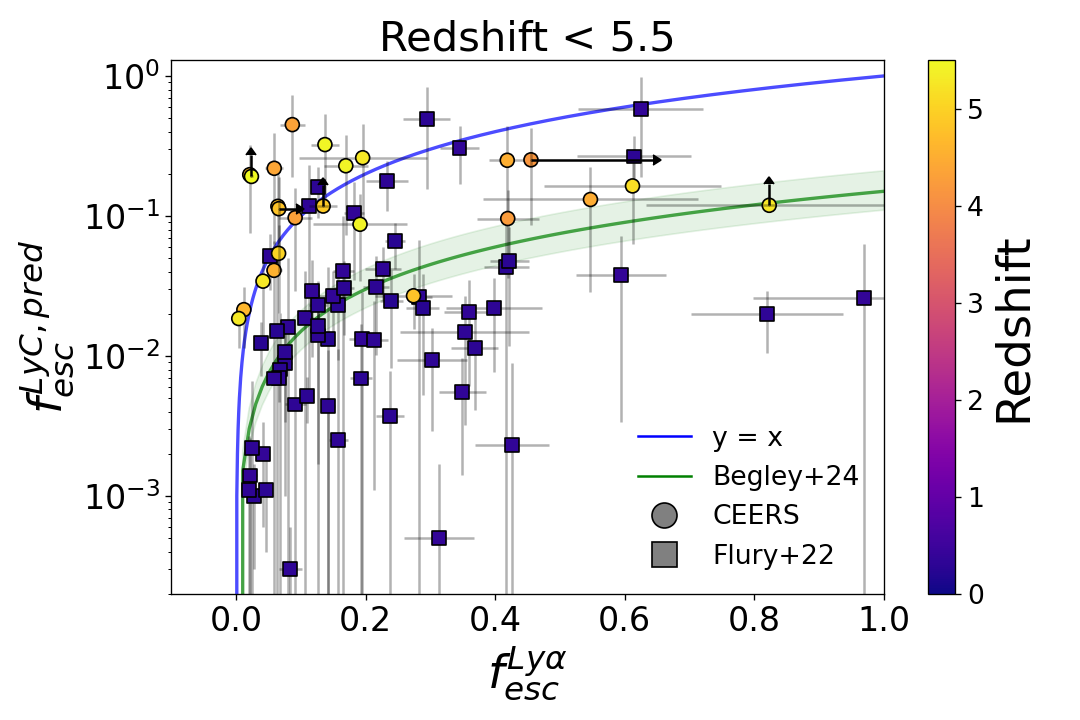}
\end{minipage}
\begin{minipage}{0.5\textwidth}
\centering
\includegraphics[width=\linewidth]{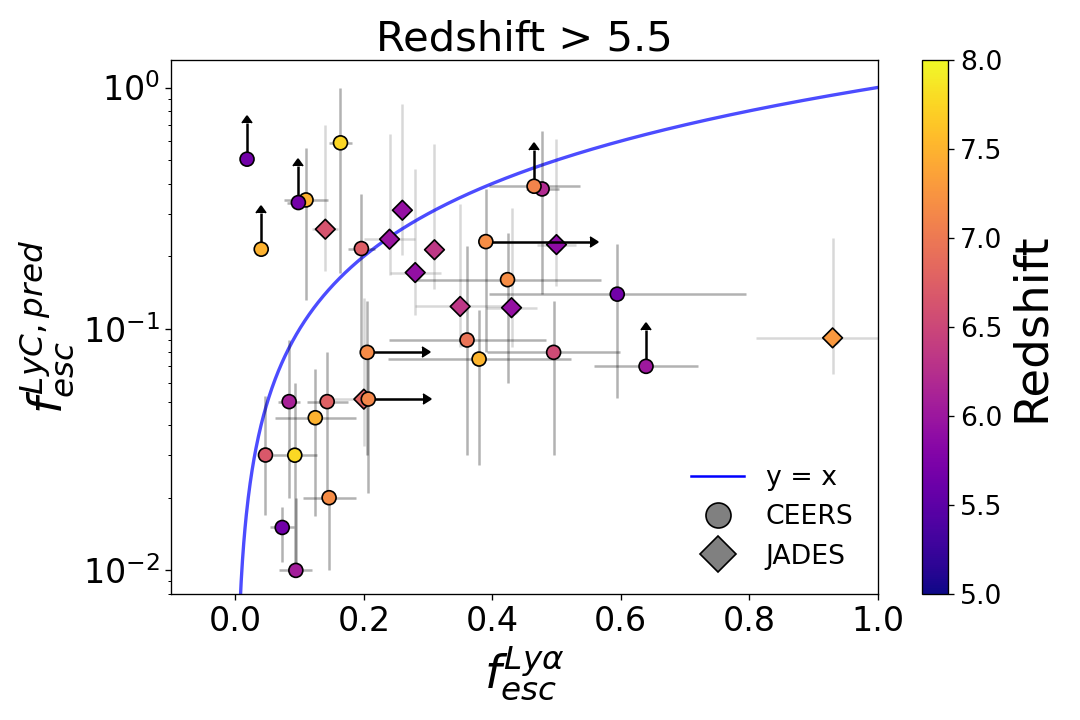}
\end{minipage}
\caption{The \lya\ escape fraction as a function of the LyC escape fraction inferred from the combination of spectroscopic properties using the relation derived by \cite{Mascia2023_CEERS}. Symbols are the same as in Fig. \ref{fig:fescVSall}. The left panel includes all post-reionization galaxies ($z<5.5$) while the right panel includes galaxies residing in a partially ionized IGM. For comparison, we report with the solid green line and region the best fitting relation found by \cite{Begley2024} for $z \sim 4-5$ emitters.}
\label{fig:fescLyAvsfescLyC}
\end{figure*}

A direct detection of the Lyman Continuum (LyC) emission escaping from the high-z galaxies is not possible given the extremely high opacity of the IGM to LyC photons at $z$ > 4.5 \citep[][]{inoue2014}. However it is essential to estimate this quantity to understand the nature of galaxies contributed mostly to the reionization process. Recently, several authors \citep[e.g.,][]{chisholm2022, Roy2023, Saxena2023} tackled this problem by deriving  empirical relations that connect key physical or observational properties to the escape of LyC photons of galaxies. The relations are either derived from simulations \citep[e.g.,][]{Choustikov2023} or from the samples of low redshift LyC emitters for which detailed derivation of the physical properties are available \citep[][]{Flury2022}.  
In this context, in \cite{Mascia2023_GLASS} and \cite{Mascia2023_CEERS} we developed two empirical relations that predict $f_{\mathrm{esc}}^{\mathrm{LyC, pred}}$ from a set of photometric and spectroscopic indirect indicators, identified in the most complete low-redshift sample of Lyman continuum emitters \citep[see][]{Flury2022}, namely the O32 ratio, the UV radius $r_e$, and the UV $\beta$ slope (or in alternative  EW(H$\beta$), the UV radius $r_e$, and $\beta$ --  see the above paper for more details and a comparison between the two methods). Similarly \cite{chisholm2022} developed an empirical relation that is based on just the UV $\beta$ slope to obtain a predicted LyC escape fraction.
For the CEERS sample of \lya\ emitters we employed the relation based on O32, $r_e$, and $\beta$ for 41 sources and EW(H$\beta$), $r_e$, and $\beta$ for the remaining 9 \lya\ emitters. We note that there are 9 unresolved sources for which $r_e$ is given as upper limit (see Sect.~\ref{sec:re}): in these cases, given the anti-correlation between $r_e$ and $f_{\mathrm{esc}}^{\mathrm{LyC, pred}}$, the derived $f_{\mathrm{esc}}^{\mathrm{LyC, pred}}$ are lower limits. 
In Table~\ref{tab:summary_data} we report the obtained values. \\
For the JADES sample, we could only calculate the predicted LyC escape fraction for 10 emitters for which the values of O32, $r_e$, and $\beta$ are reported in the literature (see Sect.~\ref{sec:re}). \\
In Fig.~\ref{fig:fescLyAvsfescLyC} we show the relation between the inferred $f_{\mathrm{esc}}^{\mathrm{LyC, pred}}$ and the  $f_{\mathrm{esc}}^{\lya}$. The left panel includes all post-reionization galaxies ($z$ < 5.5), while the right panel only shows galaxies that reside in a partially ionized IGM. 
Since \lya\ is known to correlate with the $f_{\mathrm{esc}}^{\mathrm{LyC}}$ in the local universe and at $z$ = 3 \citep[e.g.,][]{Marchi2017, Gazagnes2020, Pahl2021}, given that the photons can escape through common clear channels in the ISM \citep[e.g.,][]{Verhamme2015, Dijkstra2016, Jaskot2019}, we would expect to see a correlation between the two quantities at $z$ $\sim$ 4.5 -- 5.5. On the other hand, when galaxies start to be surrounded by a partially neutral IGM, such correlation might be lost, as the \lya\ visibility is no more driven by the galaxies properties alone, but also by the local IGM conditions. \cite{Saxena2023} and \cite{Mascia2023_CEERS} already noted the absence of any correlation in their more limited samples, with \cite{Saxena2023} also noting that the predicted LyC escape fraction is always lower than the \lya\ escape fraction. We do not see evidence for this effect, as many galaxies considered in this work actually have higher inferred LyC escape. We also note that the best fitting relation ($f_{\mathrm{esc}}^{\mathrm{LyC, pred}} \sim 0.15^{+0.06}_{-0.04} \times f_{\mathrm{esc}}^{\lya}$) found by \cite{Begley2024} from the analysis of the interstellar absorption lines in  stacked spectra at $z \sim 4-5$ from VANDELS data predicts lower values of $f_{\mathrm{esc}}^{\mathrm{LyC, pred}}$ than what we obtain in this work. 
\\
Contrary to our expectations, we do not find any secure correlations at either redshift ranges, although admittedly the lower redshift sample is rather small (24 galaxies) and the uncertainties on the inferred LyC escape fractions are significant. The sample at $z$ < 5.5 ($z$ > 5.5) has a Spearman coefficient of 0.29 (0.16) with a p-value = 0.16 (0.36).
We note however that the average $f_{\mathrm{esc}}^{\mathrm{LyC, pred}}$ of \lya\ emitting galaxies, is 0.16 i.e., slightly higher than what reported by \cite{Mascia2023_CEERS} for the general LBG population with similar $M_{\mathrm{UV}}$.
\\
For completeness we also used the \cite{chisholm2022} relation to derive alternative $f_{\mathrm{esc}}^{\mathrm{LyC, pred}}$ values, but imposing a maximum of $f_{\mathrm{esc}}^{\mathrm{LyC, pred}}$ = 1 for galaxies with extremely blue colors. On average these values are lower than the one derived with our relations, but they also do not seem to correlate with the $f_{\mathrm{esc}}^{\lya}$.

\section{The evolution of the \lya\ visibility during reionization} \label{sec:XlyA}

During the epoch of reionization, the visibility of \lya\ emission in galaxies is due to the combination of the physical properties of the galaxies, which regulate how many photons can emerge from a galaxy through the 
interstellar and circumgalactic media, and of the conditions of the surrounding IGM, whose neutral fraction $X_{\mathrm{HI}}$ determines the number of \lya\ photons that are expected to reach us. 
To take into account the first factor people usually employ $M_{\mathrm{UV}}$ matched samples, and rely on the fact that the typical timescales for galaxy evolution at very high redshift are relatively short (e.g., there are only 170 Myr between $z \sim$ 6 and $ z \sim$ 7). Thus any remaining effect on the \lya\ visibility above $z \sim$ 6 is  attributed to a changing neutral fraction in the IGM \citep[e.g.,][]{Mason2018, Pentericci2018}. In other words, the $z \sim$6 universe is usually assumed to be completely ionized, and thus the \lya\ visibility at this epoch is used as the benchmark to compare its evolution. This view has been recently questioned by the discovery of significant fluctuations in the HI optical depth in QSO spectra and the presence of extended regions of high opacity down to $z \simeq$ 5.3 which suggest an extended final phase and a late end of hydrogen reionization \citep{zhu2023,bosman22}. 
Crucially, the CEERS sample that we are presenting in this work covers observations of \lya\ emission at $z$ > 4.5 and we are thus able to assemble a solid baseline sample of \lya\ measurements in the post-reionization epoch, i.e., where no more neutral islands exists. Also contrary to previous works, this baseline post-reionization sample is observed with the same configuration as the reionization sample, while previous analysis compared $z \sim$ 6 EW distribution typically obtained by optical instruments \citep{Schenker2012,DeBarros2017} to observations at $z$ > 7 obtained with near-IR spectrographs \citep{Mason2018,Jung2022}.
In the following section we will compute the \lya\ fractions from the CEERS and JADES samples.

\subsection{The evolution of  the \lya\ emitter fractions} \label{sec:lyavis}
We define the \lya\ emitter fractions as the number of galaxies with a measured \lya\ emission with EW$_0$ > 25 \AA\ (or > 50 \AA) over the total number of galaxies, in four redshift bins, centered at $z$ = 5, 6, 7 and 8 and with $\Delta z=1$. We only include in our calculations all galaxies in the range -20.25 < $M_{\mathrm{UV}}$ < -18.75 to be consistent with most previous works, that used these values to separate bright and faint galaxies following the very first studies \citep{Stark2011, Pentericci2011, Ono2012}. A large fraction of the CEERS galaxies belong to  this magnitude range (see Fig.~\ref{fig:EWvsMuv}), with only a few sources having brighter or fainter magnitudes, whereas JADES has a higher number of sources fainter than $M_{\mathrm{UV}}=-18.75$. For all sources we take into account the rest frame EW limit (EW$_{\mathrm{0,lim}}$) that identifies the minimum rest-frame equivalent width we could possibly detect given their redshift and  continuum flux. 
This is to avoid biasing the derived visibilities, e.g., bright \lya\ emission lines detected in relatively shallow data could bias the derived visibility upwards, while the inclusion of many undetected sources for which we could not in any case detect \lya\ would bias the fraction down. We note that 115/235 galaxies, i.e. $\sim$ 49\% (131/235 galaxies, i.e. $\sim$ 56\% ) of data in the parent sample fulfills both the $M_{\mathrm{UV}}$ cut and EW$_{\mathrm{0,lim}}$ < 25 \AA\ ( < 50 \AA\ ) requirements. 
The results for the two different values of \lya\ EW are presented in Fig.~\ref{fig:XLyA} for the entire redshift range.
We show the results for the CEERS and JADES samples separately, and for their combination (for JADES only the bins at $z \geq 6$ are populated).
Given the low number of detections in each bin, the uncertainties are evaluated using the statistics for small numbers of events developed by \cite{Gehrels1986}.
We can see that while at $z$ = 6 the JADES and CEERS fractions are in perfect agreement for both EW limits, at higher redshift there is a very large discrepancy between the two fields. At $z$ = 7 for the EGS field we derive a much higher fraction, which might be due in part to the presence of 3 \lya\ emitters around $z$ = 7.18 and 4 emitters around $z$ = 7.49. We will discuss both structures in Sect.~\ref{sec:bubble}. We note that in the CEERS field another overdensity has been previously identified at $z$ = 7.7 by \cite{Tilvi2020}, \cite{Jung2022}, and \cite{Tang2023}. At this redshift we have 2 emitters in our sample, one which was previously known (ID 686 at $z$ = 7.75) and a new tentative \lya\ detection (ID 20 at $z$ = 7.77). 
On the other hand, it is well known that the GOODS-South field observed by the JADES program, despite being one of the best studied with extensive spectroscopic coverage, contains very few \lya\ emitters at $z \sim$ 7 \citep{Song2016, Pentericci2018}. 
If we remove the galaxies belonging to the two identified structures (both with and without \lya\ emission) in the CEERS data, we obtain fractions that are much lower and compatible with the GOODS-South values. These are reported with open symbols. \\
Recently, also \cite{Jones2023} and \cite{Nakane2023} inferred \lya\ visibilities during the EoR using \jwst\ data. \cite{Jones2023} present the \lya\ visibilities in the JADES sample (GOODS-South field) without applying a cut at the faint end of the $M_{\mathrm{UV}}$ range. As expected, given that a large fraction of the JADES sources are fainter than $M_{\mathrm{UV}} = -18.75$, they obtain slightly higher \lya\ visibilities than what we derived for the JADES sample, although fully consistent with our point. The recent analysis by \cite{Nakane2023} include data from JADES, GLASS, CEERS and other programs, and also medium resolution gratings data are employed. Therefore, they analyze a different set of data, with additional fields compared to our analysis. Applying our same cut both in $M_{\mathrm{UV}}$ and EW$_0$ > 25, they derive a value of $0.25^{+0.19}_{-0.13}$ at $z$ = 7 and a limit of < 0.19 at $z$ > 8. Both results are in 1-$\sigma$ agreement with what we derived.
\\
In Fig.~\ref{fig:XLyA} we also report previous results from ground based observations from classical slit spectroscopy \citep{Stark2011, Schenker2014, Tilvi2014, DeBarros2017, Pentericci_2018b, Fuller2020}, MUSE integral field spectroscopy \citep{Kusakabe2020, Goovaerts2023}, and the KMOS spectrograph \citep{Mason2019}. In the case of \cite{Goovaerts2023} we considered the result obtained after completeness correction and when considering only continuum selected sources, which reproduces more closely the LBG photometric selection. Some of the above studies \citep[e.g.,][]{Mason2019, Fuller2020, Goovaerts2023} target lensed galaxy fields, thus probe intrinsically fainter galaxy population. In  all the above studies we have selected fractions reported in the  -20.25 < $M_{\mathrm{UV}}$ < -18.75 range. The only exception is \cite{Goovaerts2023} that samples much fainter $M_{\mathrm{UV}}$ (down to -12) and it is not immediately comparable. Our JWST \lya\ fractions at $z$ = 5 are lower than all previous results, although in some cases consistent within the errors. At $z$ = 6 the comparison is harder as there is a wider range of published values: our combined  value is consistent with some previous results but significantly lower than the values reported by  \cite{Stark2011} and \cite{deBarros2016}. At $z$ = 7, the JADES fractions are consistent with previous estimates, while the CEERS fractions are larger, as already detailed above, even when removing the objects in the possible overdensities.\\
In Sect.~\ref{sec:compareXlyA} we further discuss why ground based and JWST observations might give substantially different \lya\ results.
If we take the JWST results alone at face value,  the obtained average $X_{\lya}$ (from the combination of the CEERS and JADES observations) continues to rise from $z$ = 5 to 7, dropping again at $z$ = 8. This implies that either our results are not consistent with a rapid end of reionization that was inferred by previous works, or that a rapid reionization ending is still characterized by >1 pMpc bubbles around a high fraction of modestly bright galaxies, to at least $z \sim$ 7. Indeed the large field to field variations imply a scenario in which multiple ionized bubbles can significantly alter the \lya\ visibility, when considering only a limited area and stress the need for very large surveys in multiple fields, to obtain more robust results. In the following section we further investigate this issue with the help of simulations.

\subsection{The large field-to-field variations: comparison to the CoDaII simulation}
\begin{figure*}[ht!]
\begin{minipage}{0.5\textwidth}
\centering
\includegraphics[width=\linewidth]{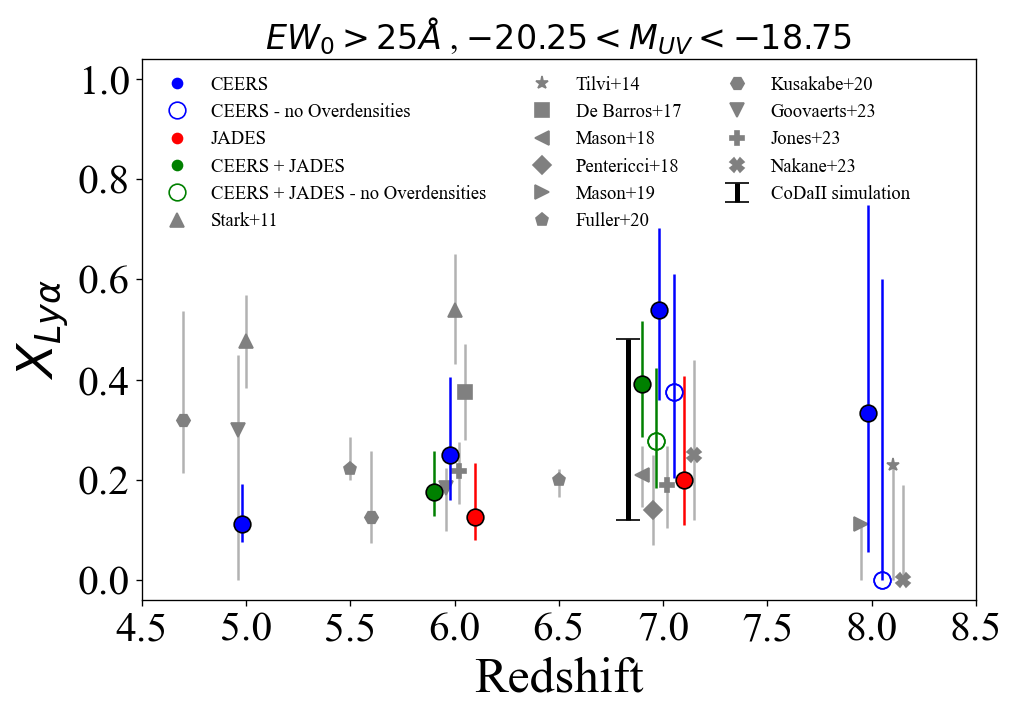}
\end{minipage}
\begin{minipage}{0.5\textwidth}
\centering
\includegraphics[width=\linewidth]{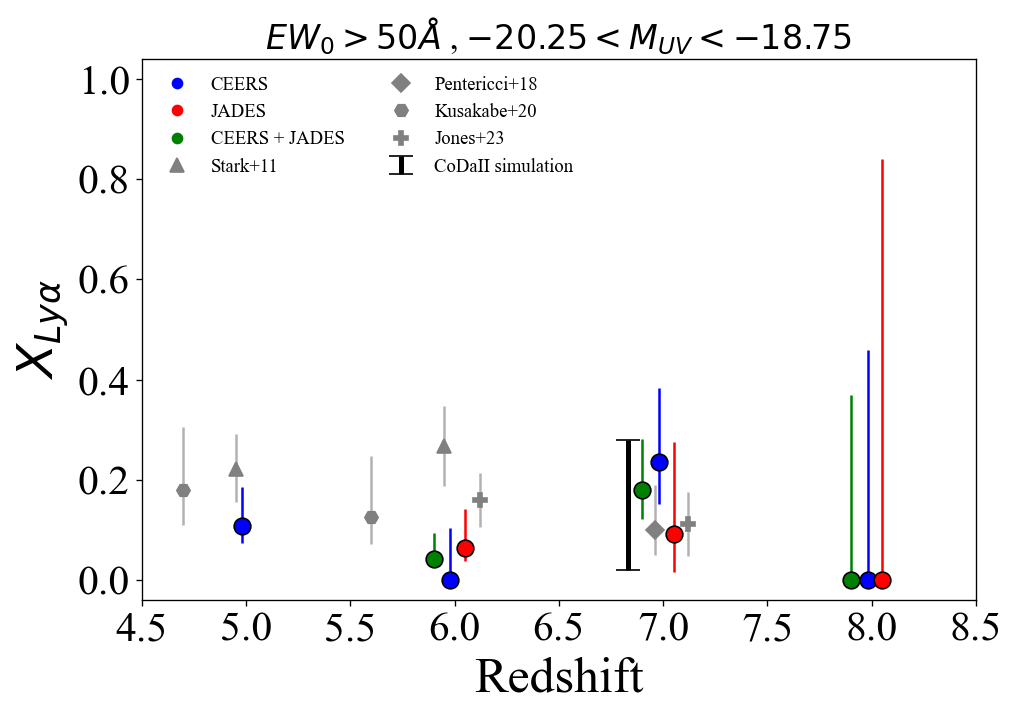}

\end{minipage}
\caption{\lya\ fraction as a function of redshift. Blue, red and green data points represent CEERS, JADES and the combined sample. The open data points are obtained when we do not consider the galaxies in the overdensity at $z \sim$ 7.18 and $z \sim$ 7.49. Error bars are calculated from the binomial statistics described in \cite{Gehrels1986}. The black bar is the result obtained with the CodaII simulation at $z$ = 7. Grey symbols are taken from literature, we provide the full list here in the same order as the legend: \cite{Stark2011}, \cite{Tilvi2014}, \cite{DeBarros2017}, \cite{Mason2018a}, \cite{Pentericci_2018b}, \cite{Mason2019}, \cite{Fuller2020}, \cite{Kusakabe2020}, \cite{Goovaerts2023}, \cite{Jones2023}, and \cite{Nakane2023}. Some of the points have been slightly shifted in redshift for an easier visualization.}
\label{fig:XLyA}
\end{figure*}

To further investigate if the large field to field variations are expected by the fundamental inhomogeneity of the reionization process, we employed calculations from the Cosmic Dawn II (CoDaII) simulation \citep[][]{Ocvirk2016}. This simulation reproduces the reionization process to capture HII bubbles forming around star-forming galaxies on a grid of ${4096}^{3}$ to resolve small-scale gas density and velocity structures, making it suitable for calculating \lya\ opacity of the IGM (see \cite{Ocvirk2016} for more details). As described in \cite{Park2021}, we calculate the \lya\ visibility of simulated galaxies using CoDaII by integrating \lya\ opacity along lines of sight. In particular, for each galaxy the transmission curve is obtained for $\pm4\AA$ around the rest-frame \lya\ and for $\sim2000$ sight-lines, to account for the sight-line variation due to the stochasticity of HI density/velocity at small scales and the diversity of HII bubble shapes. We analyze the \cite{Park2021} results at $z$ = 7 using the same -20.25 < $M_{\mathrm{UV}}$ < -18.75 cut to match the  observations and randomly select 25 galaxies out of the simulated sample, to approximately match the number of target galaxies in the observations that are employed to calculate the $z$ = 7 visibility in JADES and CEERS. We note that the simulated volume of the Universe at $z$ = 7 is $\sim$ 760,000 cMpc$^3$, that is twice the combined volume of the CEERS and JADES (GOODS-South) surveys. For each simulated galaxy, we randomly draw the intrinsic \lya\ EW assuming an exponential declining distribution of the form $P(EW) \propto e^{-EW_0/W_0}$. We determine the free parameter $W_0$ from the observed EW distribution of our sample at $z$ < 6 (i.e., the post-reionization Universe) using both the \lya\ emitting sources and the limits on the non emitting ones. We employed 
\textsc{emcee} \citep[][]{Foreman_Mackey2013} to conduct an MCMC fit and find $W_0 = (39 \pm 2) \AA$. This result is in agreement with the value obtained in \cite{Pentericci_2018b}, when analyzing a very large sample of galaxies at $z$ = 5.5 -- 6.5 in the same $M_{\mathrm{UV}}$ range. \\
The attenuation of \lya\ EW is then calculated by taking the ratio between the integrated transmitted flux at the target redshift ($z$ = 7 in this case) and the mean 55\% transmission at the post-reionization redshift $z$ = 6, which was calculated by \cite{Park2021}. Finally we obtain the fraction of galaxies with EW$_0$ > 25 (50) \AA\ to simulate the visibility measurement from observations. We repeat this process 1000 times to obtain the average and standard deviation of the $X_{\lya}$ mock measurement sampling different line-of-sights.
We find that the fractions vary in the 2$\sigma$ range [0.12,0.48] ([0.02,0.28]) for the 25 \AA\ (50 \AA) threshold. Such ranges are indeed similar to the observed difference between CEERS and JADES as shown in Fig.~\ref{fig:XLyA}. \\
We focus our attention only on the large range of the simulated visibility, and not on the central value, that  is subject to both the reionization history of CoDaII and to the intrinsic emission model adopted in this calculation (i.e., the EW distribution in the post-reionization universe) which, as discussed above,  is still uncertain. Such large uncertainty from the sight-line variation can be suppressed by enlarging the sample size in future surveys.

\subsection{The ionized regions in EGS at $z$ = 7.18 and $z$ = 7.49} \label{sec:bubble}
\begin{figure*}[ht!]
\begin{minipage}{0.4\textwidth}
\centering
\includegraphics[width=\linewidth]{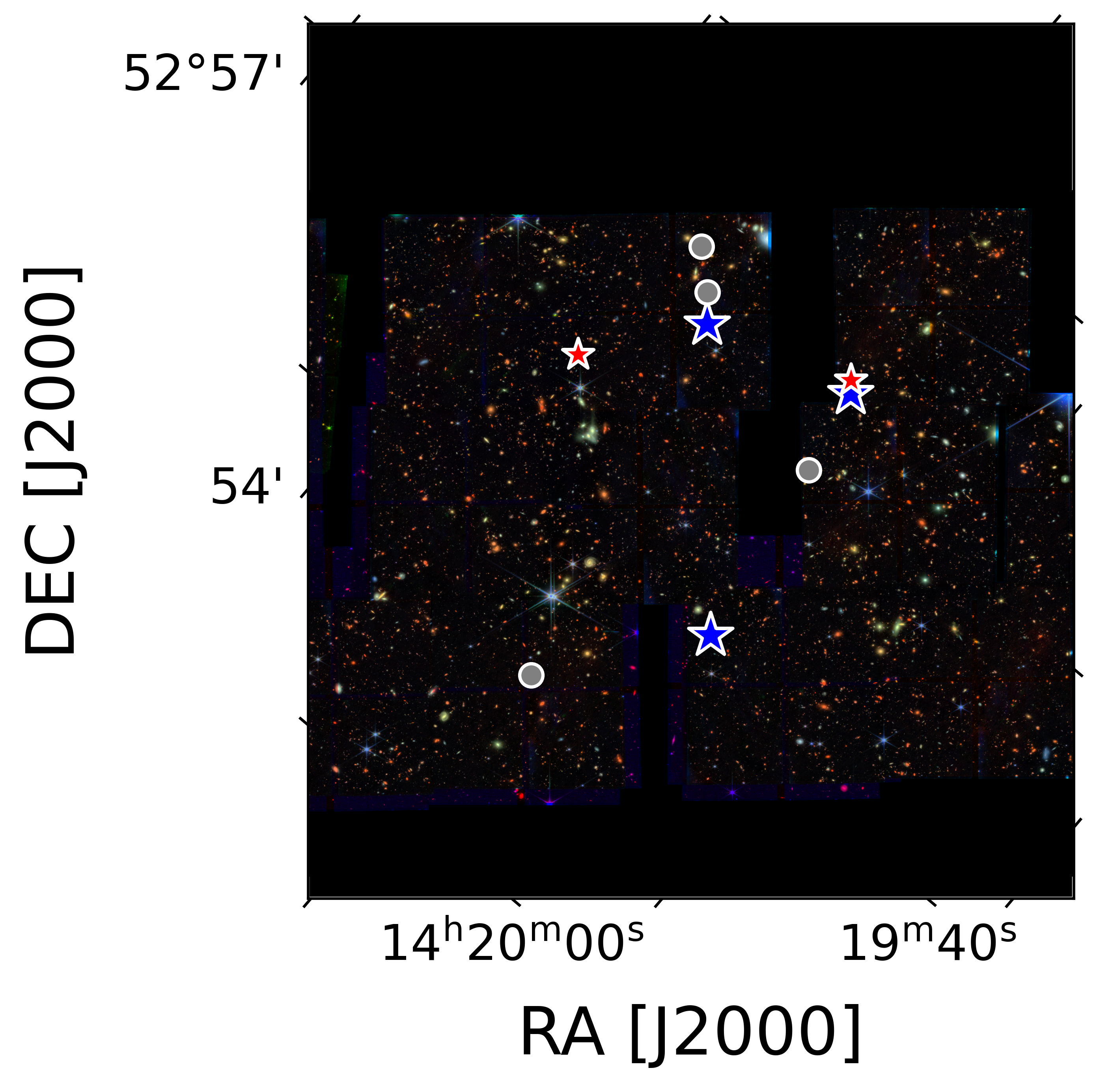}
\end{minipage}
\hfill
\begin{minipage}{0.55\textwidth}
\centering
\includegraphics[width=\linewidth]{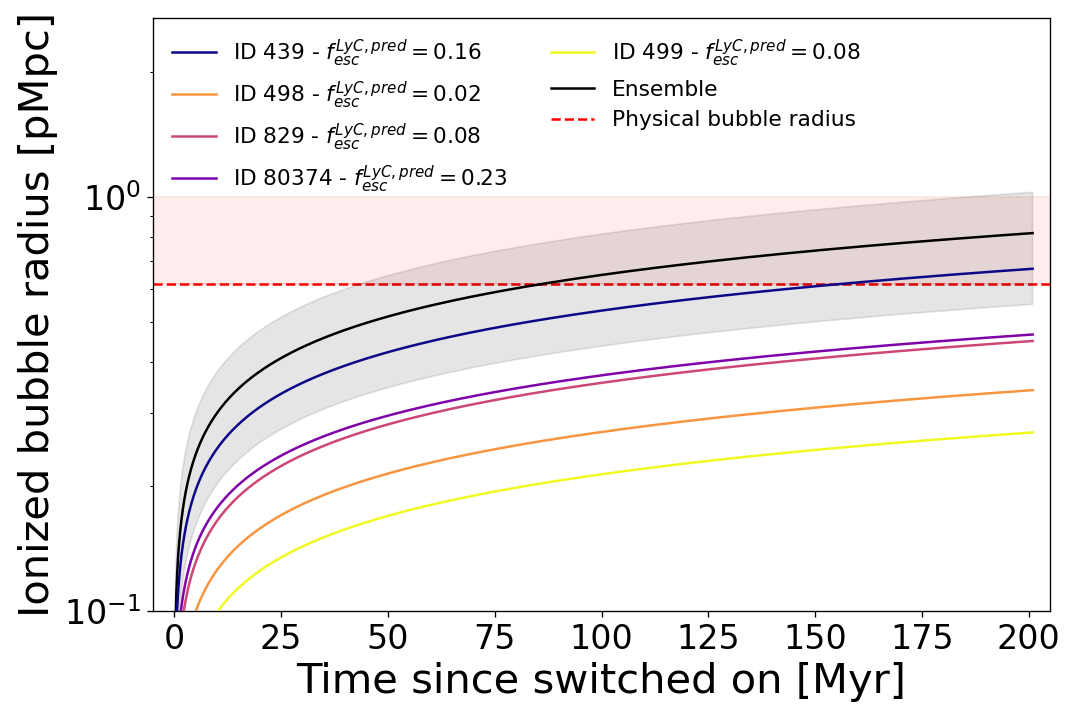}
\end{minipage}
\caption{Left: Position of the 5 spectroscopically-identified galaxies in the $z \sim$ 7.18 reionized region. Blue stars are the three \lya\ emitting galaxies with S/N > 3 and the red stars are the non-emitting ones, while we show the four photometric candidates with grey circles. The area corresponds approximately to one NIRSpec pointing or $\simeq$1.05$\times$1.10 pMpc$^2$ at $z \sim$ 7.18. Right: Predicted size of an ionized bubble as a function of time since ionizing radiation is switched on. Solid coloured lines are for individual sources while  the black solid line and shaded region represent the predicted size of the ionized bubble carved by the 5 galaxies together. The horizontal dashed line represents the physical bubble radius.}
\label{fig:newBubble}
\end{figure*}

As mentioned above, three of our \lya\ emitting galaxies are found within a very close redshift range (7.16 -- 7.18) and also very close in sky coordinates, at a maximum (minimum) distance of 209$''$ (0.3$''$) from each other (Fig.~\ref{fig:newBubble}).  
One of the sources, ID 80374 has very high EW of (171 $\pm$ 41) \AA\ and was also identified by \cite{Chen2023} as a possible indicator of a bubble ionized by an overdensity, given that it is a very faint source ($M_{\mathrm{UV}}$ = -18.37) and thus is it unlikely to have created a large ionized bubble by itself. Indeed, we find two other galaxies with \lya\ emission in the vicinity (ID 439, 498) with slightly brighter magnitudes ($M_{\mathrm{UV}}$ = -19.85, -19.96). Another galaxy at the same redshift and also very close in sky coordinates, ID 829 is listed in Table \ref{tab:summary_data} as a tentative emitter.  There are two more spectroscopically confirmed sources reported at a very similar redshift, ID 499 $z$ = 7.171 and ID 1038 at $z$ = 7.196 \citep{Mascia2023_CEERS, Tang2023}, which are not included in the present study  because they were observed with the medium-resolution grating NIRSpec configurations (G140M/F100LP, G235M/F170LP and G395M/F290LP). In particular, ID 499 is spatially very close at a distance of 0.28 pMpc (physical Mpc) from  ID 498, although its spectrum in the G140M/F100LP configuration does not show the \lya\ emission line. We consider ID 499 to be part of the structure and  
in the following, we present a detailed analysis of the possible ionized bubble origin 
by computing the size of the ionized bubble $R_{\mathrm{ion}}$ that the galaxies could carve through their ionizing photon capabilities to understand if this could justify the enhanced \lya\ visibility. We note that the separation between the most distant members along the line of sight (ID 80374 and ID 829) is 0.7 pMpc while in the transverse direction, the maximum distance between the five sources is 1.1 pMpc (between ID 80374 and ID 439). Therefore a rough estimate of the radius of the supposed ionized region leads to a value of 0.62 pMpc, which we can consider as a lower limit.   
The radius of the ionized bubble $R_{\mathrm{ion}}$ (see eq.~\ref{eq:Rion}) that each of our sources could carve either individually or as an ensemble, can be estimated following the approach detailed by \cite{Shapiro-giroux1987} and also used by other authors \citep[e.g.,][]{Matthee2018, Larson2022, Saxena2023B}, taking into account the ionizing photon output $\dot N_{\mathrm{ion}}$ (in units of $s^{-1}$), the ionizing escape fraction of the sources and the time since the source switched on: 
\begin{equation} \label{eq:Rion}
    R_{\mathrm{ion}}(t)  \propto \left( \frac{\dot{N}_{\mathrm{ion}} {f_{\mathrm{esc}}} {t}} {{H_0}{\Omega_b}{(1+z)^3}} \right)^{1/3} 
\end{equation}
the quantity $\dot{N}_{\mathrm{ion}}$, can be derived following eq (9) from \cite{Mason2020}, that we report here:
\begin{equation} \label{eq:Nion}
    \dot{N}_{\mathrm{ion}} = \frac{3.3 \times 10^{54}}{\alpha} 10^{-0.4 (M_{\mathrm{UV}}+20)} \left( \frac{912}{1500} \right)^{\beta +2} \text{s}^{-1}
\end{equation}
and depends on the $M_{\mathrm{UV}}$ and UV $\beta$ slope, assuming the spectral slope of the ionizing continuum $\alpha=1$, as found for galaxies at high $z$ with massive stars \citep[e.g.,][]{Steidel2014, Feltre2016}. 
For the LyC escape fraction we employ the inferred quantities $f_{\mathrm{esc}}^{\mathrm{LyC, pred}}$ derived in Sect.~\ref{sec:fesc_lyCmeasure} for the emitters and the value reported in Table 1 of \cite{Mascia2023_CEERS} for ID 499.
In Fig.~\ref{fig:newBubble} we plot the derived bubble radius as a function of time elapsed since the ionizing sources have switched on. The latter is of course unknown and, therefore, we let it vary in the range 0 to 200 Myr. 
Results are obtained both individually for the  5 sources (3 of which are \lya\ emitting galaxies)  and for the combined output. The grey area represents the 1$\sigma$ confidence region which takes into account the uncertainties in the $M_{\mathrm{UV}}$, $\beta$ and $f_{\mathrm{esc}}^{\mathrm{LyC, pred}}$ values, with the largest source of uncertainty being that on the $f_{\mathrm{esc}}^{\mathrm{LyC, pred}}$.\\
Considering the joint ionizing output of the five confirmed sources, the size of the region that is ionized becomes compatible with the measured physical radius of the region for ages larger than $\sim 50$ Myr, and with 1 pMpc for ages larger than $\sim$190  Myr. Therefore the bubble could have been produced by these 5 sources if they have all switched on earlier than this time, of if they have been "helped" by additional sources of ionizing radiation in the vicinity. Indeed the photometric redshift catalog for the CEERS field (Finkelstein et al. in preparation) includes 4 more sources with photometric redshifts in the range [7.03-7.27] and in the same NIRSpec pointing (see Fig.~\ref{fig:newBubble}). Three of them were placed in the MSA but no features or continuum are detected, due possibly to their faint magnitude (F115W $\sim$ 28.4), while the fourth one (which is slightly brighter having F115W = 26.75) was not selected for the spectroscopic observations. Even if these four sources had relatively modest ionizing production and $f_{\mathrm{esc}}^{\mathrm{LyC, pred}}$ as the five confirmed ones, their additional contribution would help explain the formation of the large observed ionized bubble, that would allow the \lya\ to be observed. \\
At variance with previous early reionized regions \citep[e.g.,][]{Castellano2018, Matthee2018, Tilvi2020, Endsley2022, Leonova2022}, this bubble is not dominated by a UV bright source, whose large photon production might clear the path for the \lya\ emission to become visible also from fainter nearby sources.
In this case all galaxies are relatively faint (-18 < $M_{\mathrm{UV}}$ < -20) and have also modest inferred Lyman continuum escape fractions (0.02 -- 0.23), and it is their combined output that creates a large enough ionized region. 
%
%
%
%
Similarly, four \lya\ emitting galaxies (ID = 80432, 80372, and 80445 with S/N > 3 and the  ID = 80239 with S/N = 2) are found in the redshift range (7.48 -- 7.51)  withing a maximum  physical distance of 2.5 pMpc. Another galaxy in the CEERS field was confirmed at $z$ = 7.473 from grating spectra \citep{Mascia2023_CEERS}, but it is located several pMpc away from the four emitters. This structure was also noted by \cite{Chen2023} and it appears indeed much more extended than the one  at $z$ = 7.18. The brightest source in the structure is ID 80432 with a $M_{\mathrm{UV}}$ = -20.22 $\pm$ 0.10 and a high predicted escape fraction of $f_{\mathrm{esc}}^{\mathrm{LyC, pred}}$ = 0.34. Following the same approach detailed before, we find that the ionizing photons coming from ID 80432 dominate over the contribution of the other three sources, which are all  fainter and have very low inferred $f_{\mathrm{esc}}^{\mathrm{LyC, pred}}$. 
We also checked the CEERS photometric redshift catalog (Finkelstein et al. in preparation) to search for possible photometric candidates in the region, and although there are several sources with photometric redshift compatible with $z$ $\sim$ 7.5, they seem to be spatially offset from the confirmed emitters. It is therefore unclear if the sources reside in a unique very large ionized region ($\sim$ 2 pMpc), or alternatively if they are in separate smaller bubbles (< 1 pMpc). High resolution spectroscopy could shed light on this point, by providing the missing information on the \lya\ velocity offset: if small offsets are observed then the line emission is consistent with a single large ionized bubble \citep[e.g., see Figure 1 from][]{Mason2020}.


\subsection{Comparison with ground based observations}
\label{sec:compareXlyA}
The \lya\ visibilities we have derived in Sect.~\ref{sec:XlyA} are somewhat at odds with previous ground based derivations. In particular our fractions are lower than most previous determinations at $\sim$ 5 and 6 and higher at $z$ $\sim$ 7.
We first remark that prior to the advent of \jwst\ most samples were identified on the basis of photometric redshifts whenever \lya\ was not present, while in this study all the galaxies used to compute \lya\ visibility are spectroscopically confirmed, thanks to the identification of optical lines in the spectra. This, together with the lack of telluric lines from space improves dramatically  our ability to identify \lya\ even at low SNR, and to probe secure limits when the line is not detected.\\
Also, as noted in Sect.~\ref{sec:lyavis}, the discrepancy is clearly driven by the field-to-field variations and by the inhomogeneity of the IGM conditions and the possible presence of peculiar ionized regions in some fields. However, the discrepancy might also be due to the different \lya\ flux that is measured in NIRSpec MSA spectra compared to ground based slit spectroscopy.
The different EW values recovered have already been noted for few individual objects that boast both JWST and ground based slit spectroscopy observations (including some of the CEERS galaxies) \cite[e.g., by][]{Chen2023, Larson2023, Jiang2023, Tang2023}. In most cases the derived \lya\ fluxes are smaller when measured from the NIRSpec MSA observations. Recently \cite{Jiang2023} also discussed the non detection by deep JWST NIRSpec PRISM spectroscopy of a previously known \lya\ emitting galaxy in the A2744 field at $z$ = 5.66.  
\\
The MSA pseudo slits are indeed  much smaller than the slits employed in ground based programs, which are typically matched to the seeing, and vary from 0.7$''$ to 1.0$''$ \citep{Schenker2012, Pentericci2014}. The MSA might therefore miss part of the \lya\ flux due to various effects including  i) the presence of spatial offsets between the UV and \lya\ emission, with MSA typically placed on the rest-frame UV emission barycenter; ii) a spatially extended diffuse \lya\ emission, due to resonant scattering of the \lya\ photons on the neutral hydrogen inside the galaxies; iii) differential dust extinction and inhomogeneous neutral ISM structure within the galaxies. 
\\
\cite{Hoag2019} presented the first systematic study of \lya\ vs UV offsets finding that they are in general small, of the order of 0.2$''$ at $z \sim$ 5 (corresponding to 1.3 kpc) and  decrease towards high redshifts. This trend is also consistent with the most recent results obtained by \cite{Ning2023} on a small sample of bright \lya\ emitting sources in the COSMOS-Web survey also observed with NIRCam, for which they find a median offset of 0.12$''$ at $z \sim$ 6, and by \cite{Lemaux2021}, who studied a large sample of emitters, including lensed sources and found an average offset of 0.11$''$ at 5 < $z$ < 7. Therefore, the predicted offset at $z$ = 7 would be smaller than the MSA size but comparable to the MSA size at lower redshift. This could imply that the discrepancy between the ground based results and JWST ones is more significant at $z<6$ than at higher redshift, which could partially explain our lower \lya\ fractions at $z$ = 5 and 6.
\\
Even in the absence of significant offset, the \lya\ emission could be more spatially extended than the UV emission, due to the fact that \lya\ photons are resonantly scattered by the neutral hydrogen atoms inside the galaxies. This was first shown by narrow band imaging  capturing the \lya\ emission \citep{steidel2011} and more recently by MUSE observations \citep{leclercq2017, Kusakabe2022} indicating  that a large fraction (around 80\%) of high redshift star forming galaxies show diffuse \lya\ halos, whose scale length is up to 10 times larger than the UV emission (although with a very large scatter). Such halos seem to be more important for very bright objects (see also \cite{matthee2020}). The contribution of the halo to the total \lya\ flux is at least a factor of 2, and this quantity  does not seem to be related to other galaxy properties.
The median scale length of the halos of $\sim$ 4 -- 5 kpc \citep{leclercq2017} means that the large slits employed in ground based observations, and the IFUs, can still recover most of the flux, while the same would not be true for the much smaller MSA pseudoslits. Again, this could be a redshift dependent effect, with the discrepancy becoming less evident at high redshift when galaxies become more compact.
Finally, dust produced in the star forming events suppress the \lya\ emission and, depending on its spatial distribution, it could differentially attenuate some regions more than others. Additionally, the uneven structure of the neutral ISM within the galaxy, as highlighted by \cite{Hu2023} for local galaxies, influences \lya\ visibility. These combined effects also contribute to produce the mismatch between ground-based and space observations.
\\
As demonstrated by \cite{Maiolino2023b} for the AGN GN-z11, NIRSpec-IFU observations in the future could help us to  interpret this mismatch and reconcile space and ground based \lya\ observations. By capturing flux from larger regions, we could investigate the spatial extent of \lya\ emission, as previously done for low-z \lya\ emitters from ground based MUSE observations and understand  how it impact the derivation of the \lya\ visibility evolution.

\section{Summary} \label{sec:Conclusion}
We have presented the results of a comprehensive study of \lya\ emission from 4 $< z <$ 8.5 galaxies in the NIRSpec/PRISM dataset of  the CEERS survey targeting the EGS field. The sample consists of 144 galaxies, each with secure spectroscopic redshifts identified through multiple optical line detection. We identify 43 secure (S/N > 3) \lya\ emitting galaxies and 7 tentative ones (2 < S/N < 3), while  94 galaxies have no \lya\ emission. 
We supplemented the CEERS dataset with 91 additional galaxies with published data observed with the same instrument configuration from the JADES GOODS-South survey, of which 15 have \lya\ in emission. 
\\
We summarize our main results in the following:
\begin{itemize}
\item We compute \lya\ escape fractions ($f_{\mathrm{esc}}^{\lya}$) and we explore the correlations between $f_{\mathrm{esc}}^{\lya}$ and the  physical properties of the galaxies. 
We measure low values of $f_{\mathrm{esc}}^{\lya}$ in dusty, massive, UV bright galaxies. These sources probably have larger amounts of hydrogen and dust in their ISM and, due to the resonant scattering of \lya\ photons, these characteristics attenuate the emission we observe. We find only a marginal anti-correlation with the half-light radius. 
The correlations do not vary much with redshift.
\item We predict the escape fraction of LyC photons ($f_{\mathrm{esc}}^{\mathrm{LyC, pred}}$) using two empirical relations, based on well-explored indirect indicators derived at $z \sim 0.3$. 
Contrary to the expectation from lower redshift results and from theoretical models of LyC escape, we do not find any solid correlations between $f_{\mathrm{esc}}^{\lya}$ and $f_{\mathrm{esc}}^{\mathrm{LyC, pred}}$ either in the ionized universe ($z$ < 5.5) or during the epoch of reionization, although the sample considered is rather small and the uncertainties on the LyC escape fractions are still significant.
\item We determine the redshift evolution of the \lya\ visibility (X$_{\lya}$), i.e. the fraction of galaxies with a \lya\ rest frame equivalent width EW$_0$ > 25 \AA\ (and > 50 \AA) in four redshift bins at 4.5 $< z <$ 8.5, for the CEERS and the JADES subsets separately and as a whole. We find significantly lower \lya\ fractions at $z$ = 5 and 6 compared to previous ground-based observations. At $z$ = 7, while the JADES fraction is consistent with previous results, the visibility in the CEERS field appears much enhanced, probably due to two  ionized regions at $z$ = 7.18 and $z$ = 7.49. We highlight that the 
average $X_{\lya}$ derived from the combined CEERS+JADES data continues to rise from $z$ = 5 to 7, dropping again at $z$ = 8. This implies that either our results are not consistent with a rapid end of reionization, or that a rapid reionization ending is still characterized by >1 pMpc bubbles around a high fraction of modestly bright galaxies, to at least $z \sim$ 7.
\item We further investigate the effect of cosmic variance whose substantial effect on the observational data is suggested by the large difference of X$_{\lya}$ between CEERS and JADES. We employed the Cosmic Dawn II (CoDaII) simulation to calculate the predicted \lya\ visibility at $z$ = 7 in a $\sim$ 760,000 cMpc$^3$ volume. The uncertainty range of the mock measurement is similar to the difference between the two fields and stresses the paramount importance of acquiring more data on a larger number of independent fields, to obtain a more robust redshift evolution of the \lya\ emitter fraction.
\item We further characterize the two ionized regions in EGS at $z$ = 7.18 and $z$ = 7.49, which respectively have three and four \lya\ emitting galaxies. We compute the radius of the ionized bubble $R_{\mathrm{ion}}$ that each of the spectroscopically-confirmed members of the ionized region could carve either individually or as an ensemble. For the $z$ = 7.18 bubble, a radius $R_{\mathrm{ion}} \sim$ 1 pMpc can be achieved by sustained star-formation activity over $\sim$ 190 Myr, or less if helped by additional ionizing sources.
\item We discuss the possible effects that can impact the measurement of \lya\ emission through the NIRSpec MSA pseudo slits, and how they can differ from the ground based slit measurements. The well studied presence of spatial offsets between \lya\ and UV emission, the spatially diffuse nature of \lya\ emission and  differential dust extinction all have substantial effect, which might change with redshift. Consequently, it is not possible to study the evolution of the \lya\ visibility combining low redshift results from ground based telescopes with z $\geq$ 7 measurements from JWST. 

\end{itemize}

Essential insights could emerge from future NIRSpec-IFU observations, aiding in harmonizing space and ground-based \lya\ observations. The potential to capture flux from larger regions opens avenues to explore the spatial extent of \lya\ emission, akin to previous studies on low-z \lya\ emitters using ground-based MUSE observations, offering valuable insights into how it impacts the derivation of the \lya\ visibility evolution. 
Future larger surveys covering additional fields will be needed to  reduce the chances of the samples being biased to overdense or underdense regions (as in the present study), overcome cosmic variance, and provide a number of sources large enough to compute the spatial variation of $\chi_{\mathrm{HI}}$ with high statistical significance.

\begin{acknowledgements}
We acknowledge support from the PRIN 2022 MUR project 2022CB3PJ3 - First Light And Galaxy aSsembly (FLAGS) funded by the European Union – Next Generation EU. L.N. thanks Arianna Favale for providing valuable insights into the MCMC fitting procedure.
\end{acknowledgements}

\bibliographystyle{aa}
\bibliography{biblio.bib}

\appendix
\section{\lya\ Emission lines fitted}\label{sec:app_figures}
In this appendix, we show the \lya\ line profiles fitted for the 50 \lya\ emitting galaxies reported in Table \ref{tab:summary_data} (including the 7 tentative ones). We present the best fit profiles in order of descending redshift. \\
\begin{table}[h]
\centering
\begin{tabular}{ccc}
\includegraphics[width=0.3\textwidth, height=0.25\textheight]{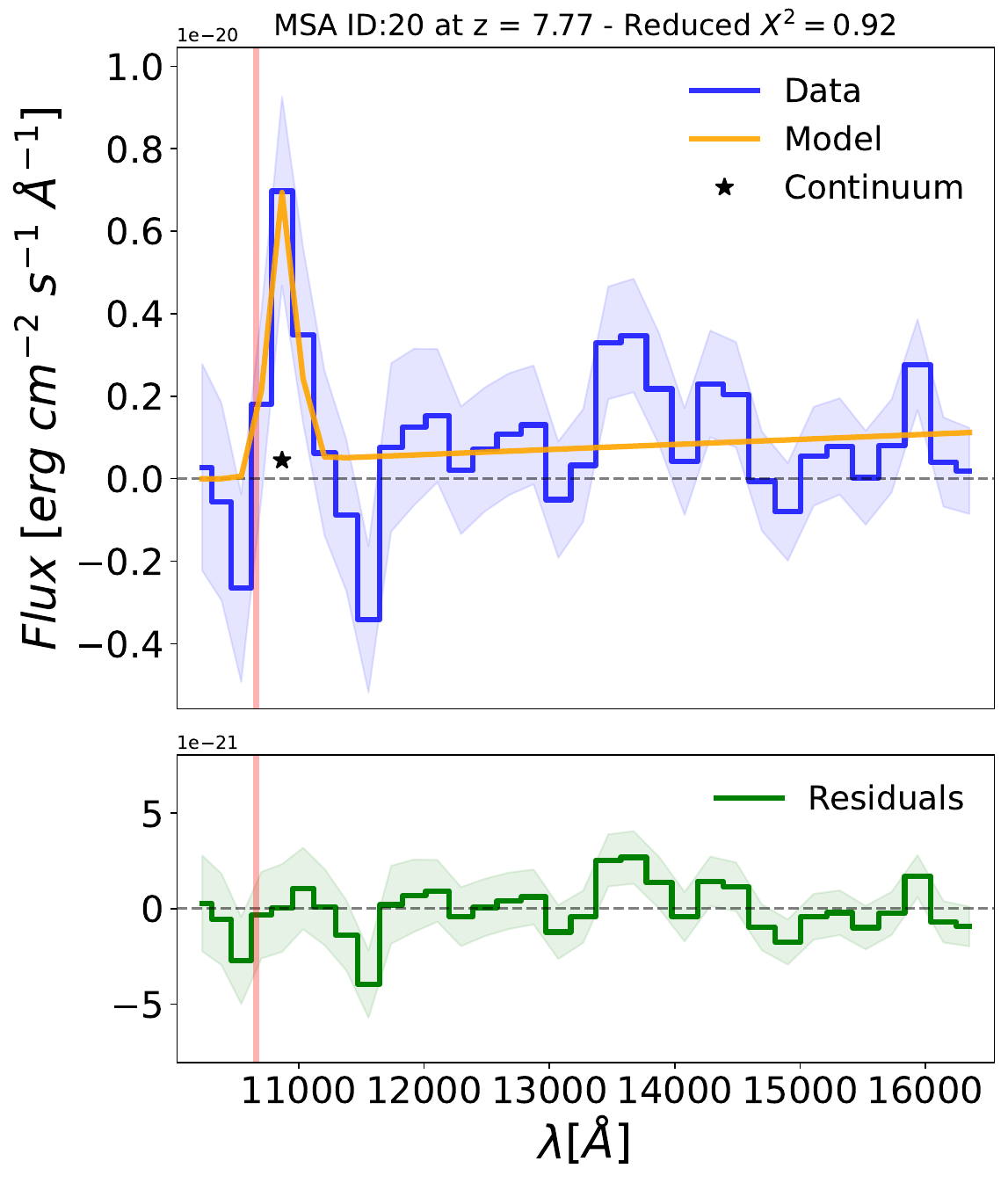} &
\includegraphics[width=0.3\textwidth, height=0.25\textheight]{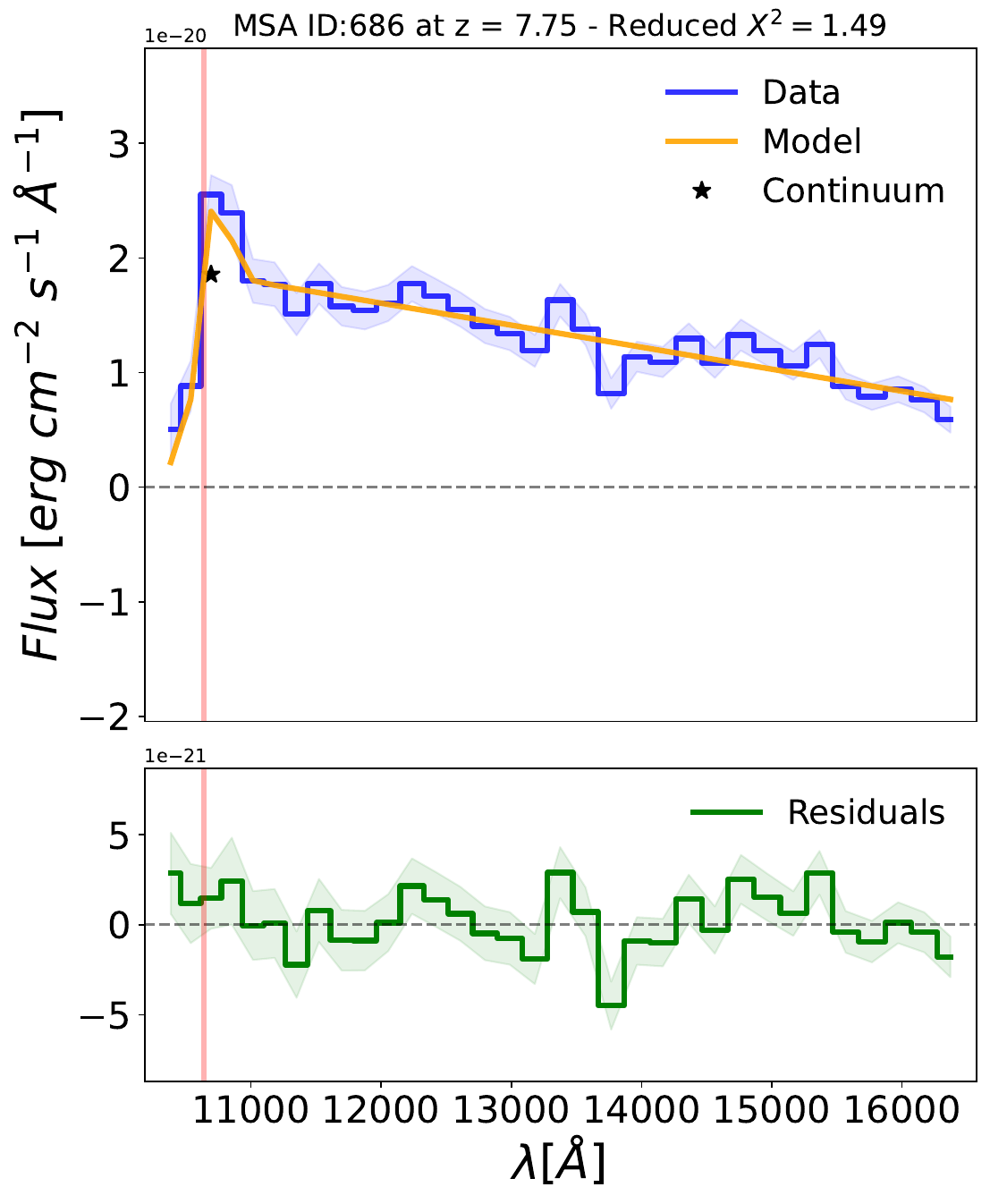} &
\includegraphics[width=0.3\textwidth, height=0.25\textheight]{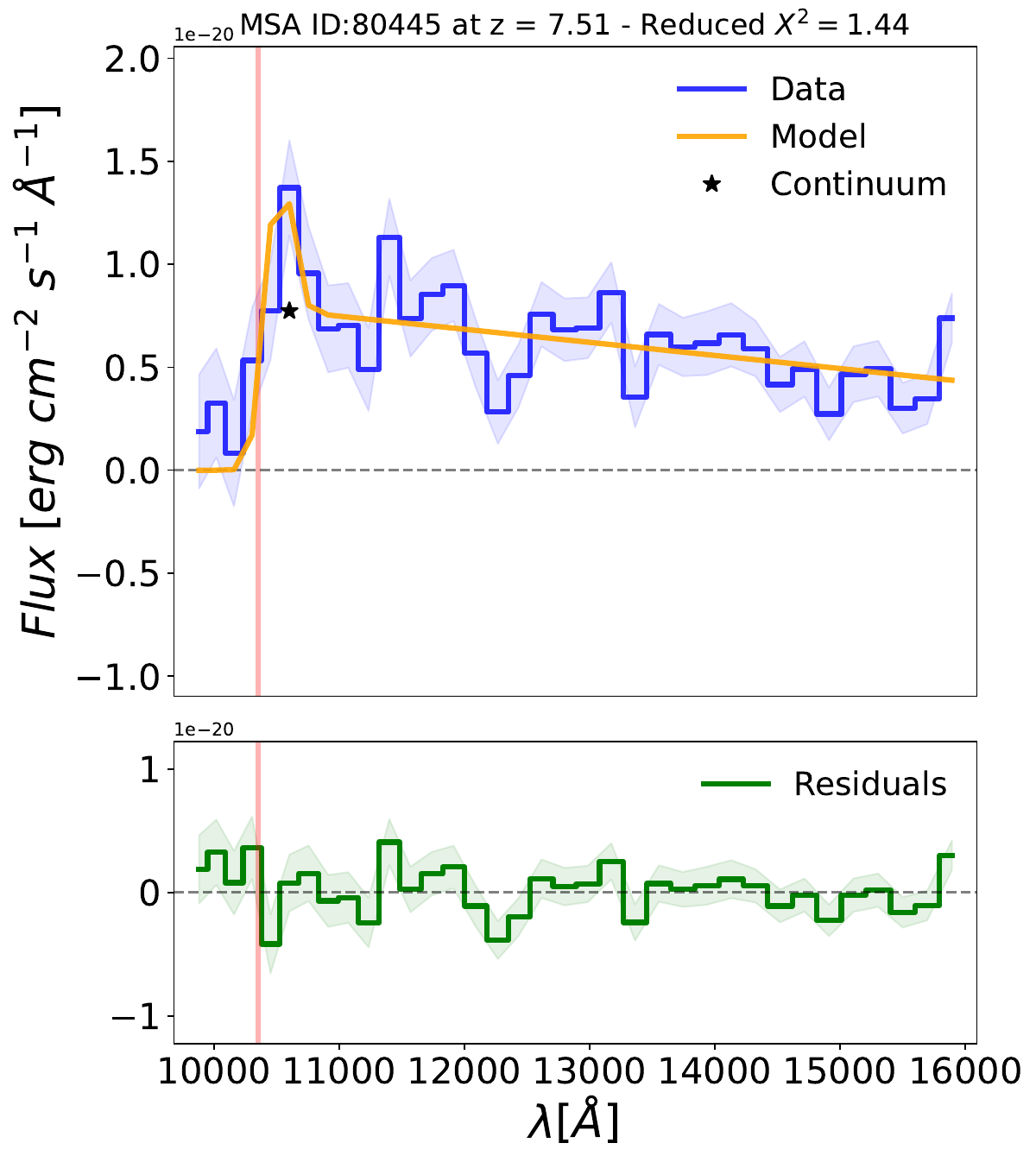} \\
\includegraphics[width=0.3\textwidth, height=0.24\textheight]{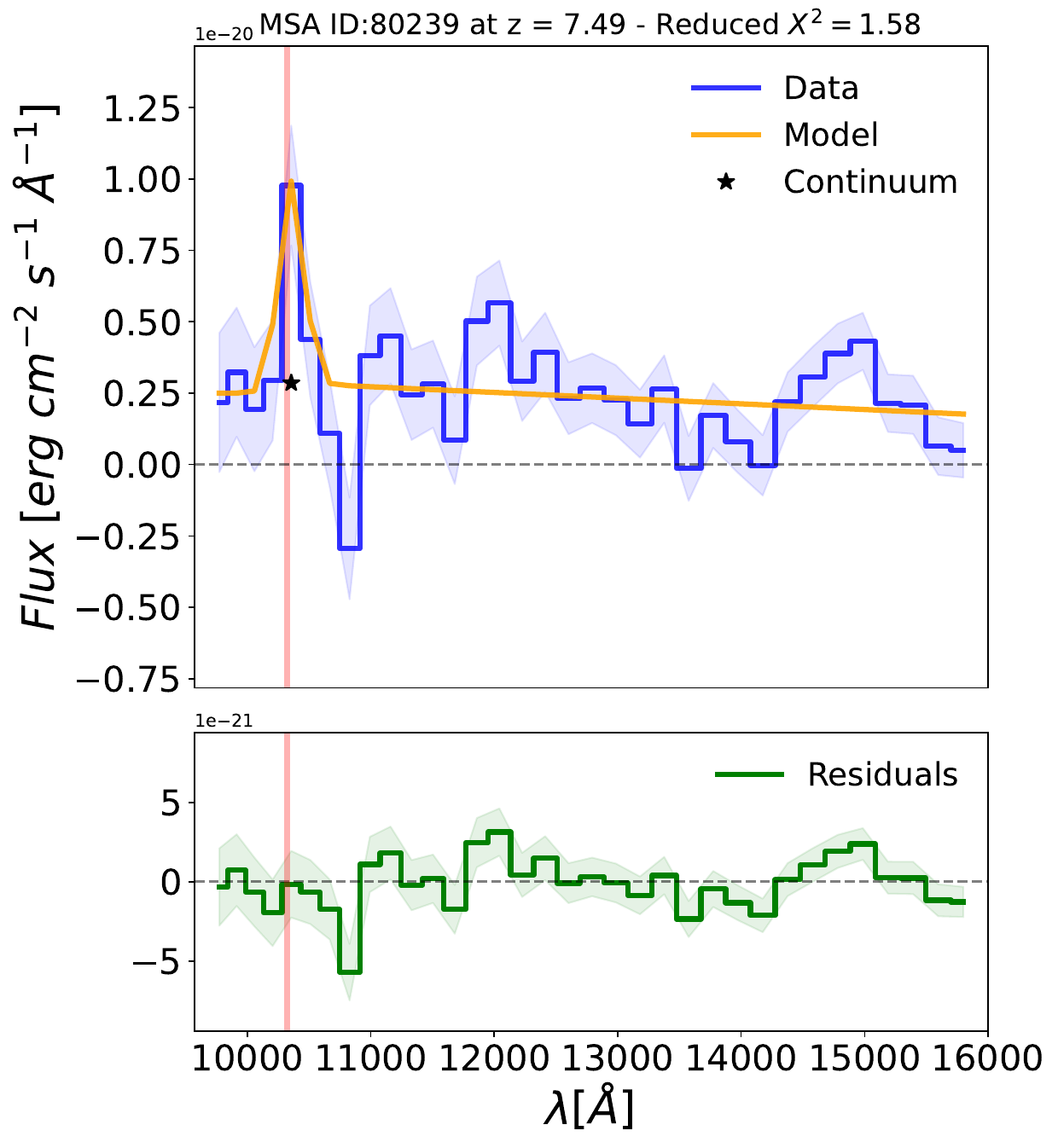} &
\includegraphics[width=0.3\textwidth, height=0.24\textheight]{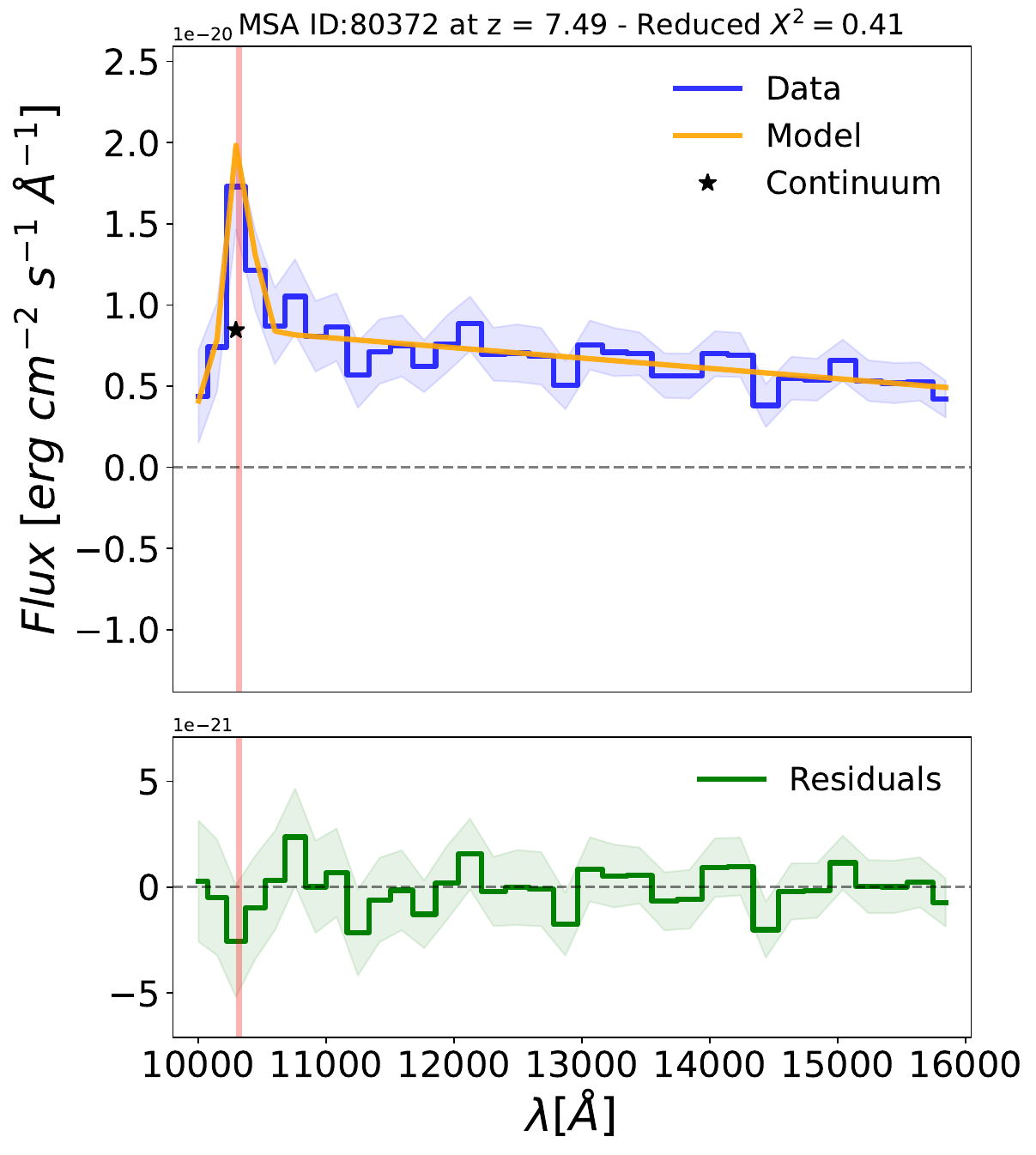} &
\includegraphics[width=0.3\textwidth, height=0.24\textheight]{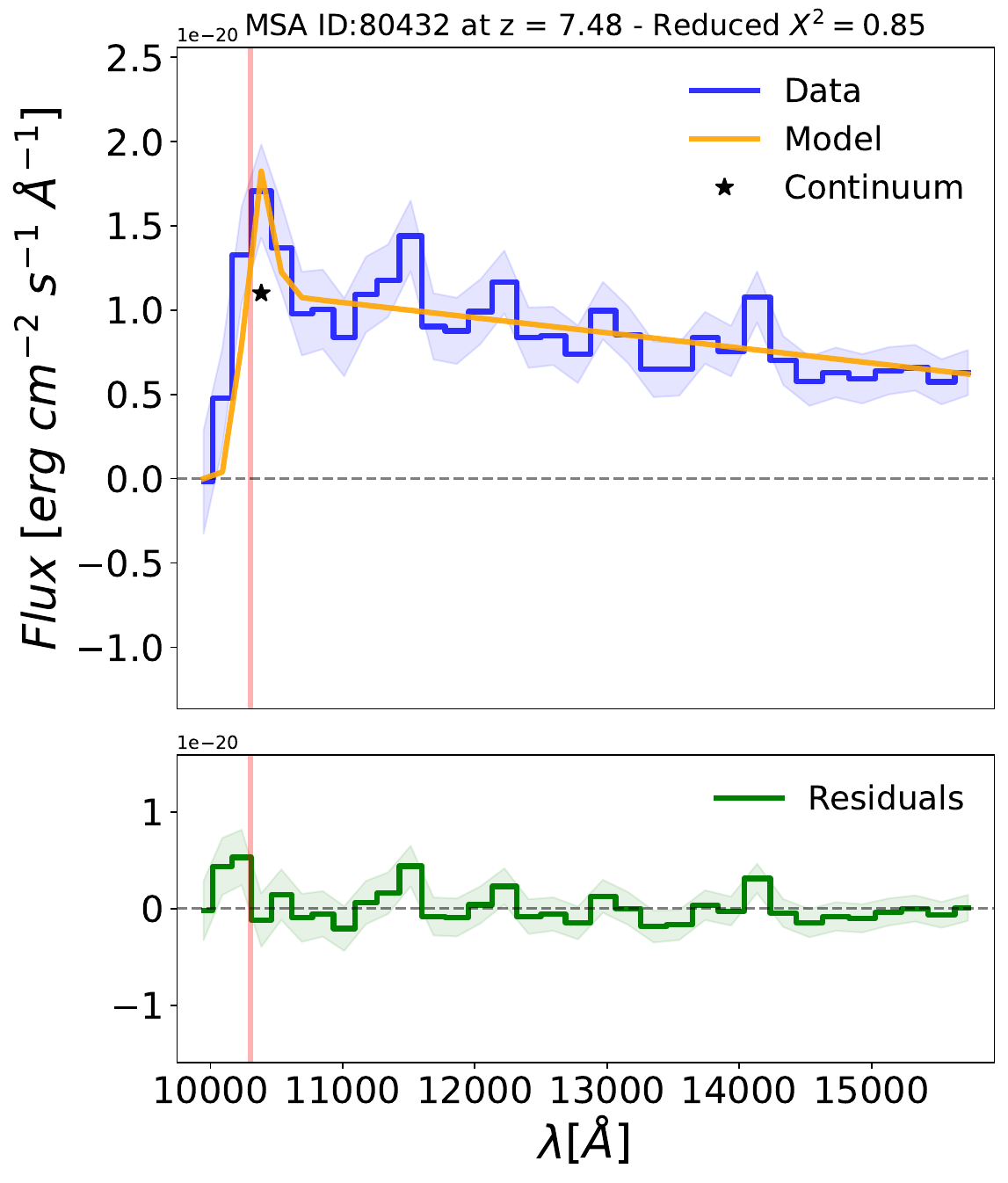} \\
\includegraphics[width=0.3\textwidth, height=0.24\textheight]{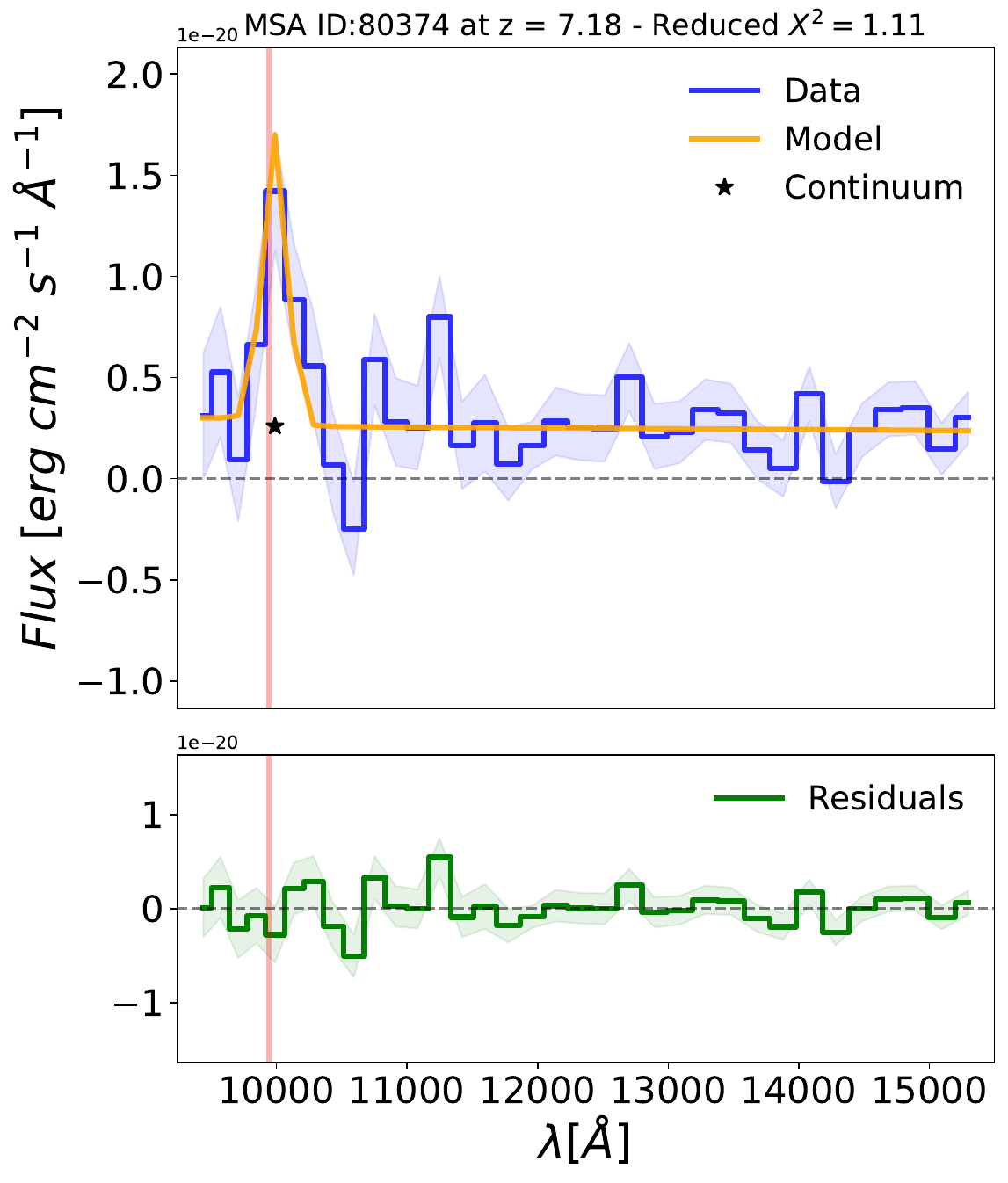} &
\includegraphics[width=0.3\textwidth, height=0.24\textheight]{images/P_439.pdf} &
\includegraphics[width=0.3\textwidth, height=0.24\textheight]{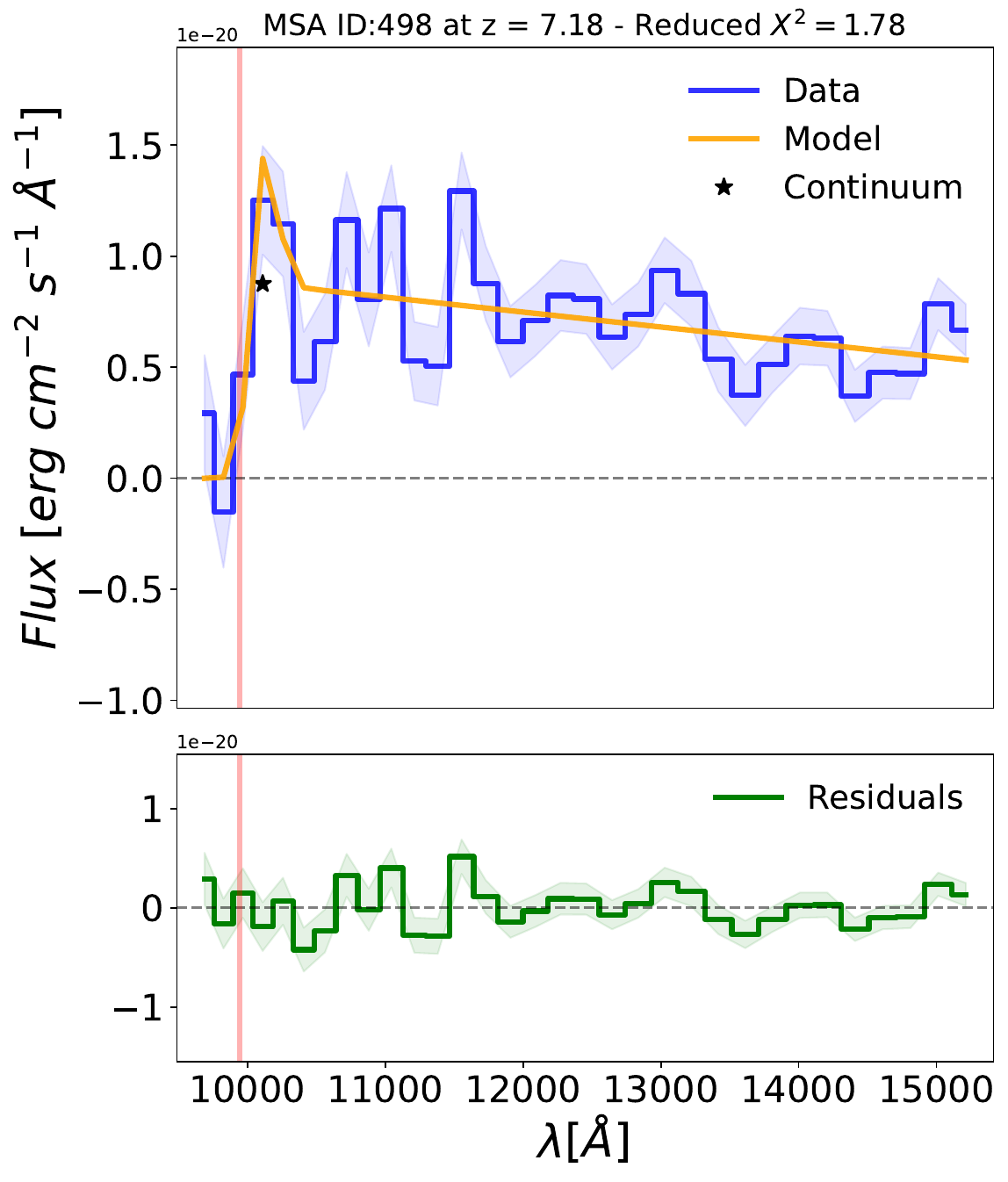} 
\end{tabular}
\caption{See the description of Fig.~\ref{fig:EW_fit}.}
\label{fig:EW_fit_new1}
\end{table}

\begin{figure*}[ht!]
\begin{minipage}{\textwidth}
\centering

\includegraphics[width=0.3\textwidth, height=0.24\textheight]{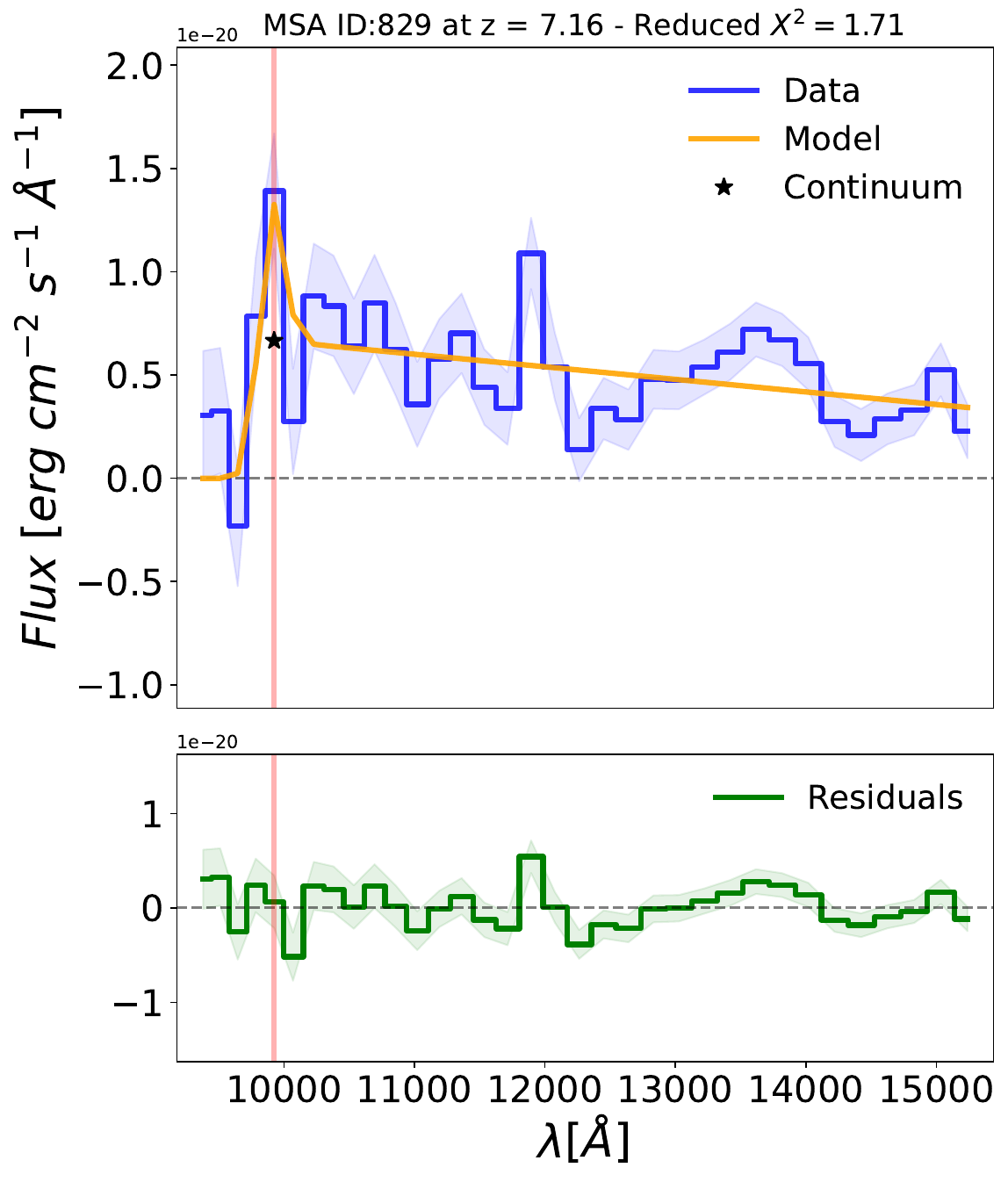}%
\hfill%
\includegraphics[width=0.3\textwidth, height=0.24\textheight]{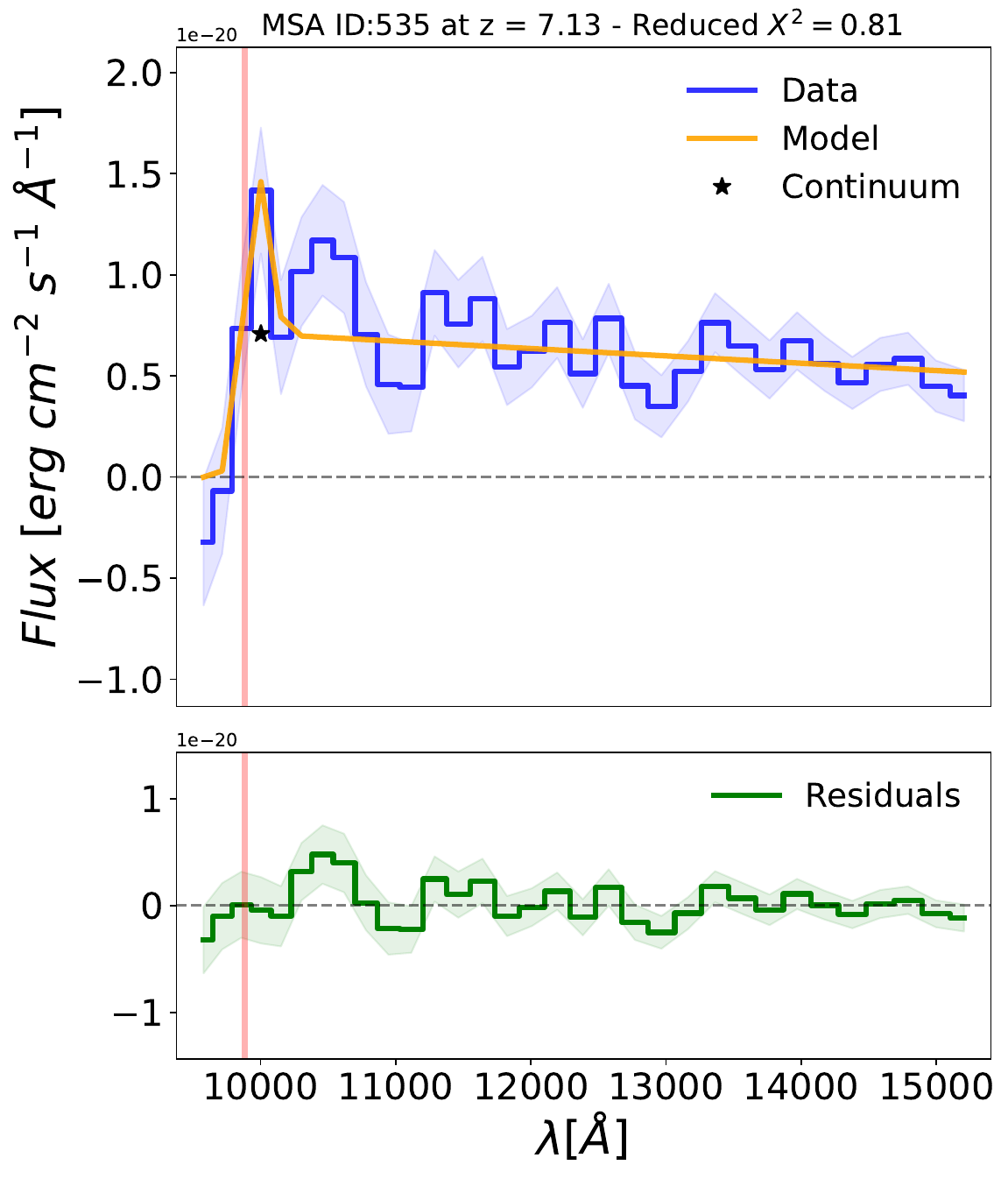}%
\hfill%
\includegraphics[width=0.3\textwidth, height=0.24\textheight]{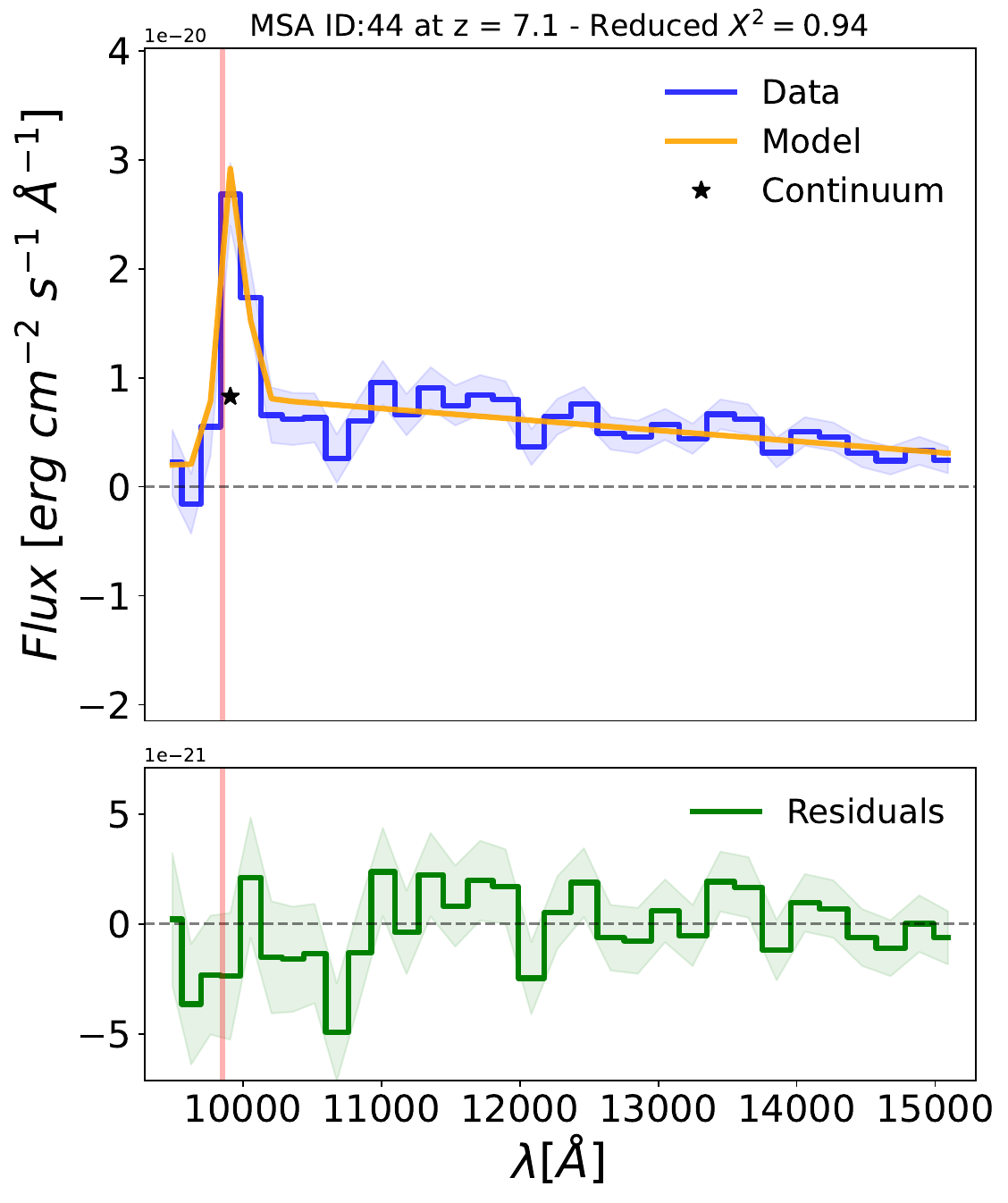}

\medskip

\includegraphics[width=0.3\textwidth, height=0.25\textheight]{images/P_1142.pdf}%
\hfill%
\includegraphics[width=0.3\textwidth, height=0.25\textheight]{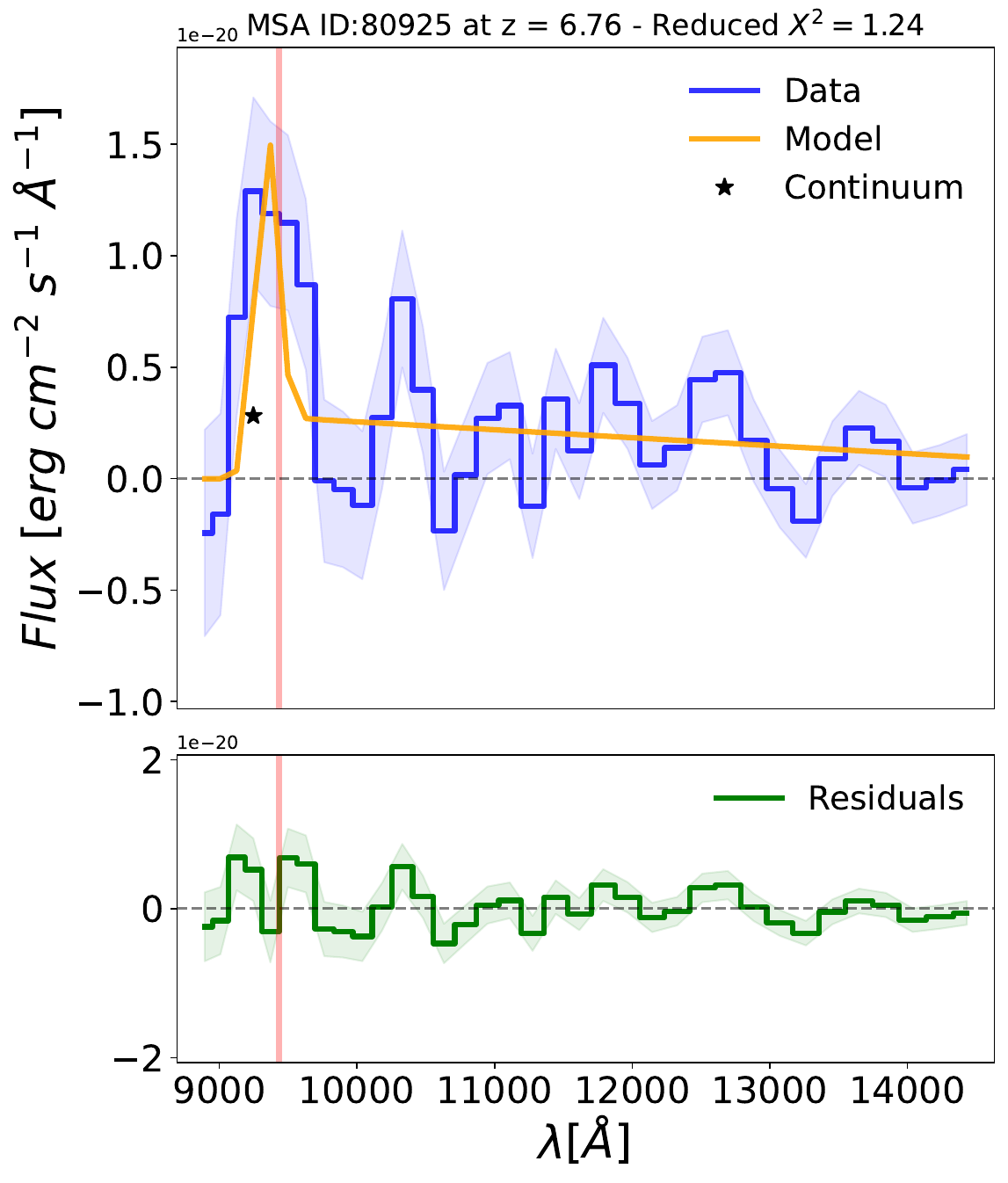}%
\hfill%
\includegraphics[width=0.3\textwidth, height=0.25\textheight]{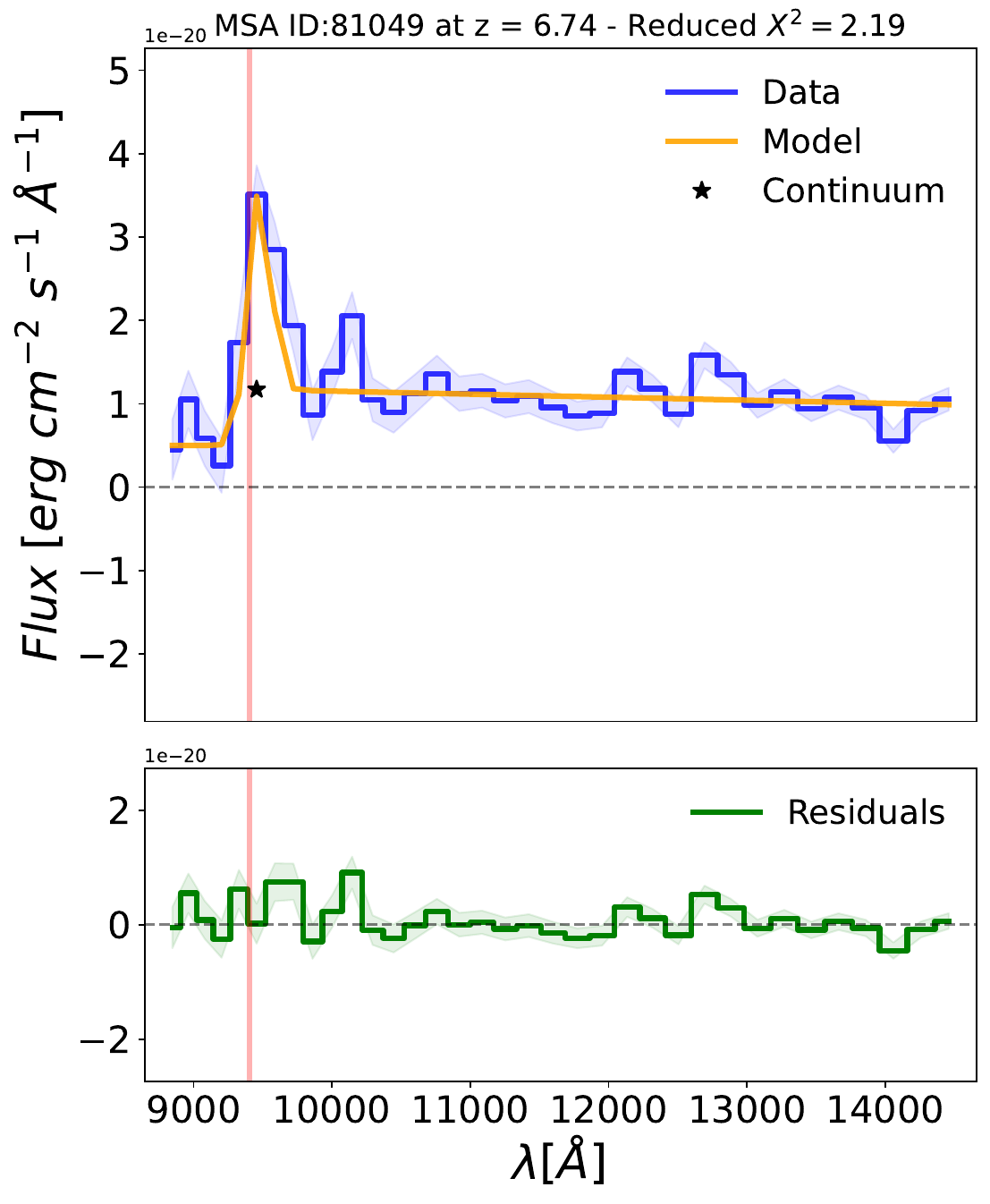}

\medskip

\includegraphics[width=0.3\textwidth, height=0.24\textheight]{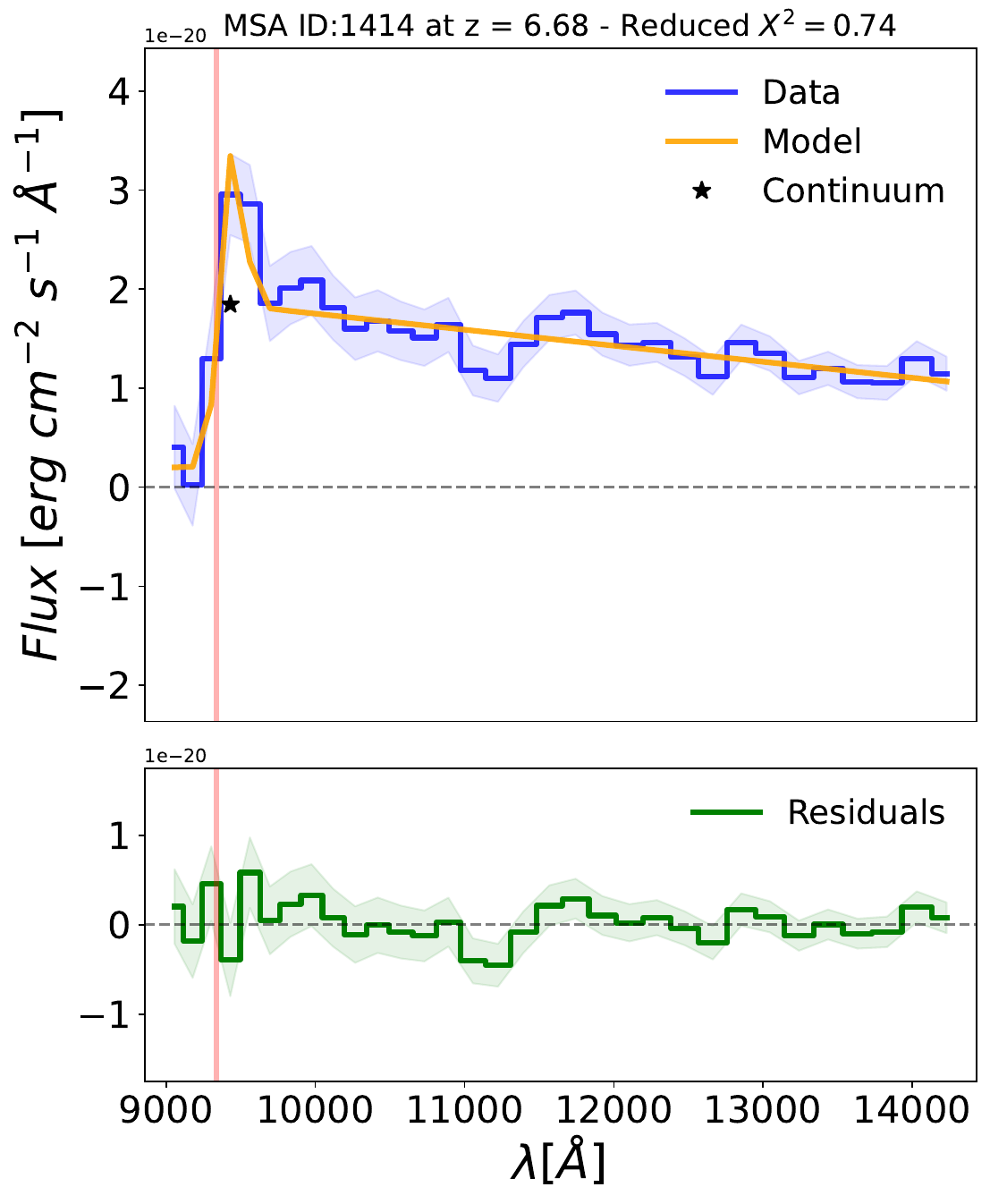}%
\hfill%
\includegraphics[width=0.3\textwidth, height=0.24\textheight]{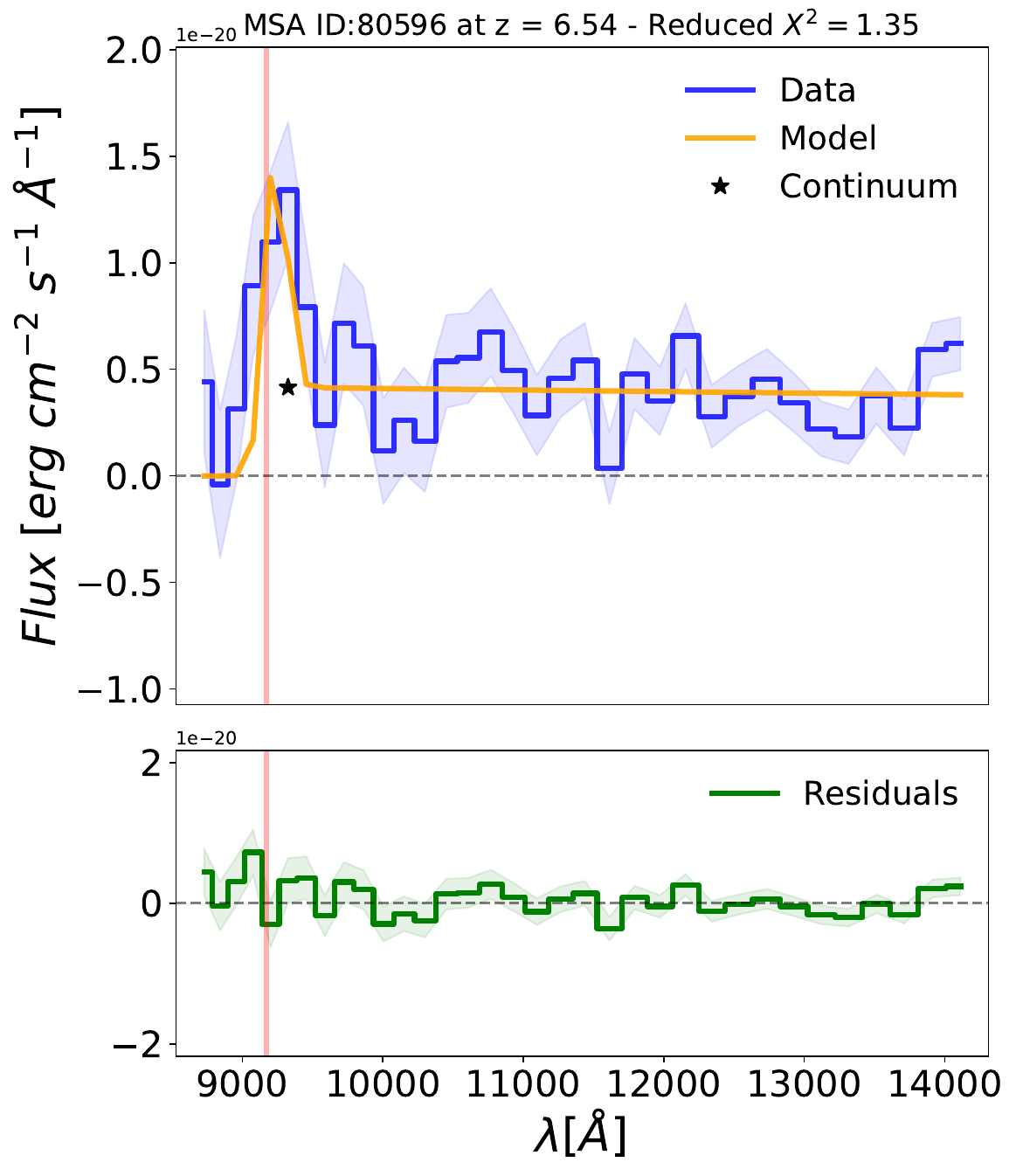}%
\hfill%
\includegraphics[width=0.3\textwidth, height=0.24\textheight]{images/P_1561.pdf}

\medskip

\includegraphics[width=0.3\textwidth, height=0.24\textheight]{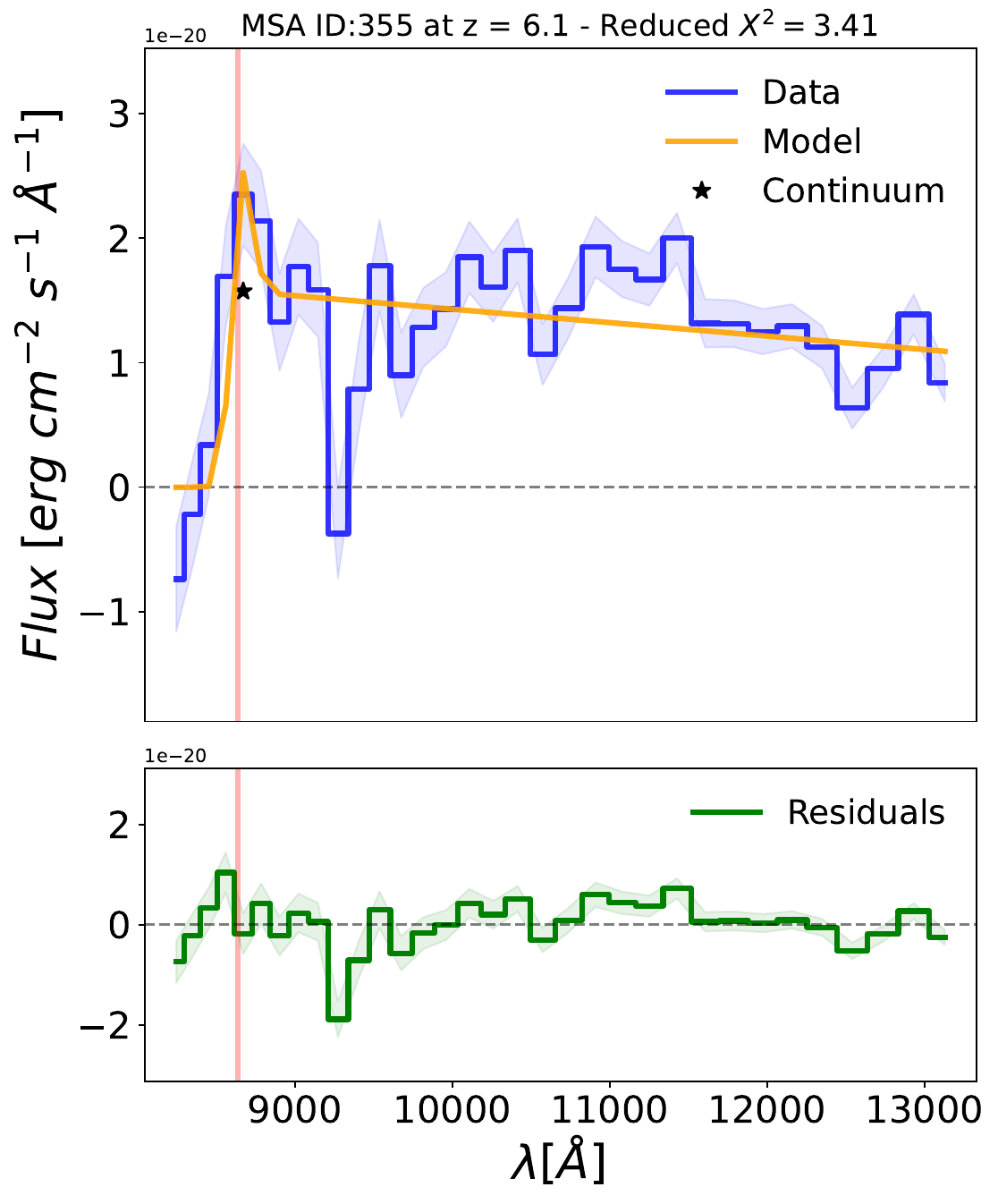}%
\hfill%
\includegraphics[width=0.3\textwidth, height=0.24\textheight]{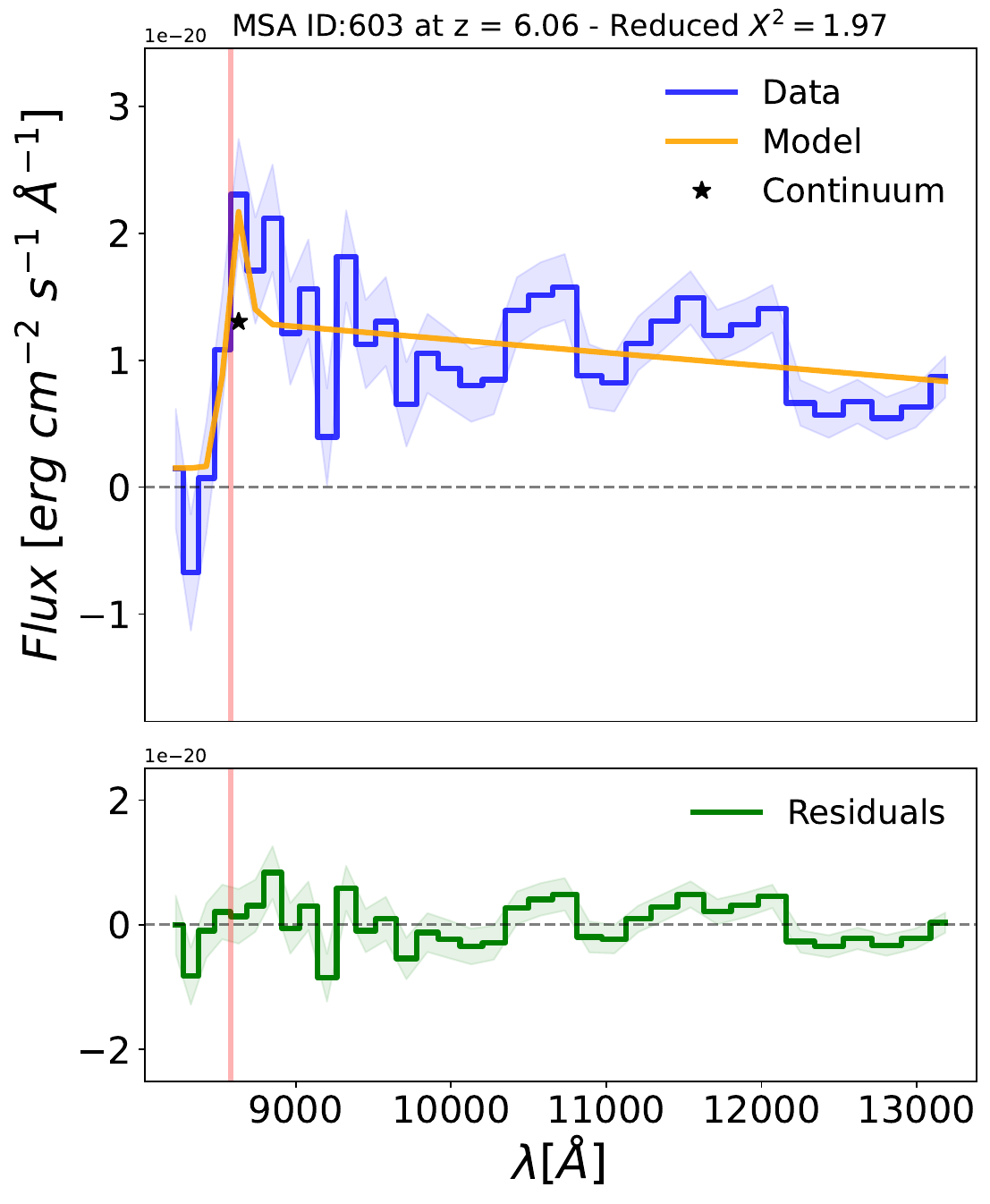}%
\hfill%
\includegraphics[width=0.3\textwidth, height=0.24\textheight]{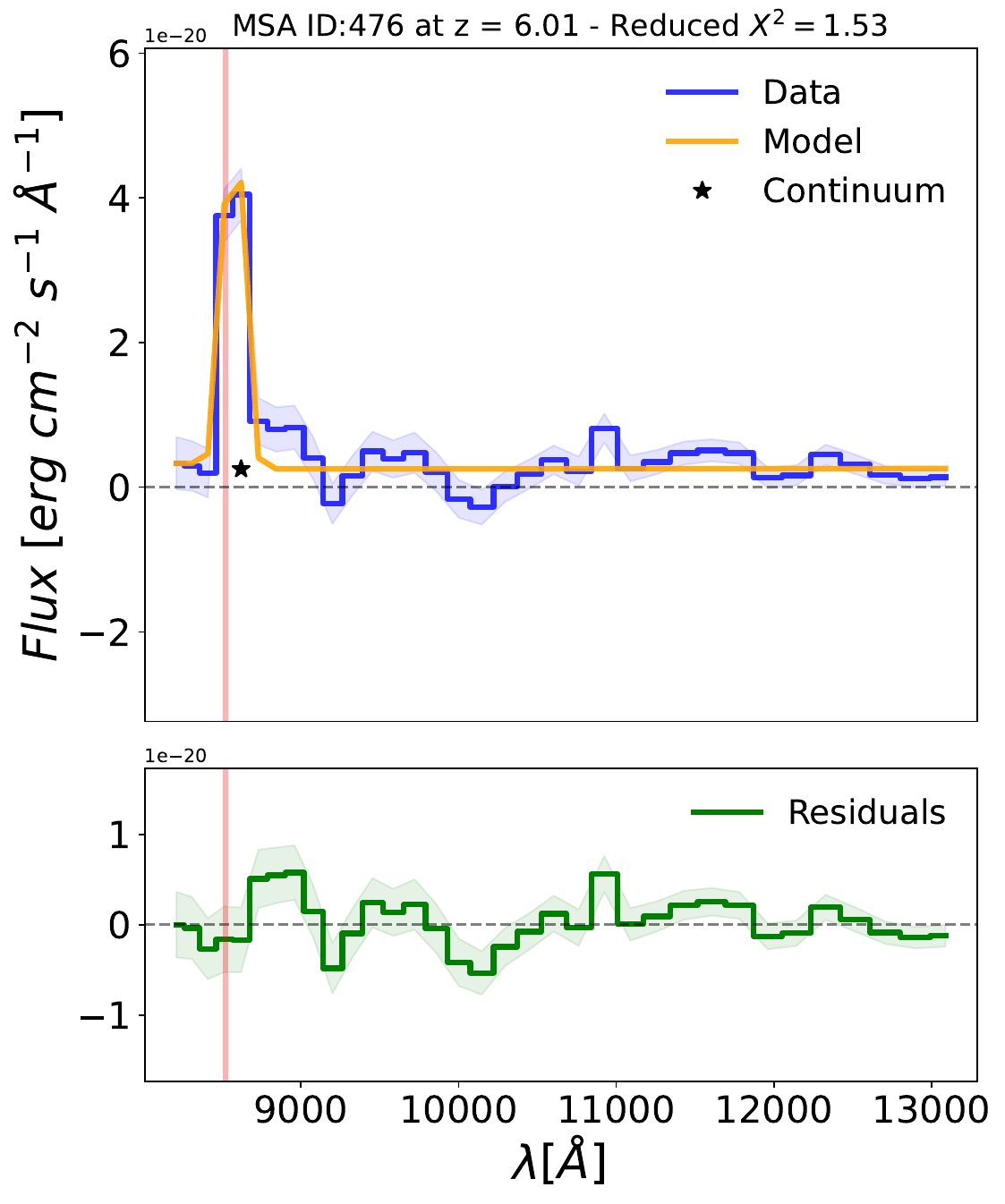}

\caption{See the description of Fig.~\ref{fig:EW_fit}.}
\label{fig:EW_fit_new2}
\end{minipage}
\end{figure*}
\begin{figure*}[ht!]
\centering

\includegraphics[width=0.3\textwidth, height=0.24\textheight]{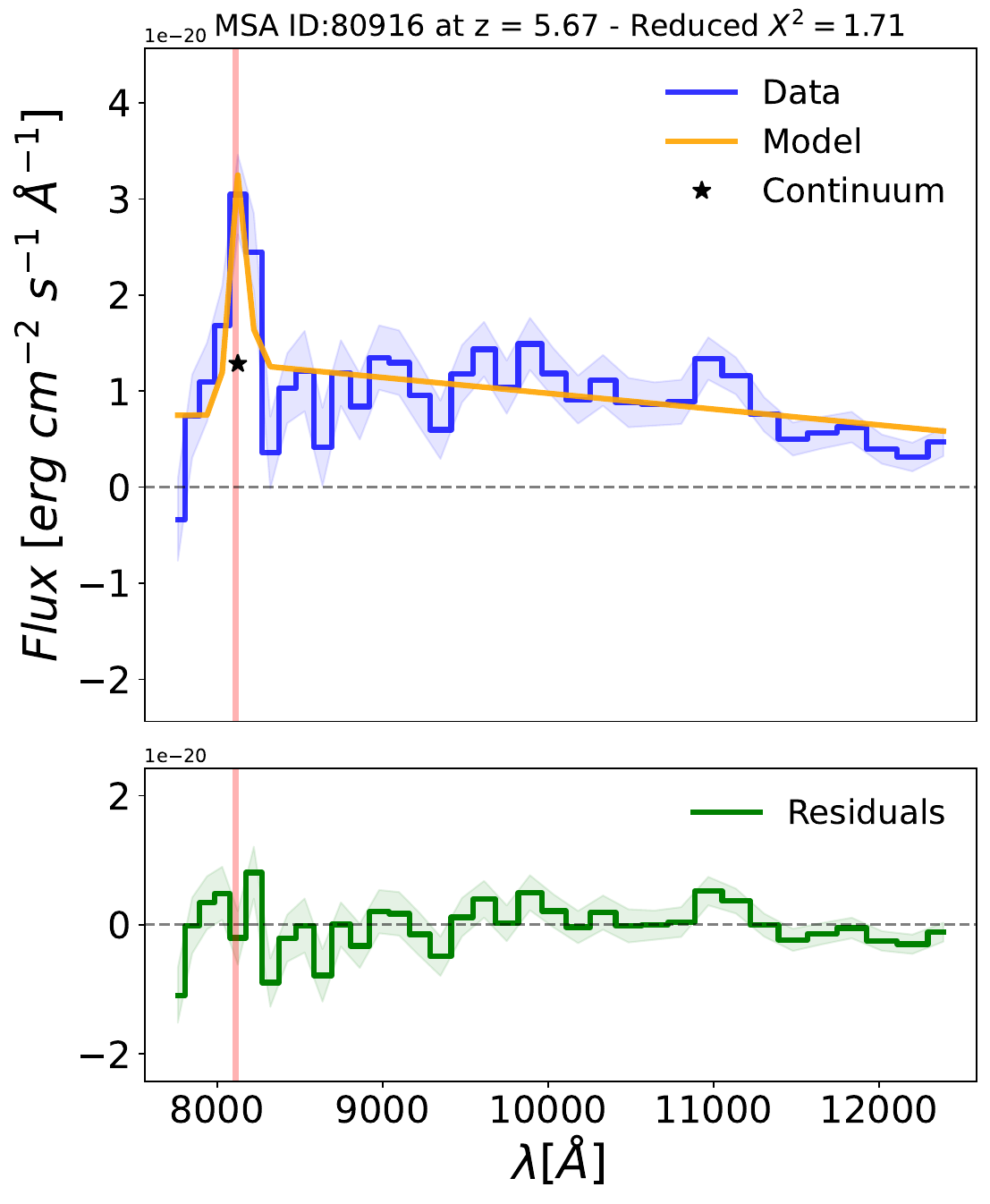}%
\hfill%
\includegraphics[width=0.3\textwidth, height=0.24\textheight]{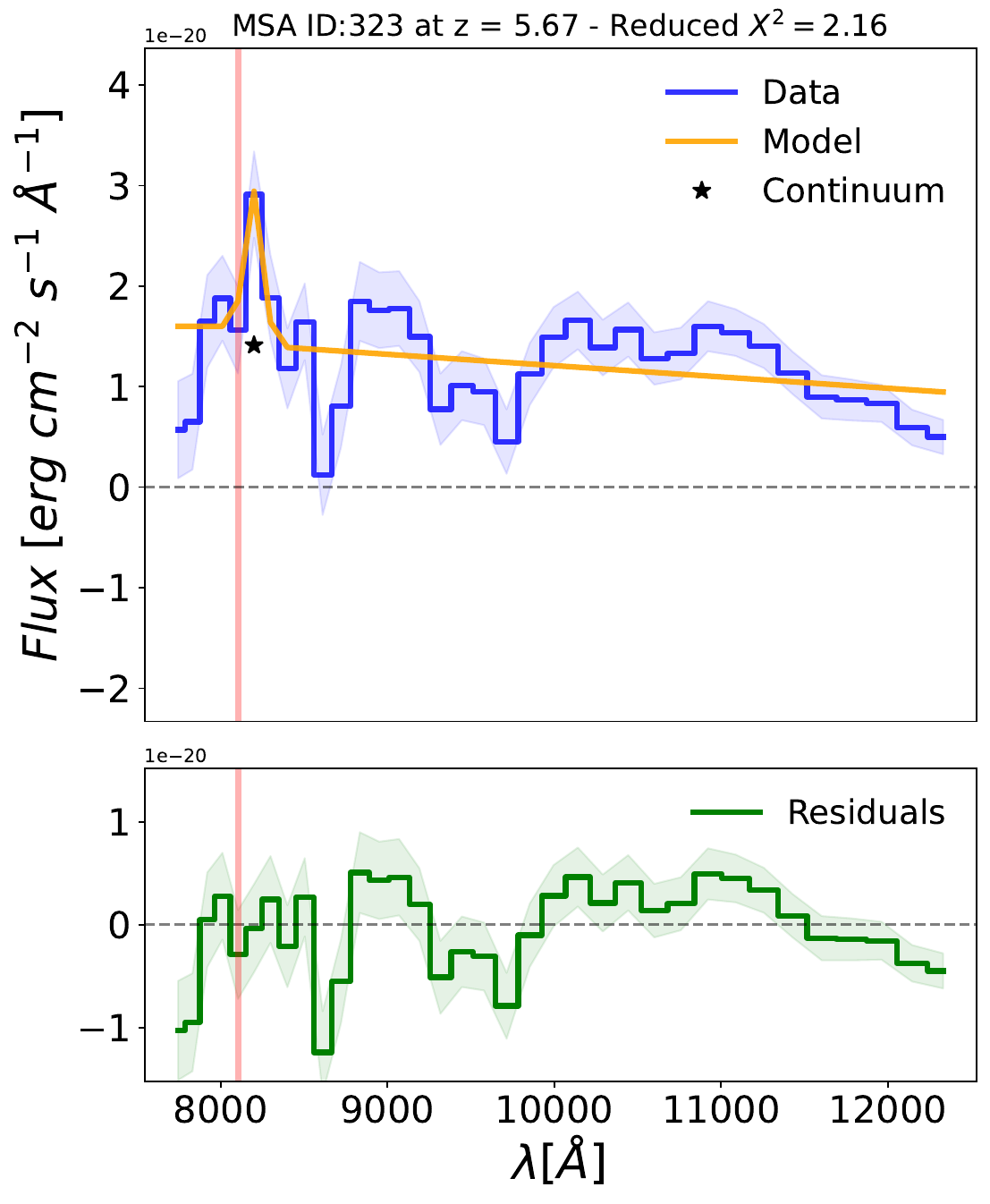}%
\hfill%
\includegraphics[width=0.3\textwidth, height=0.24\textheight]{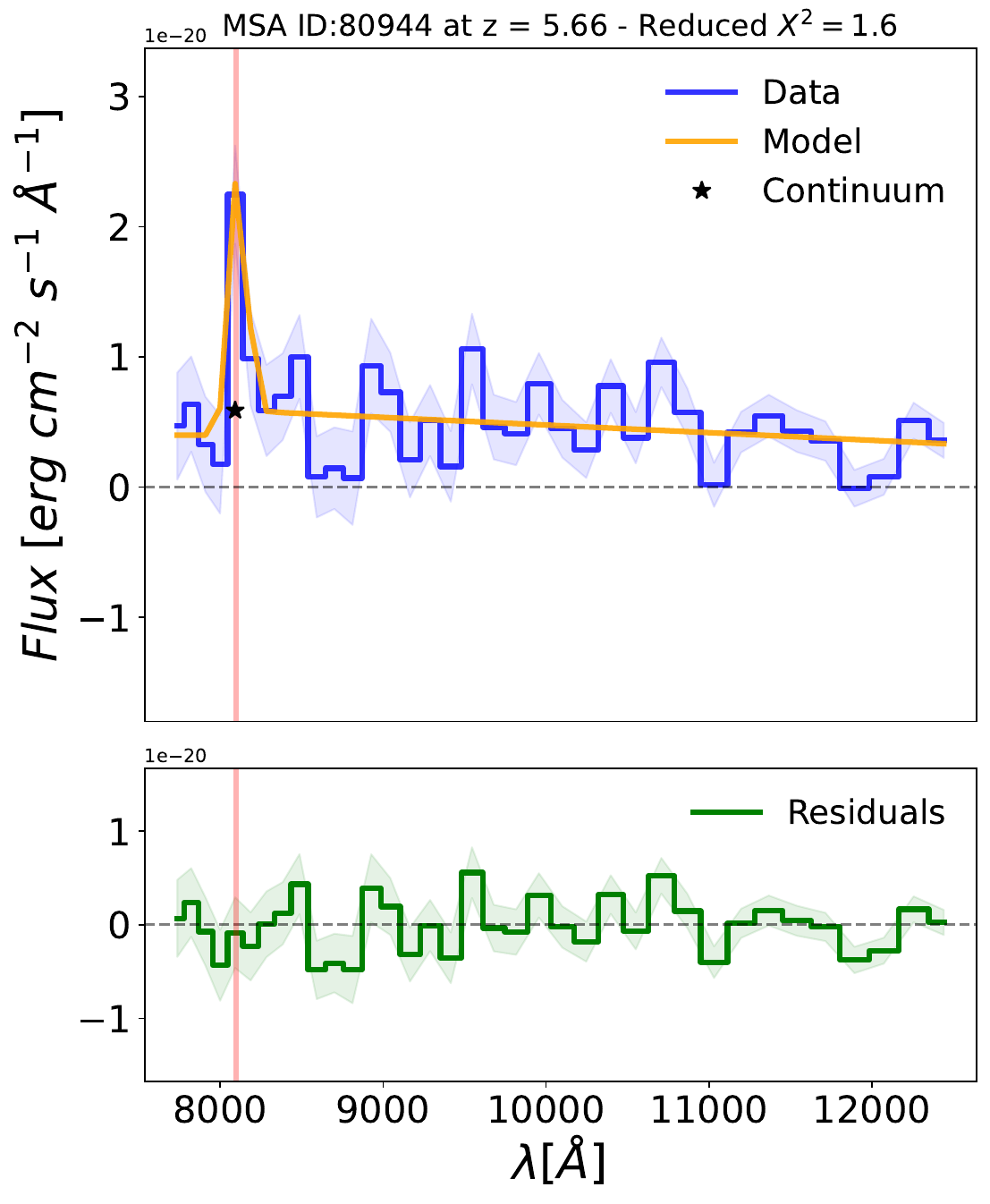}

\centering

\includegraphics[width=0.3\textwidth, height=0.25\textheight]{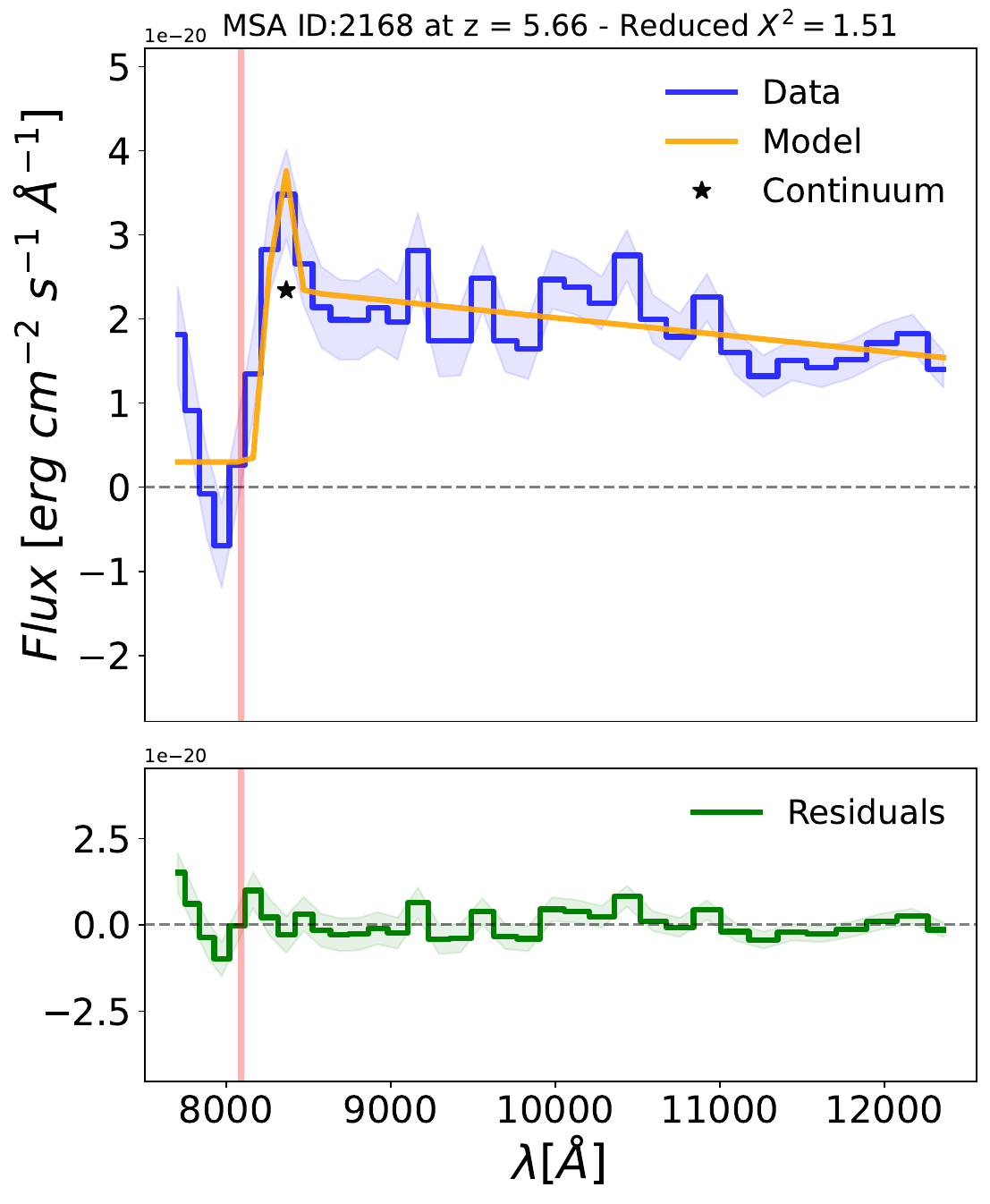}%
\hfill%
\includegraphics[width=0.3\textwidth, height=0.25\textheight]{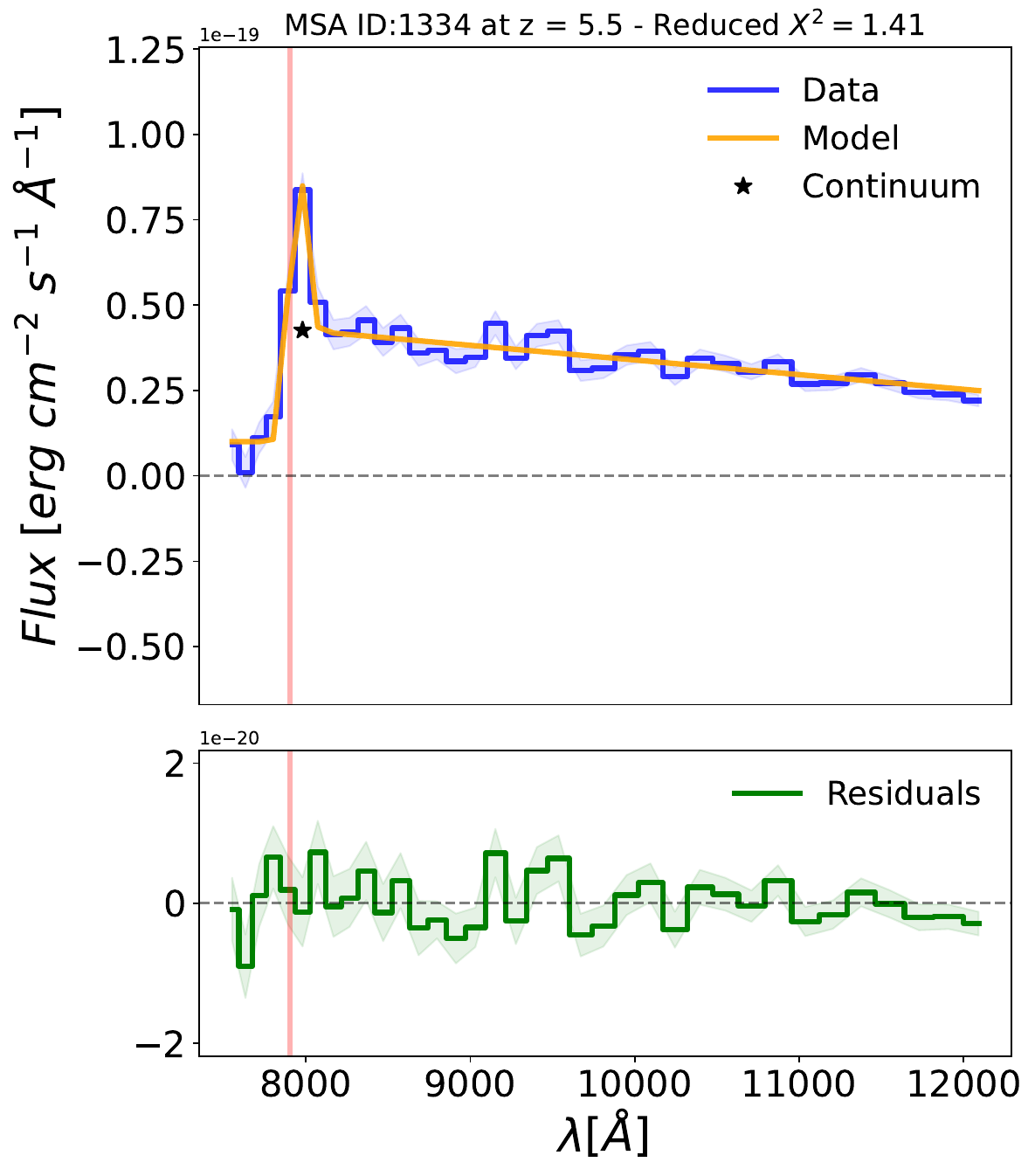}%
\hfill%
\includegraphics[width=0.3\textwidth, height=0.25\textheight]{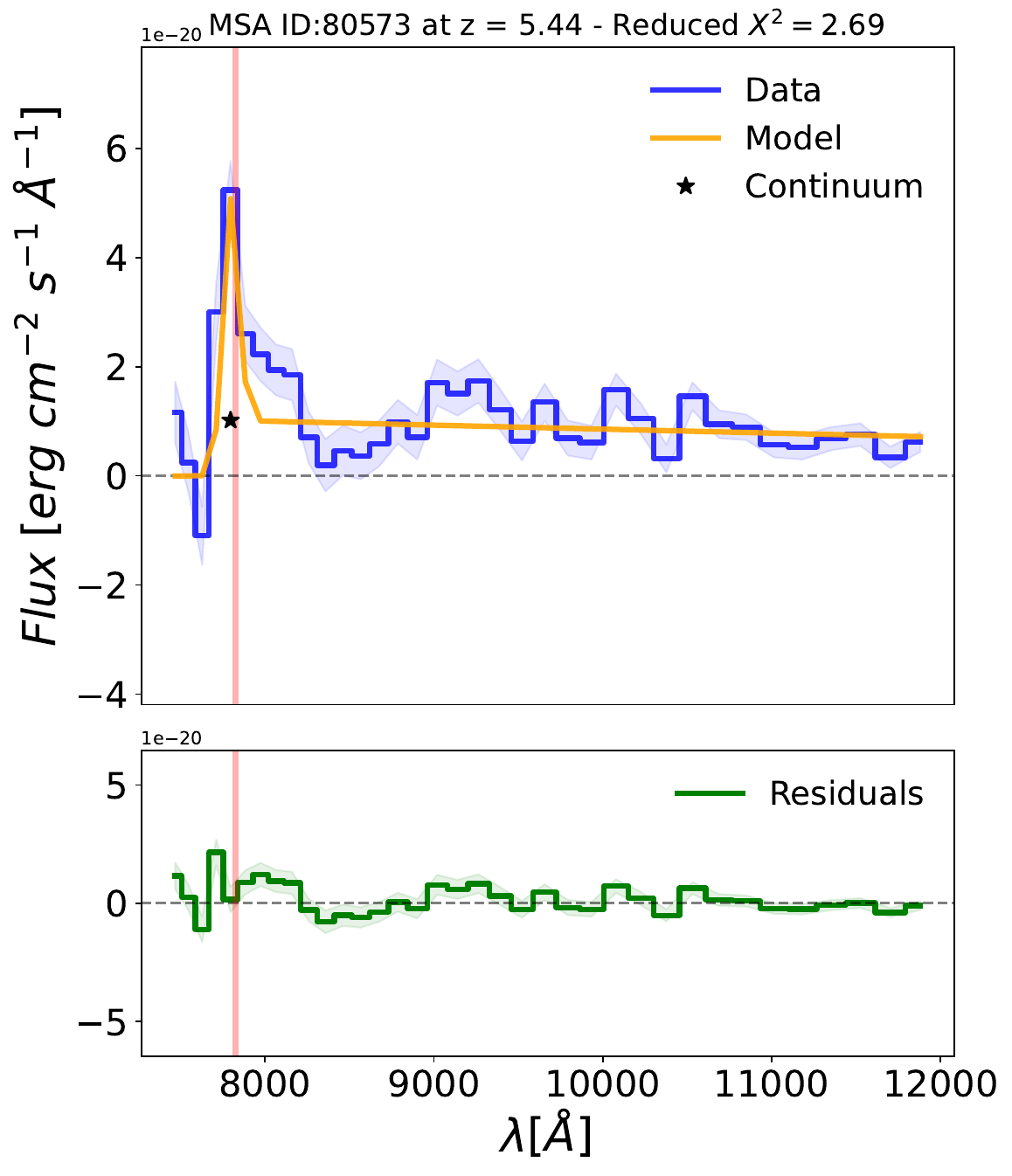}

\medskip

\includegraphics[width=0.3\textwidth, height=0.24\textheight]{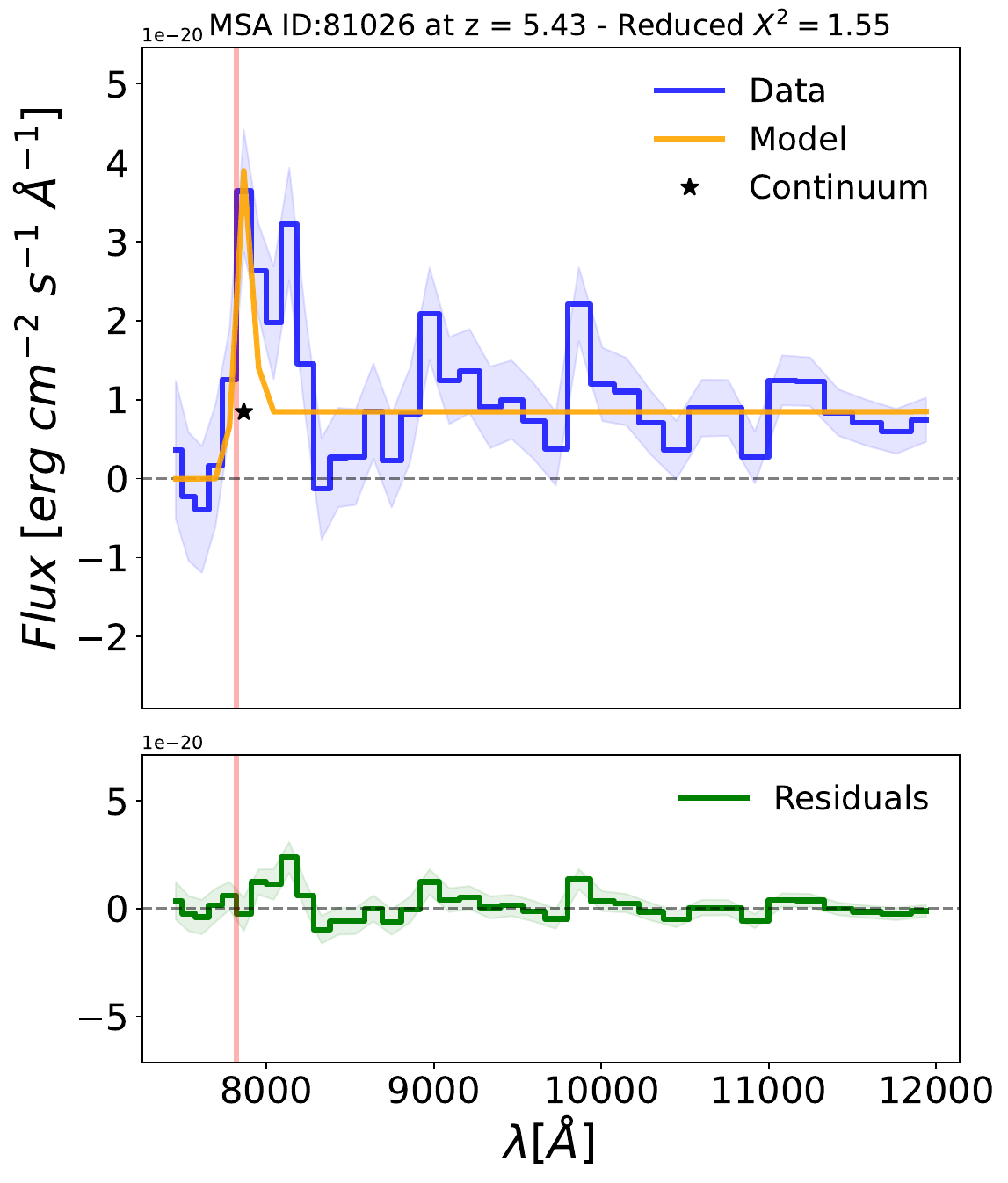}%
\hfill%
\includegraphics[width=0.3\textwidth, height=0.24\textheight]{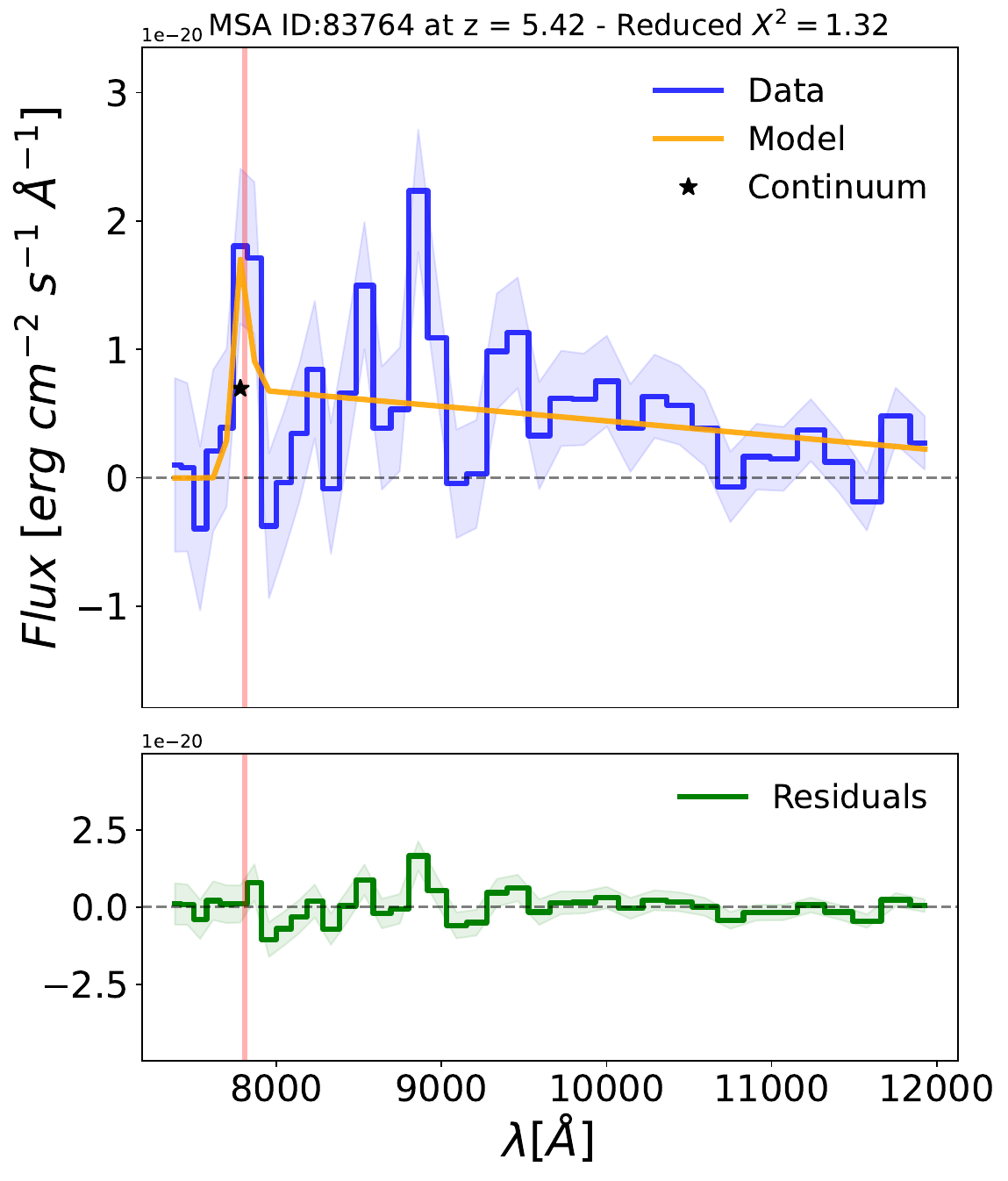}%
\hfill%
\includegraphics[width=0.3\textwidth, height=0.24\textheight]{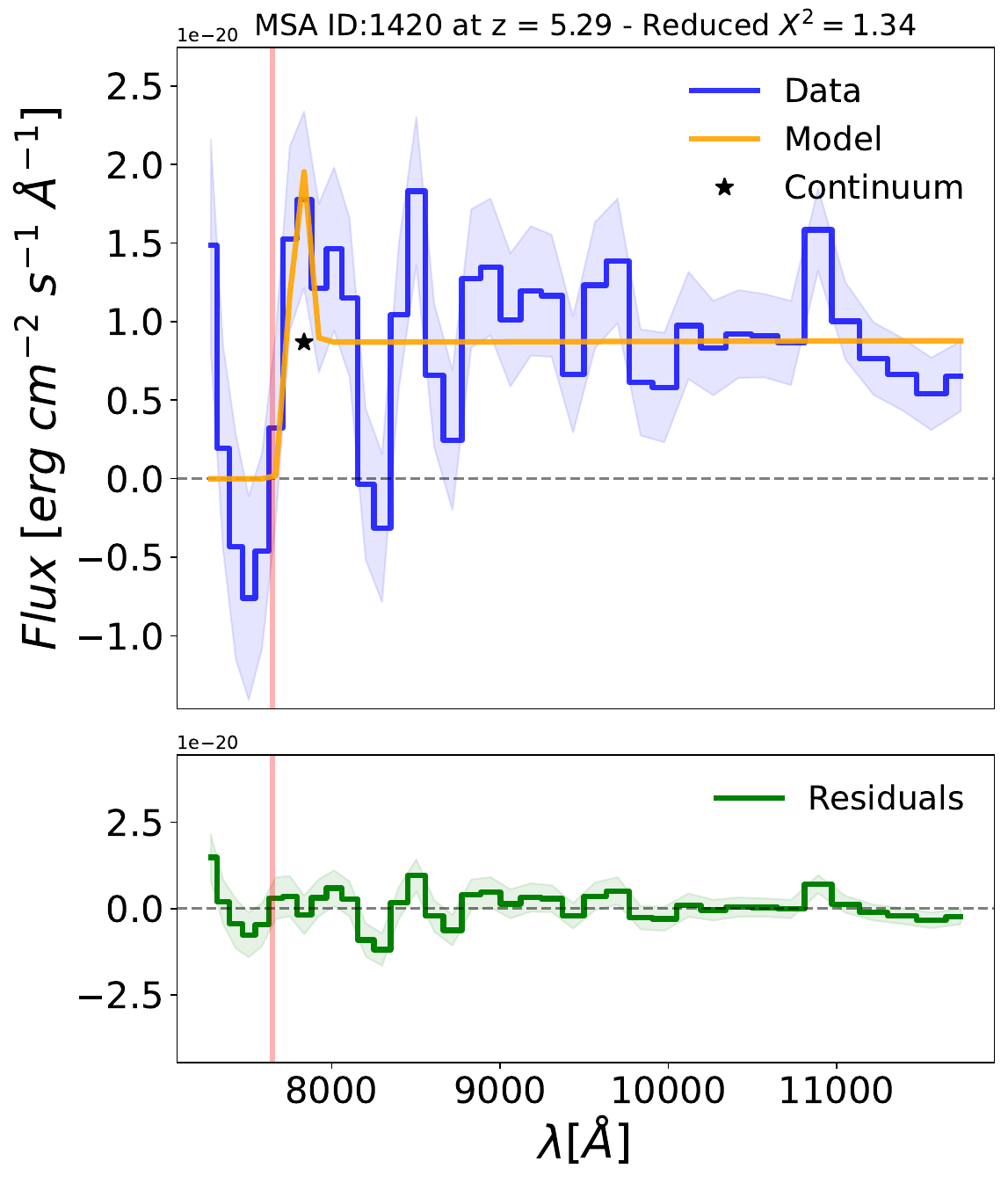}

\medskip

\includegraphics[width=0.3\textwidth, height=0.24\textheight]{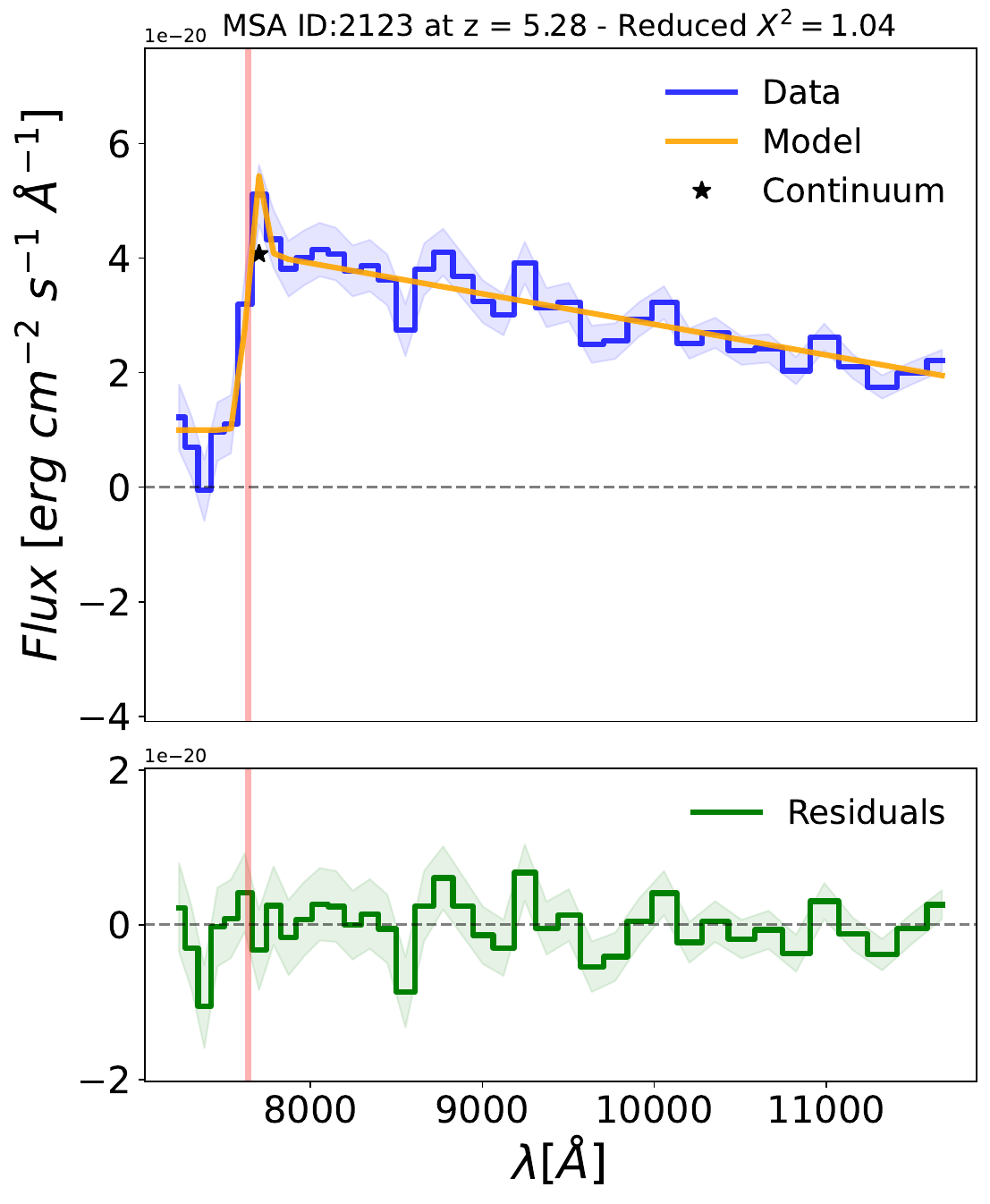}%
\hfill%
\includegraphics[width=0.3\textwidth, height=0.24\textheight]{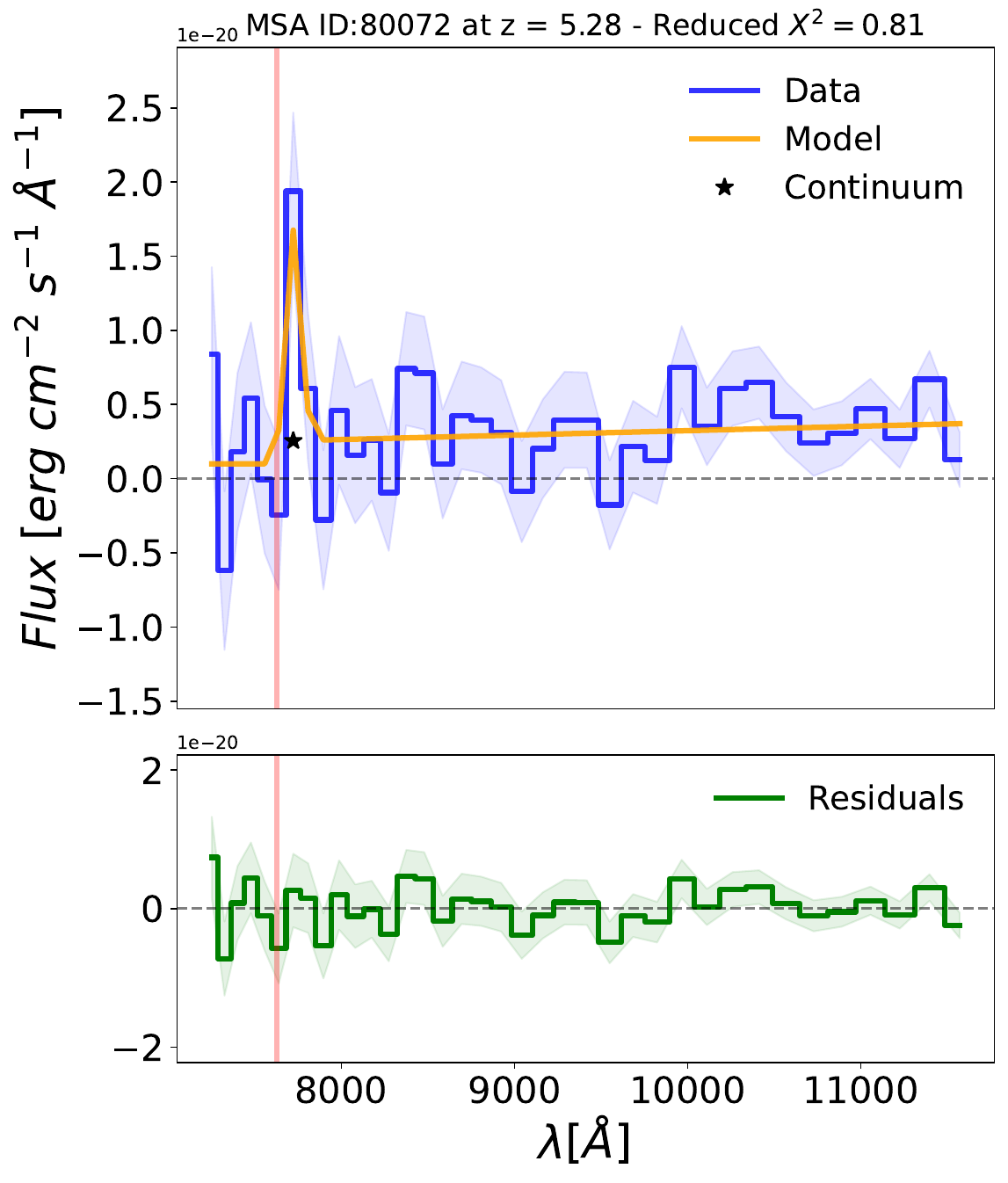}%
\hfill%
\includegraphics[width=0.3\textwidth, height=0.24\textheight]{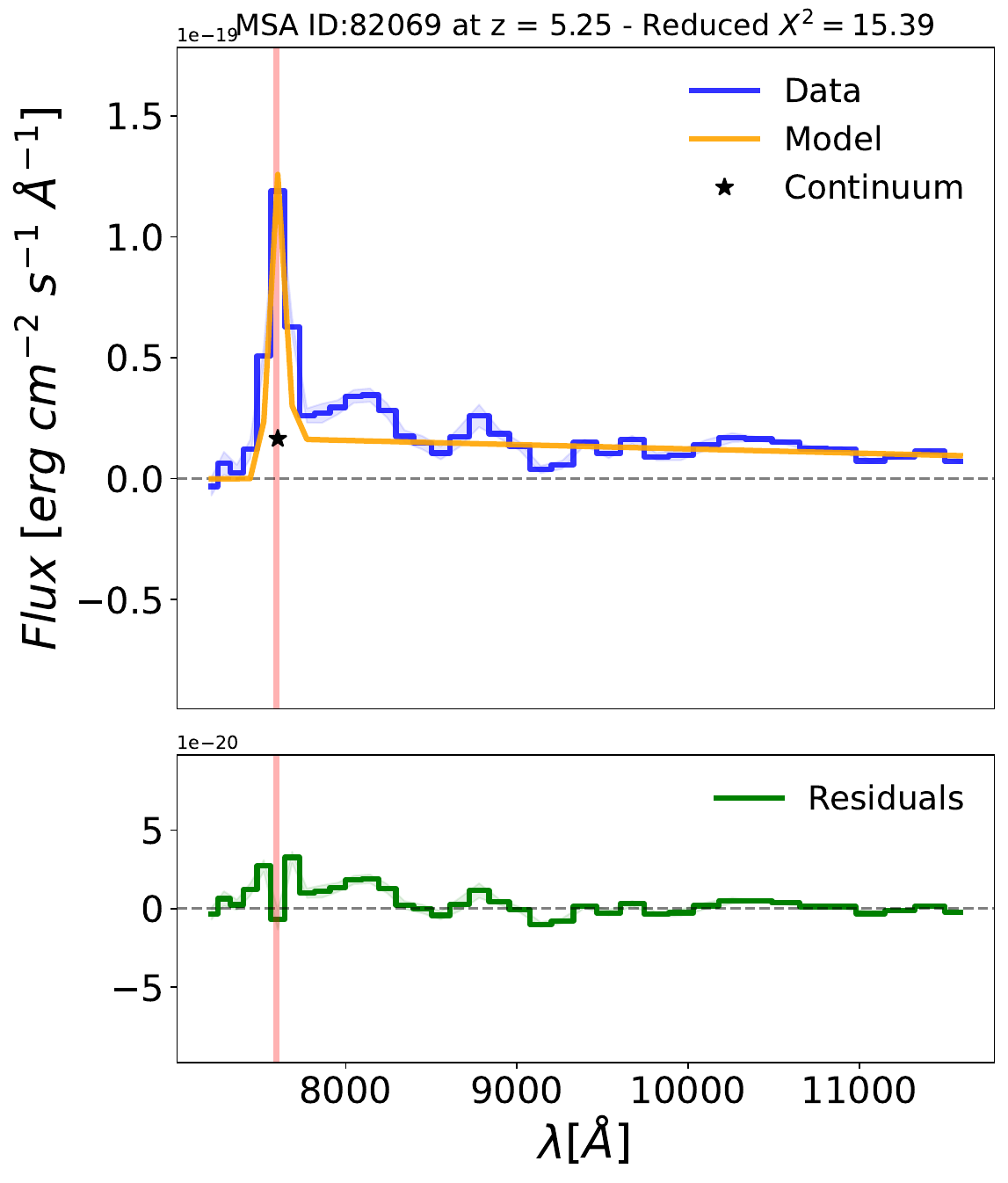}

\caption{See the description of Fig.~\ref{fig:EW_fit}.}
\label{fig:EW_fit_new3}
\end{figure*}
\begin{figure*}[ht!]

\includegraphics[width=0.3\textwidth, height=0.24\textheight]{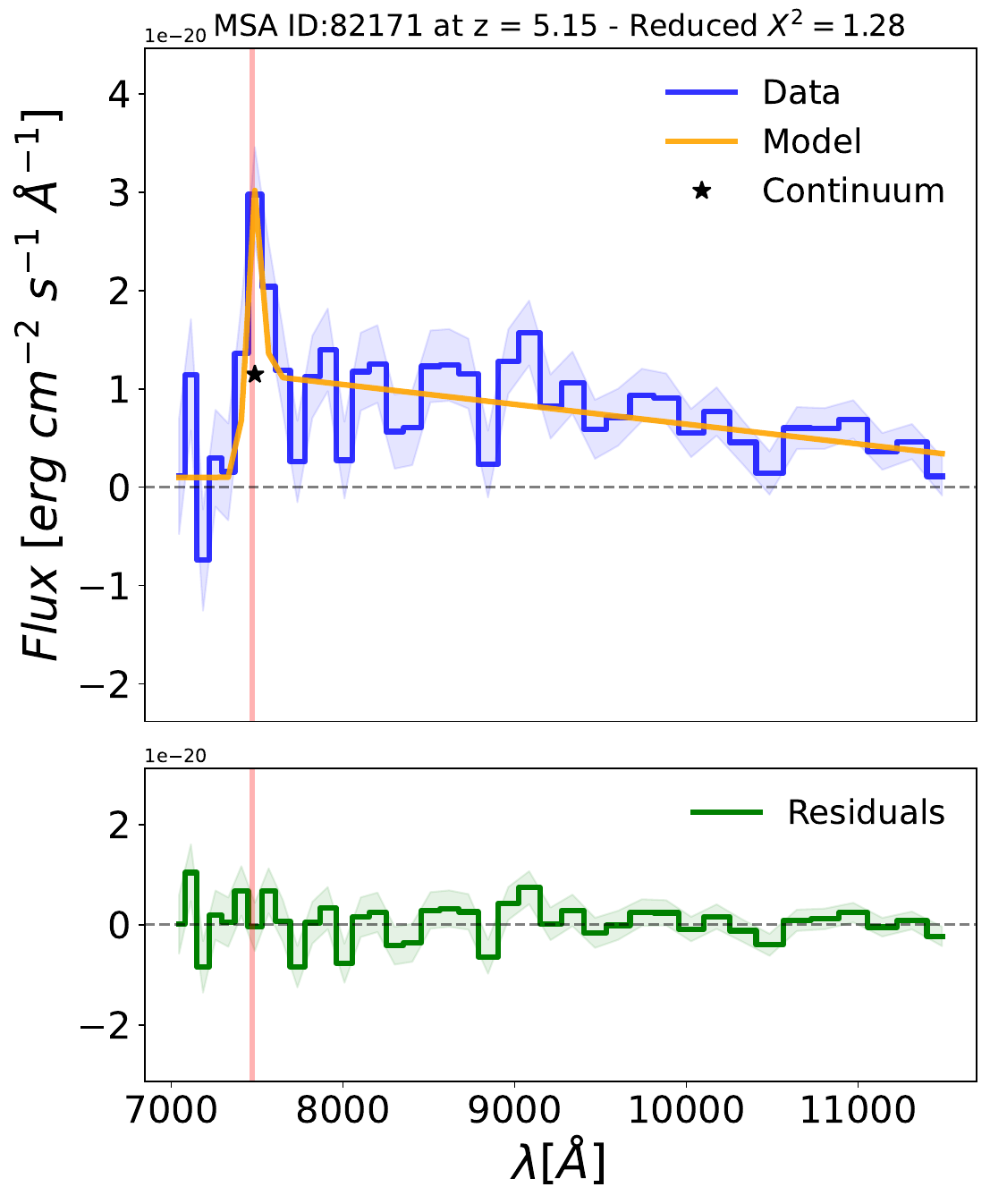}%
\hfill%
\includegraphics[width=0.3\textwidth, height=0.24\textheight]{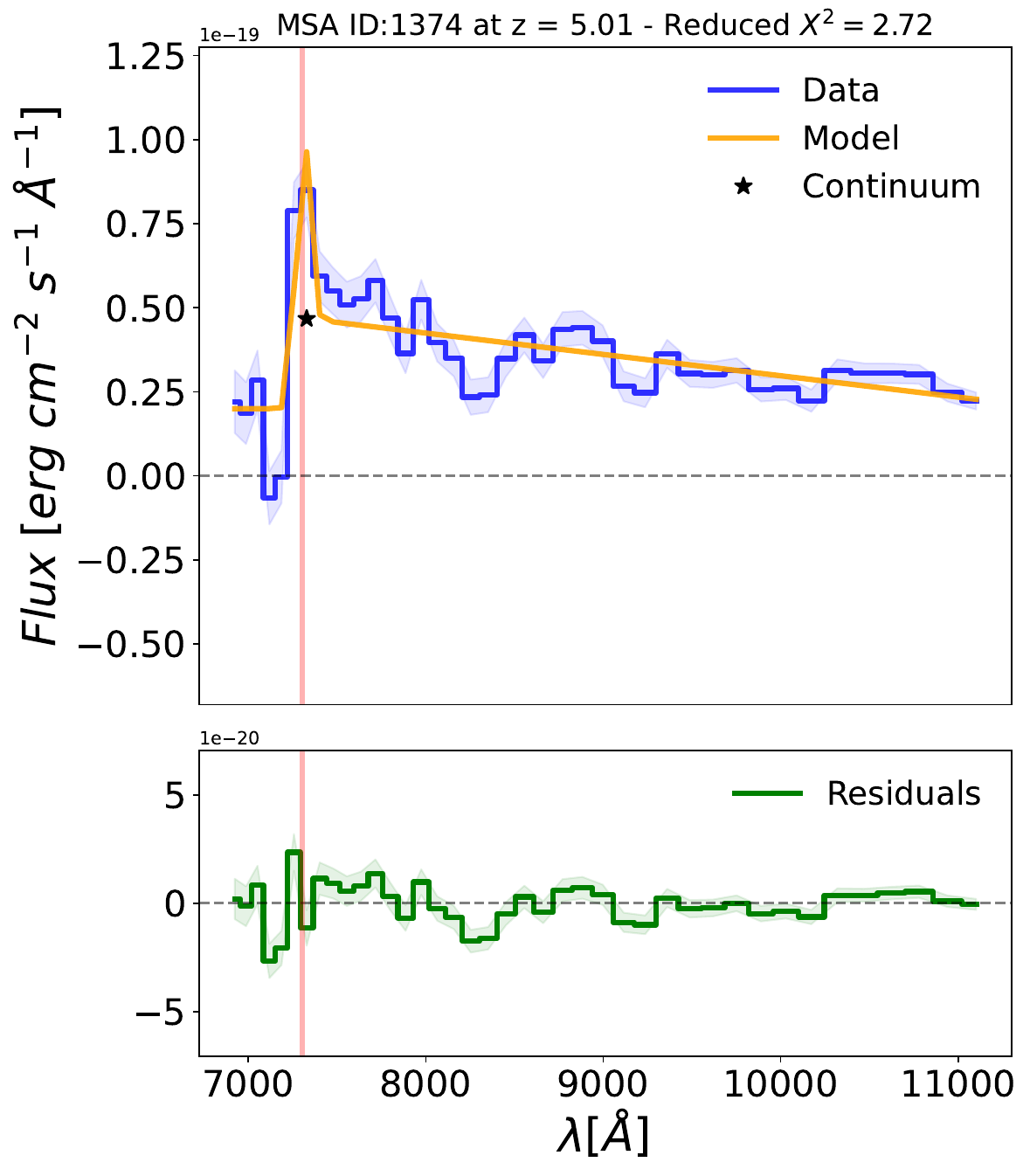}%
\hfill%
\includegraphics[width=0.3\textwidth, height=0.24\textheight]{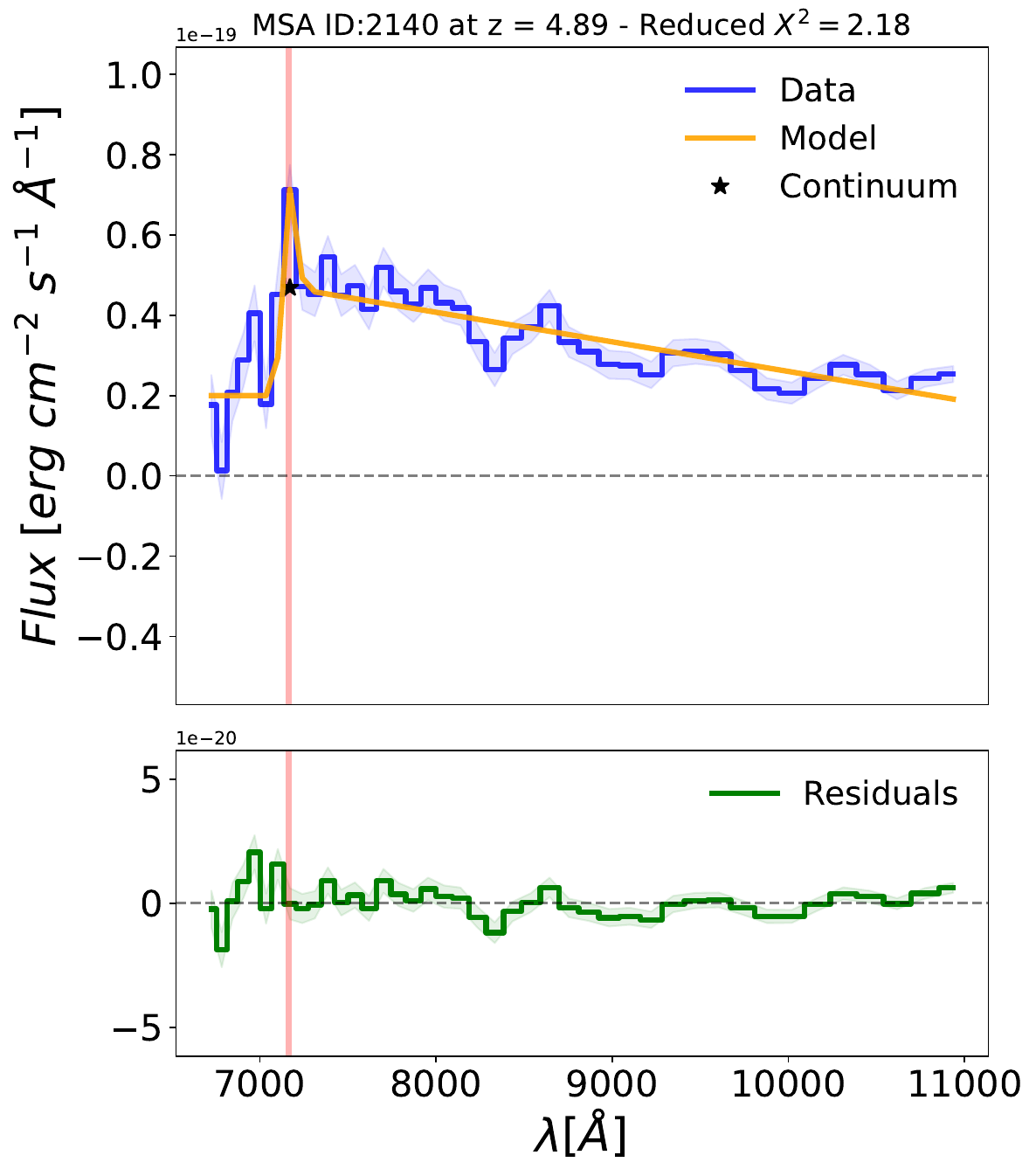}

\medskip

\includegraphics[width=0.3\textwidth, height=0.25\textheight]{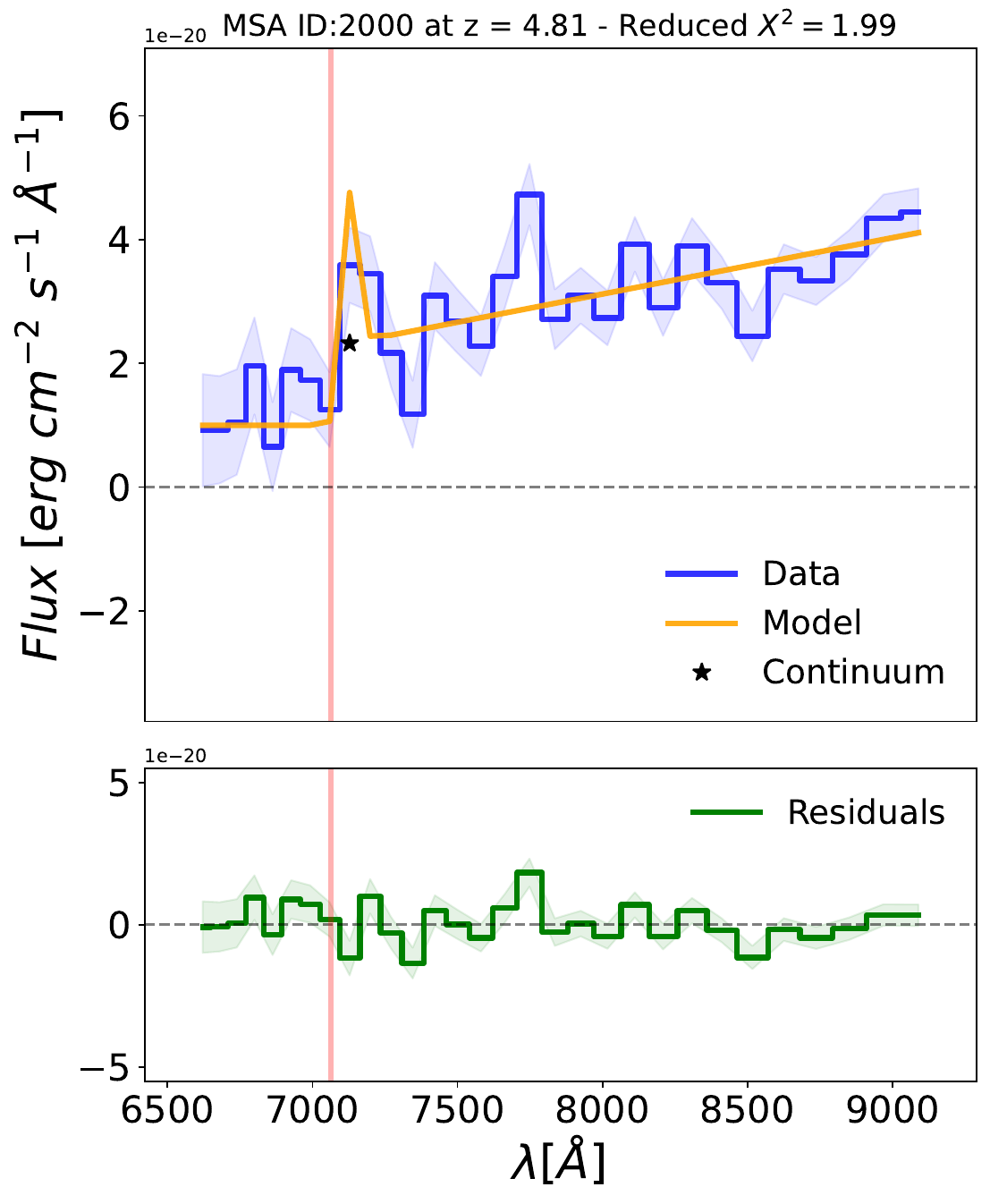}%
\hfill%
\includegraphics[width=0.3\textwidth, height=0.25\textheight]{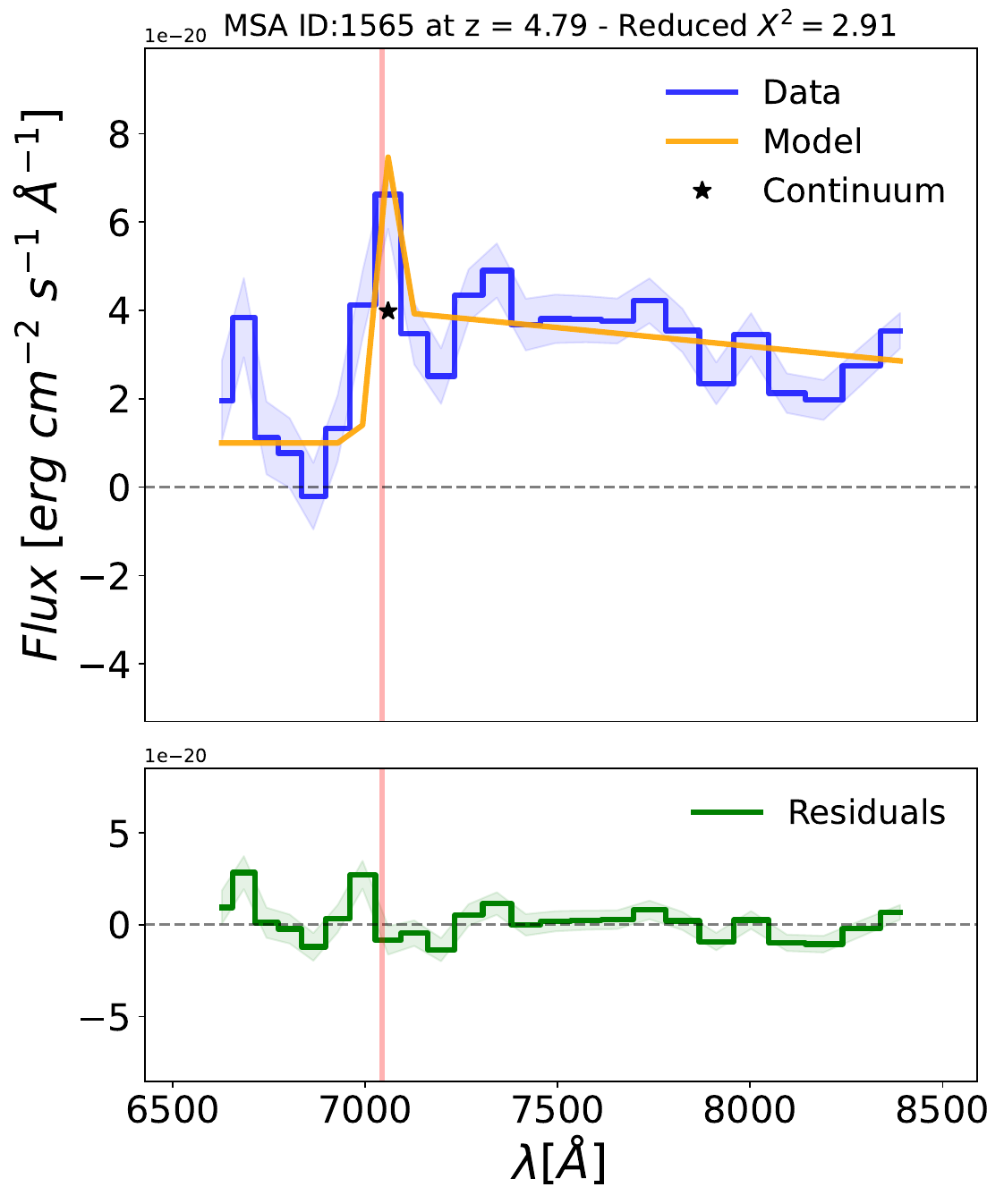}%
\hfill%
\includegraphics[width=0.3\textwidth, height=0.25\textheight]{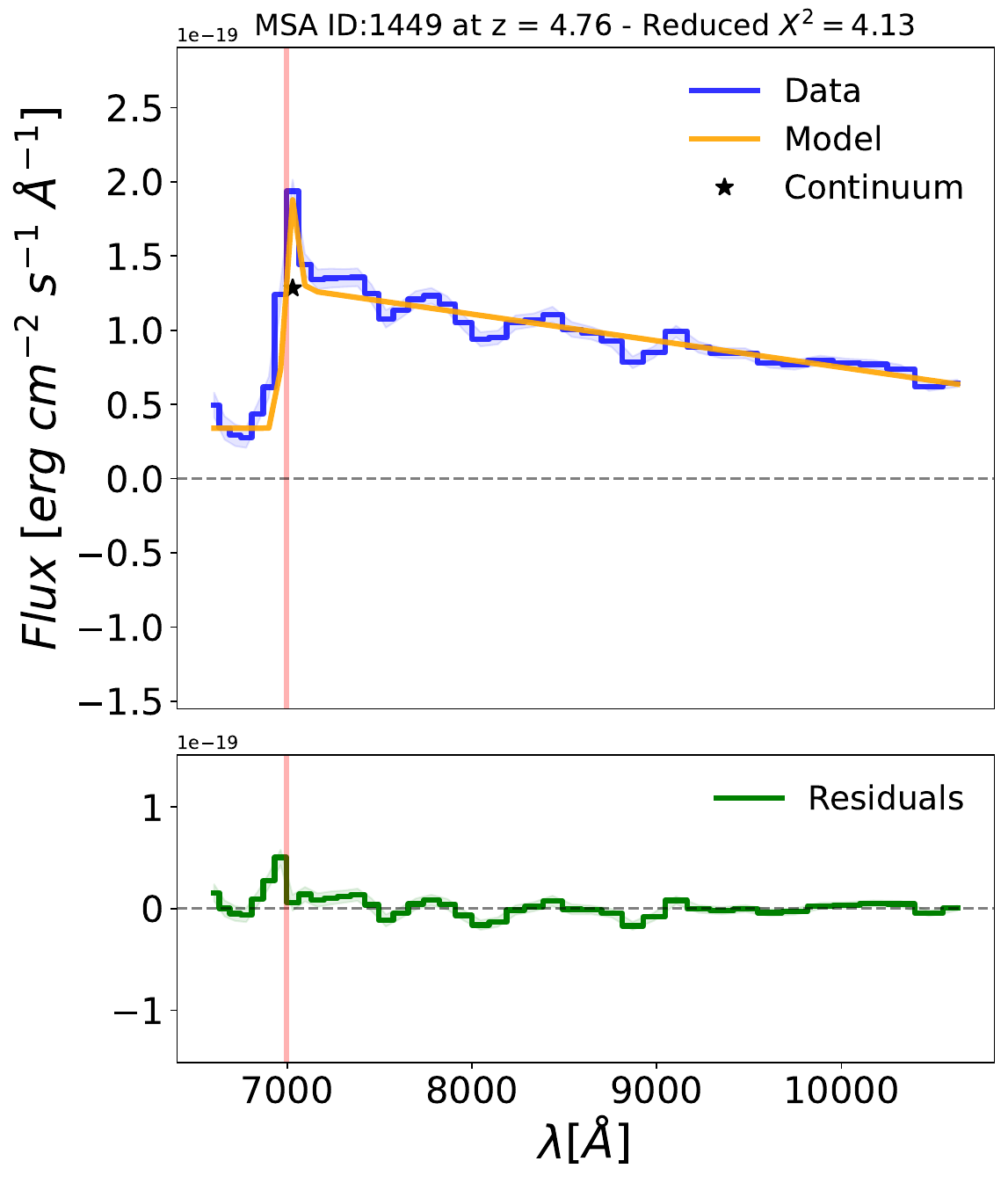}

\medskip

\includegraphics[width=0.3\textwidth, height=0.24\textheight]{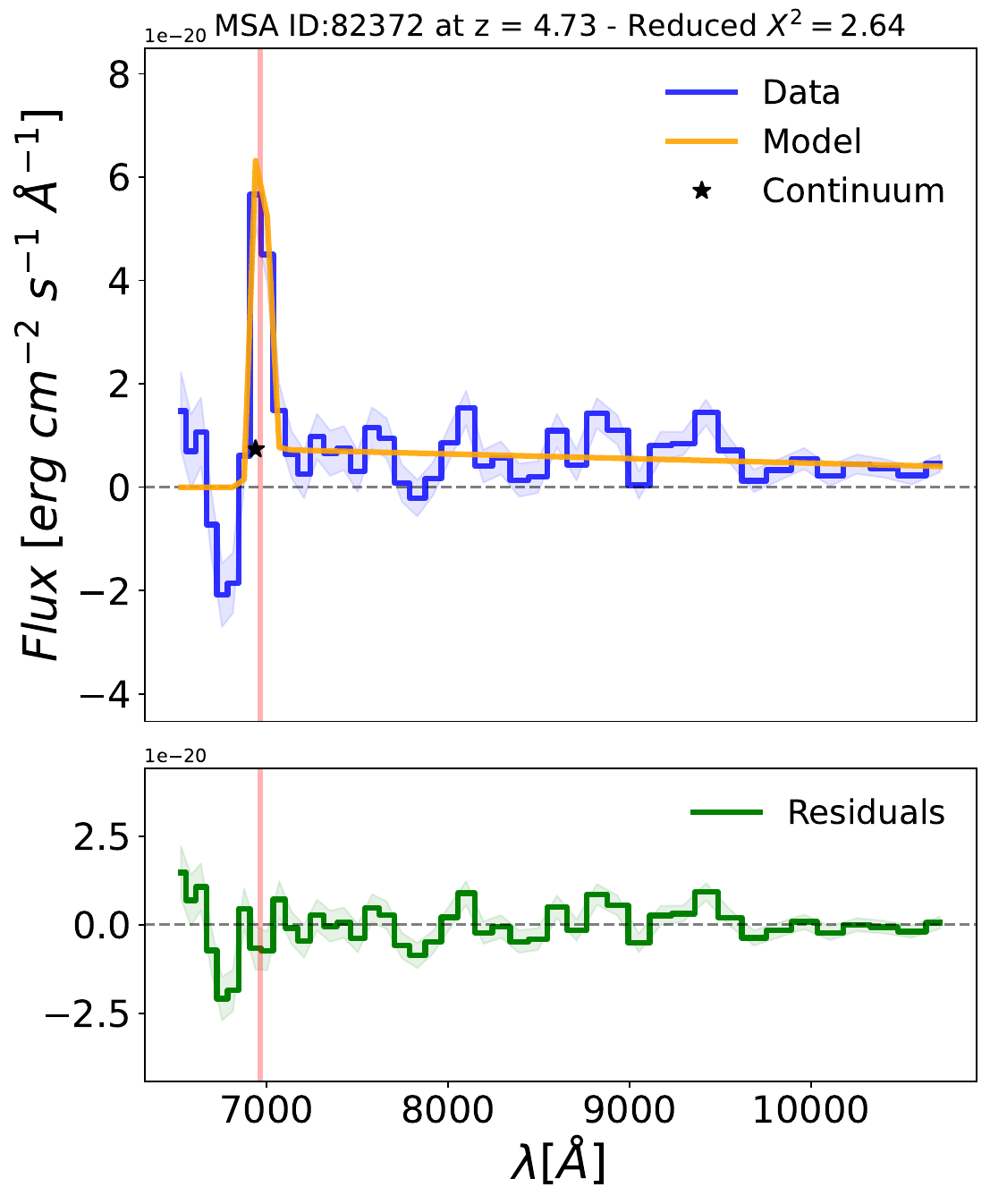}%
\hfill%
\includegraphics[width=0.3\textwidth, height=0.24\textheight]{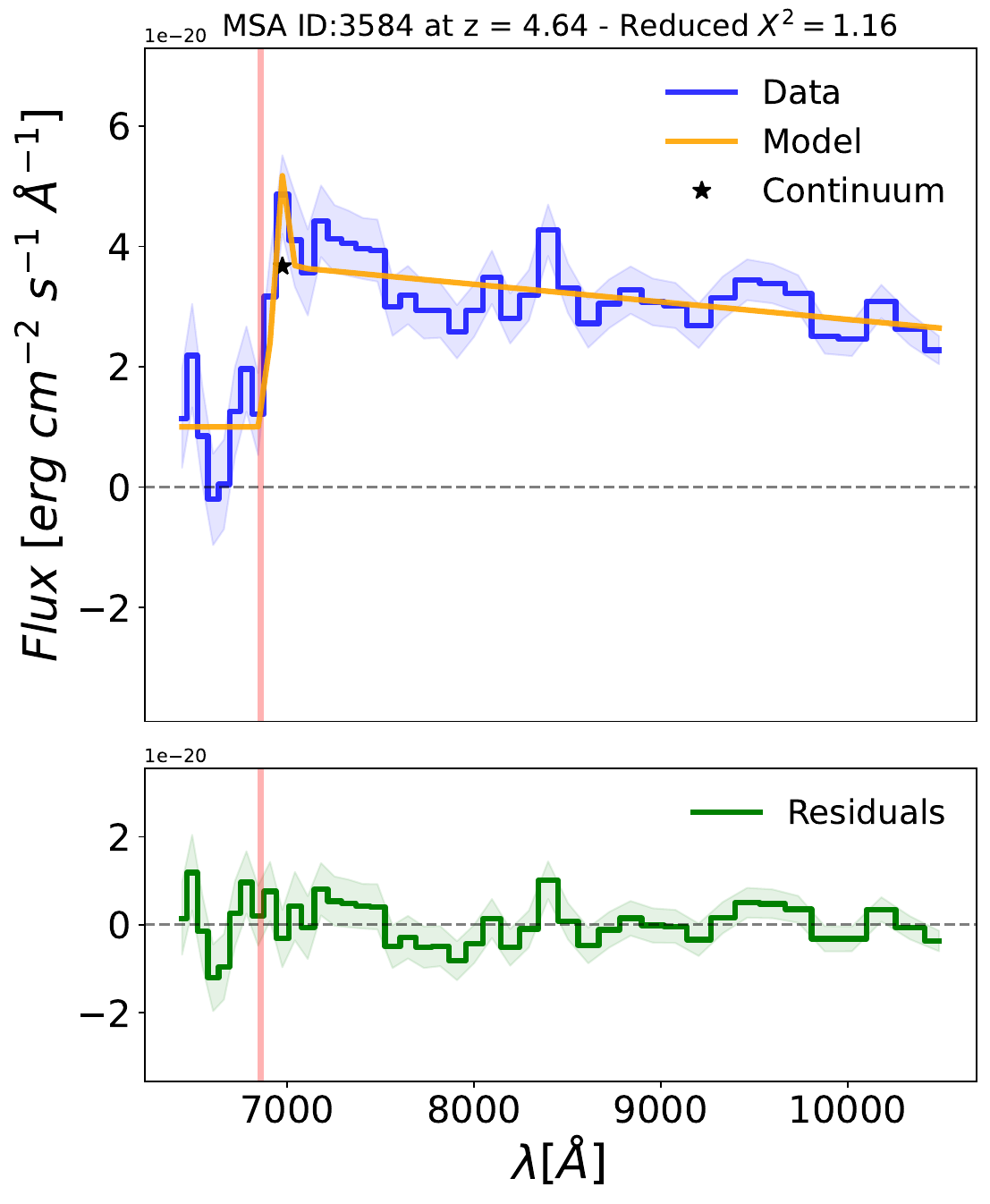}%
\hfill%
\includegraphics[width=0.3\textwidth, height=0.24\textheight]{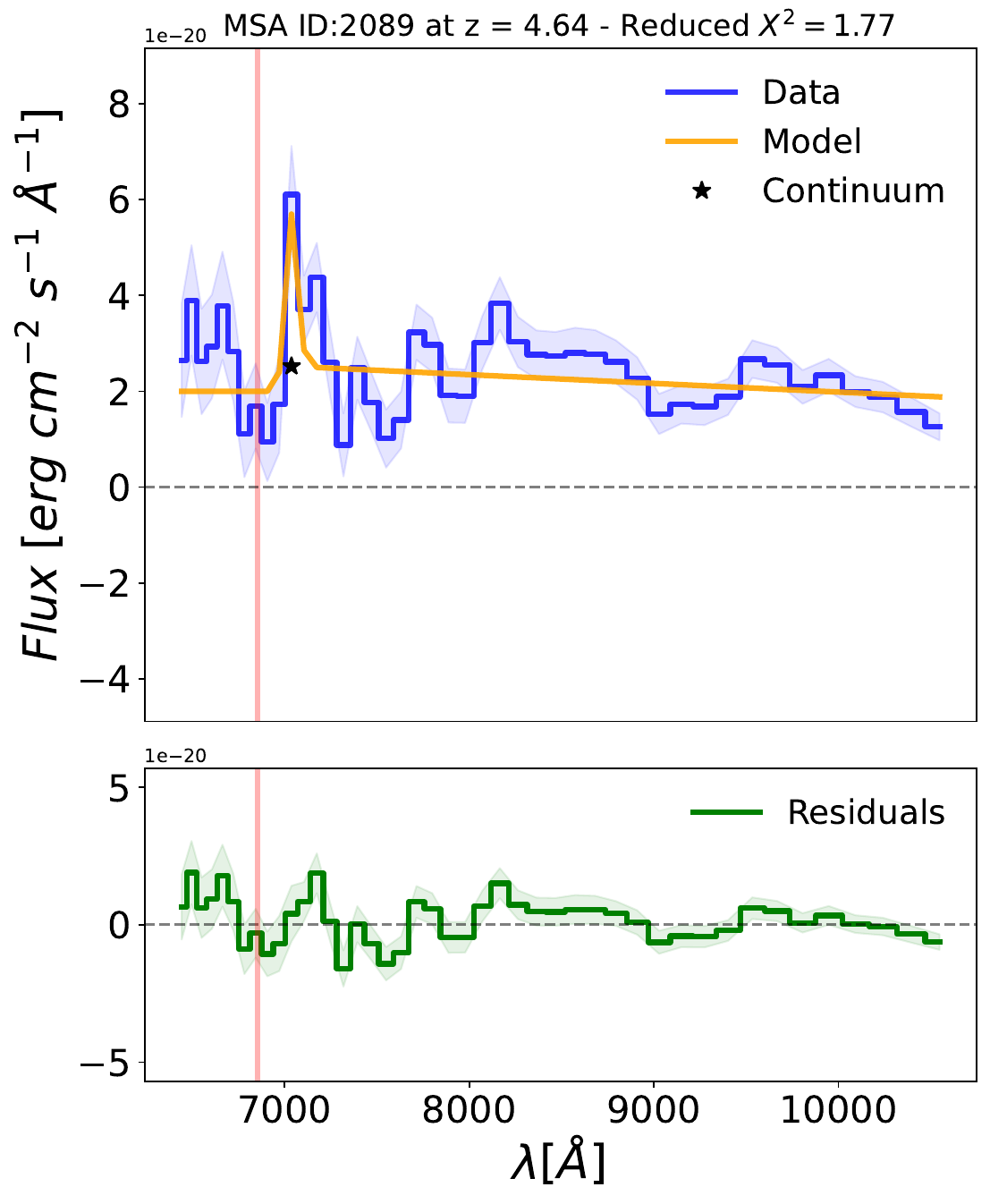}

\medskip

\includegraphics[width=0.3\textwidth, height=0.24\textheight]{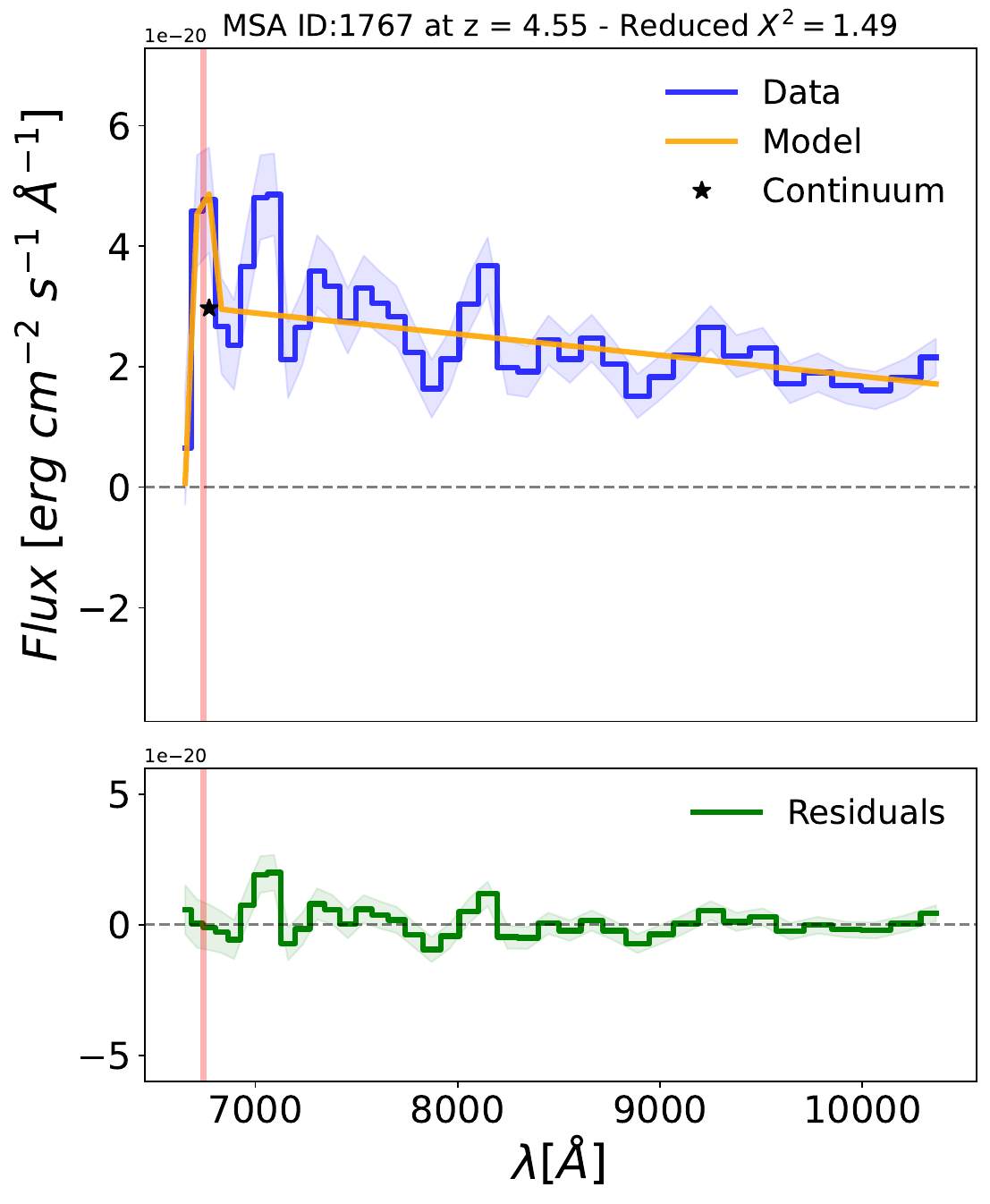}%
\hfill%
\includegraphics[width=0.3\textwidth, height=0.24\textheight]{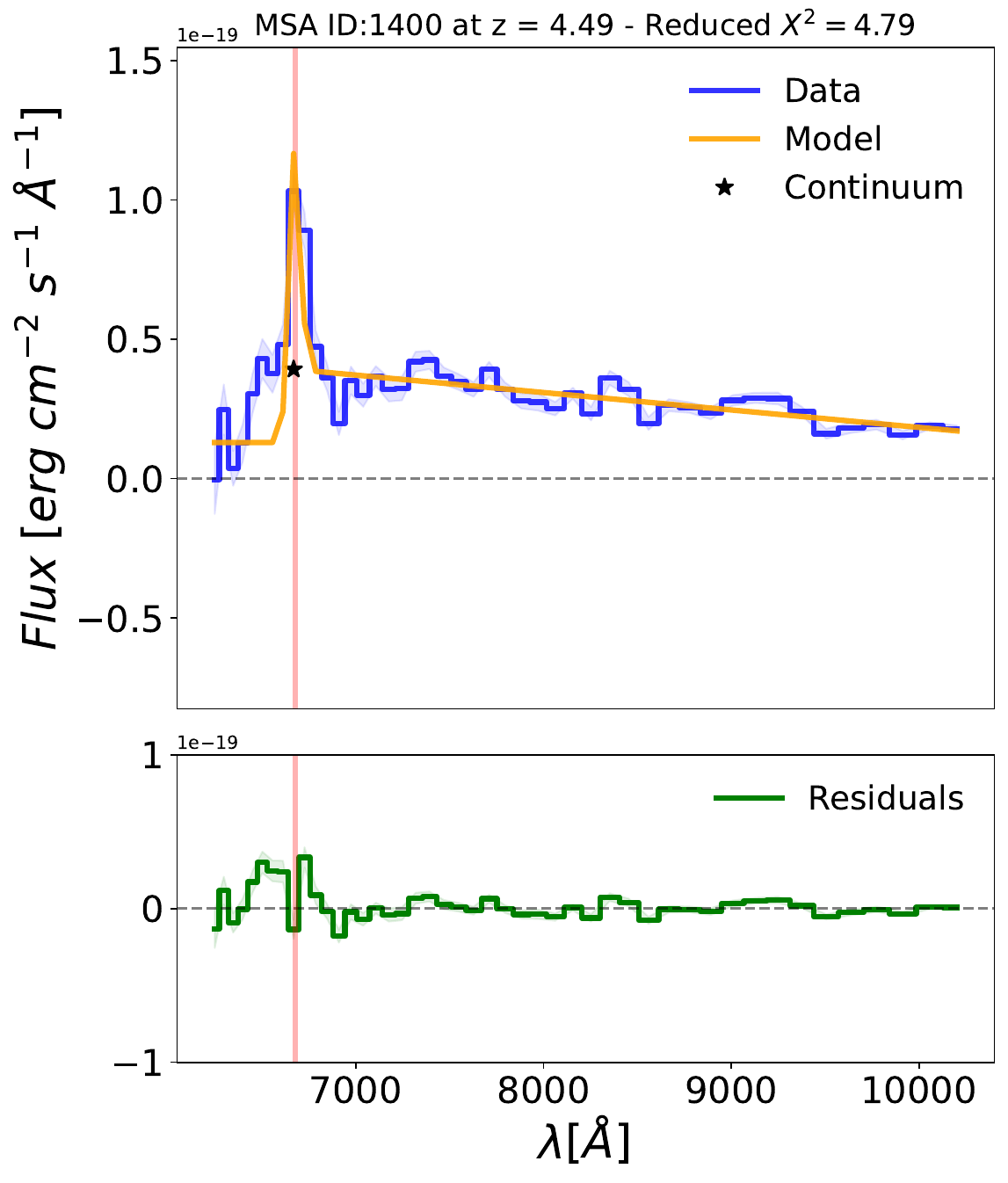}%
\hfill%
\includegraphics[width=0.3\textwidth, height=0.24\textheight]{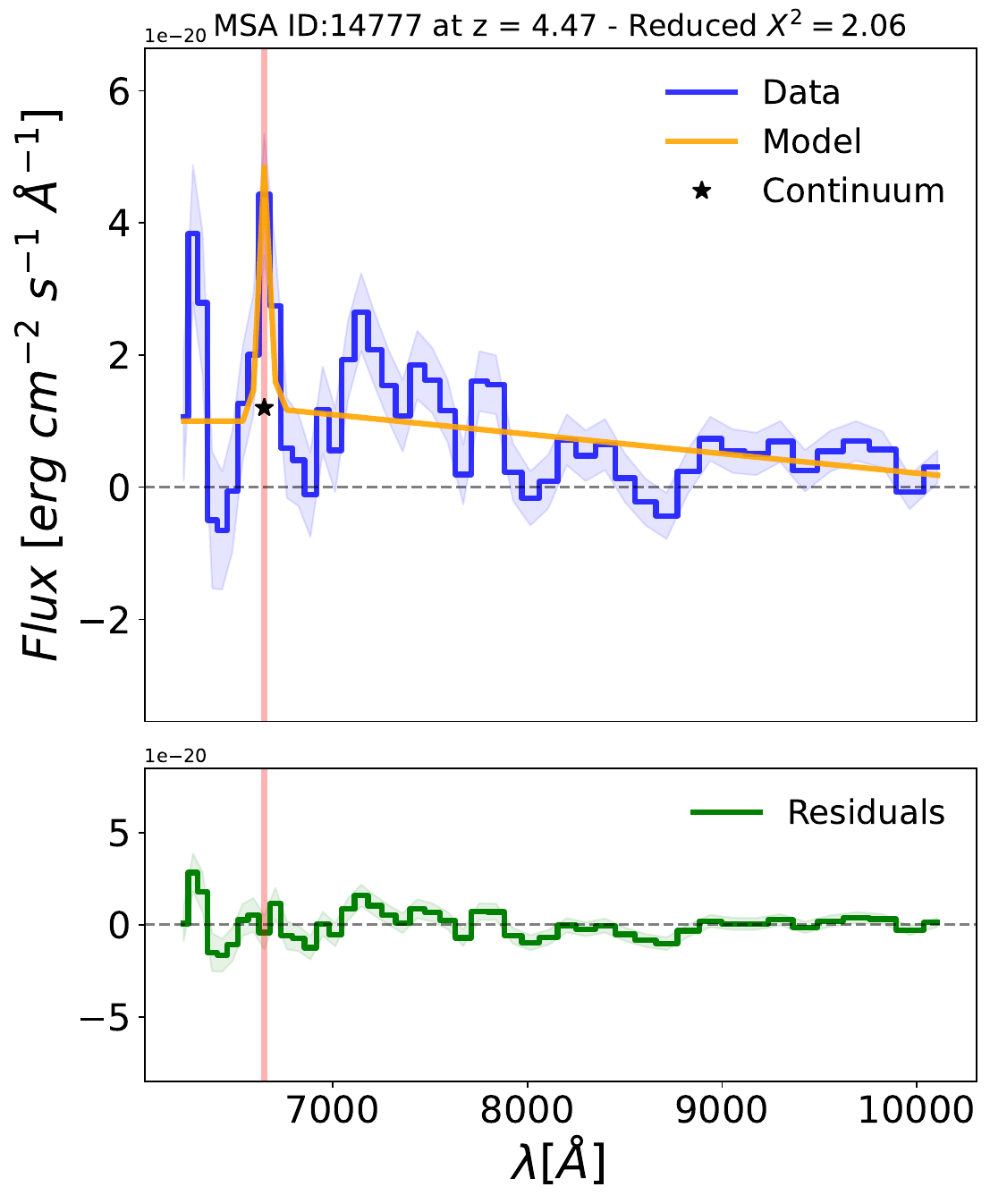}

\caption{See the description of Fig.~\ref{fig:EW_fit}.}
\label{fig:EW_fit_new4}
\end{figure*}
\begin{figure*}[ht!]
\centering

\includegraphics[width=0.3\textwidth, height=0.24\textheight]{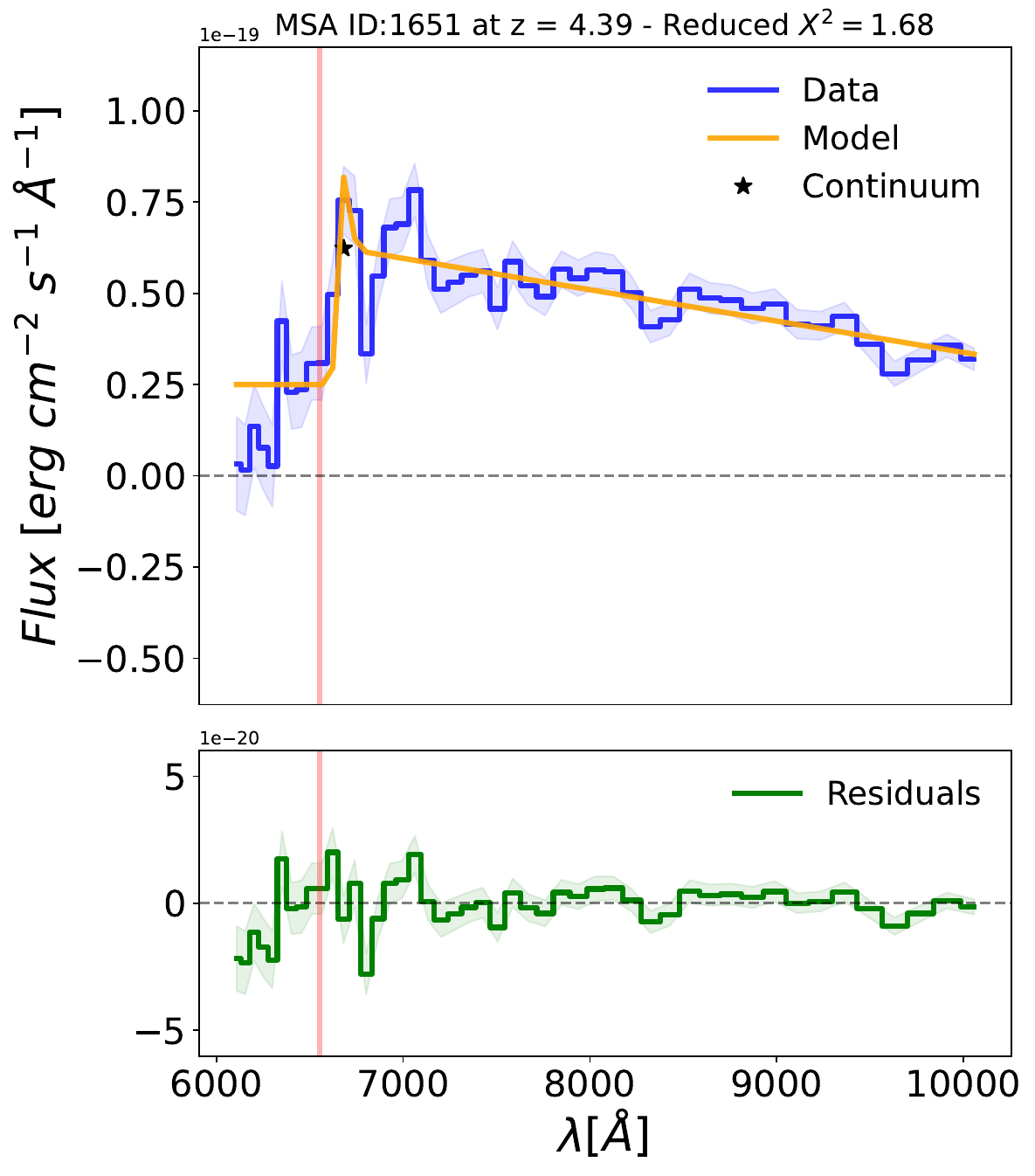}%
\hfill%
\includegraphics[width=0.3\textwidth, height=0.24\textheight]{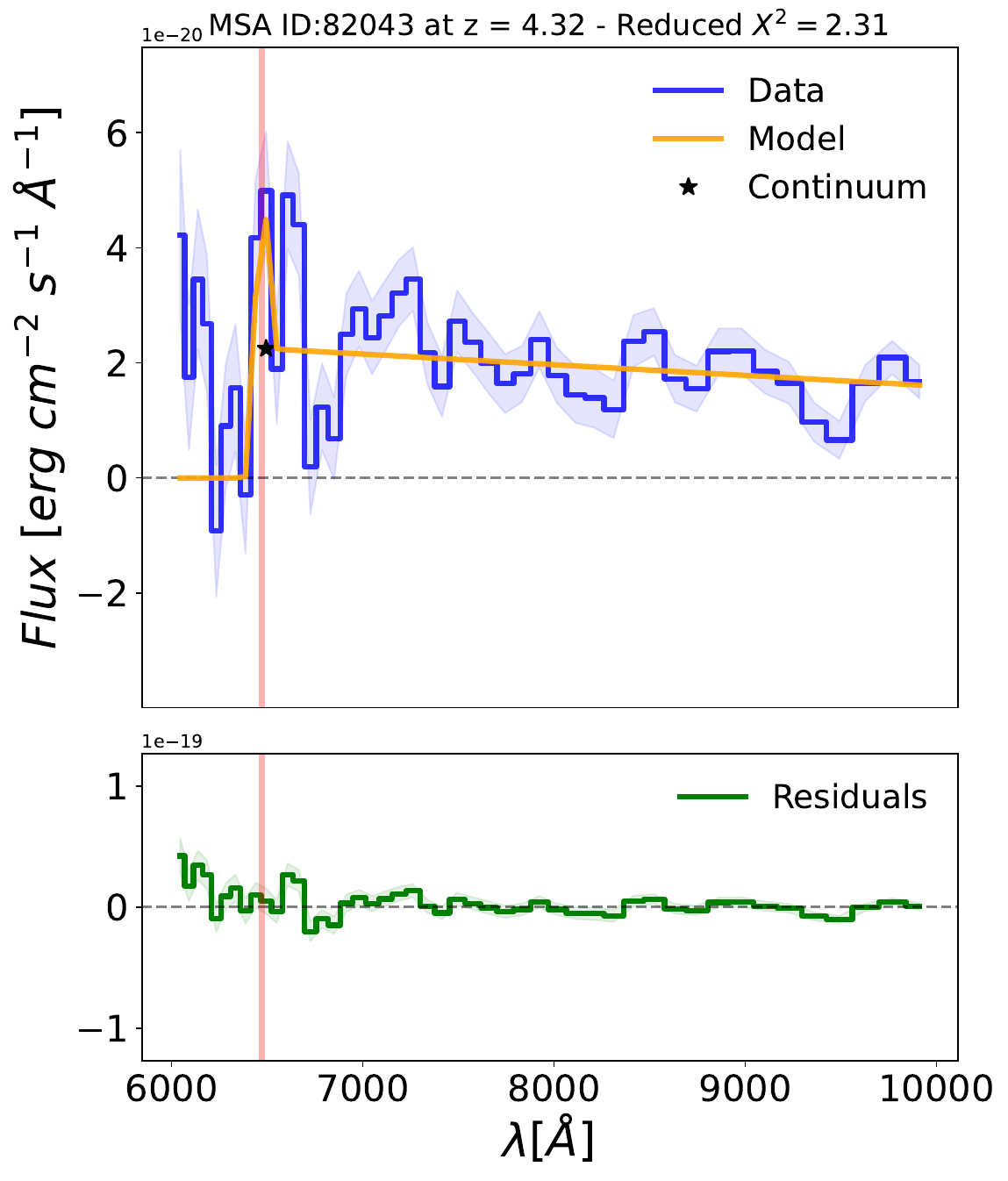}%
\hfill%
\includegraphics[width=0.3\textwidth, height=0.24\textheight]{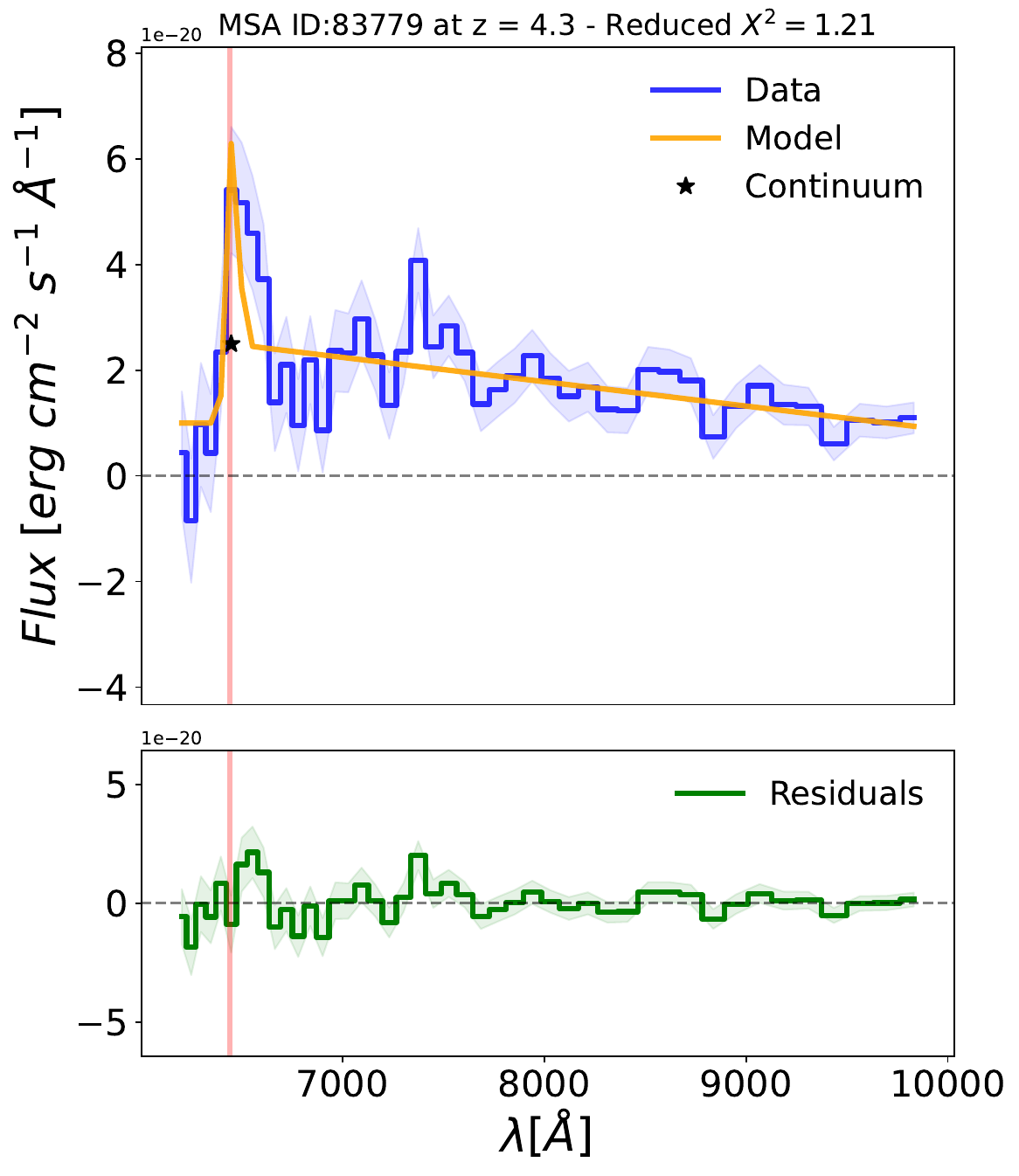}

\medskip

\includegraphics[width=0.3\textwidth, height=0.25\textheight]{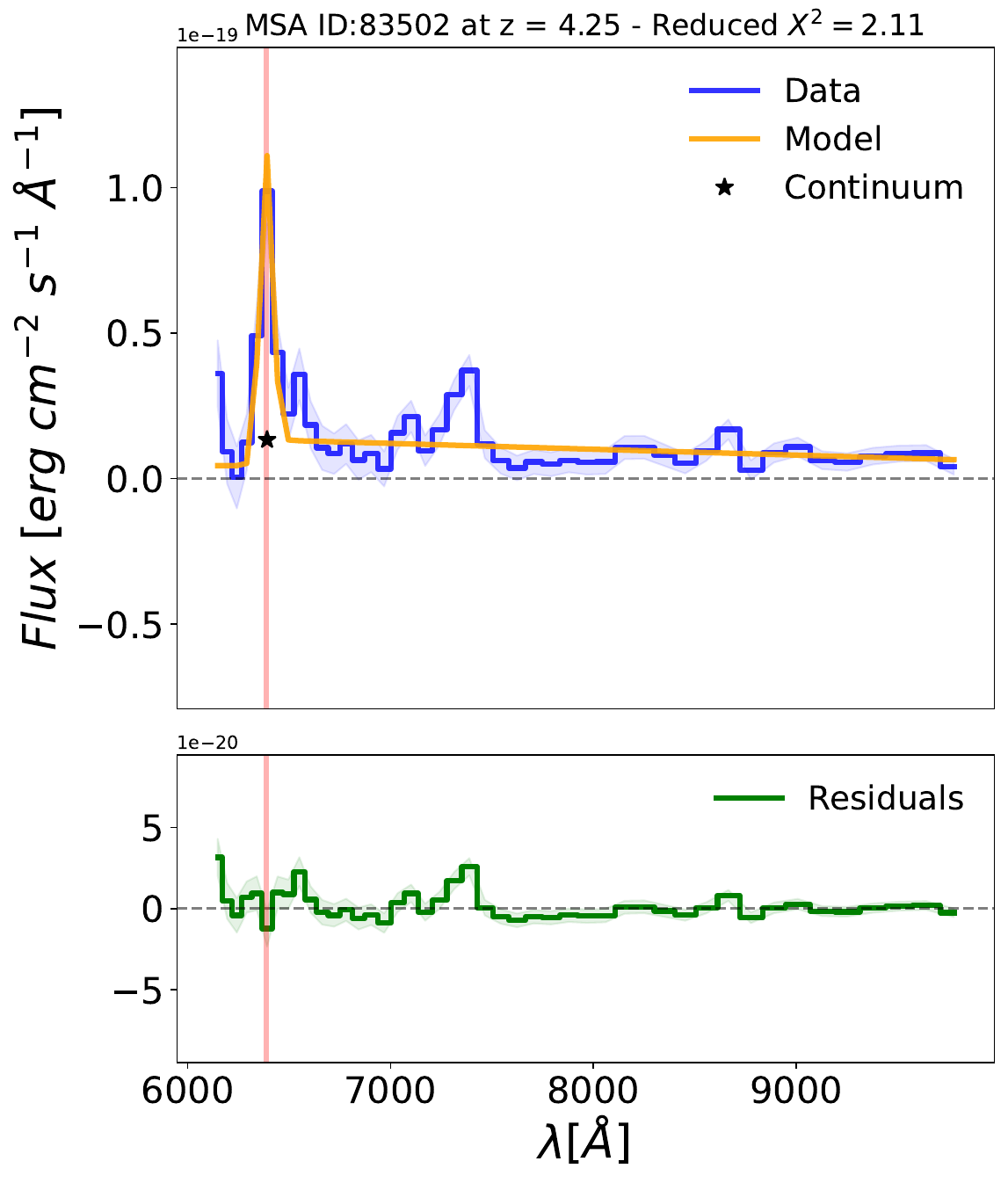}%
\hfill%
\includegraphics[width=0.3\textwidth, height=0.25\textheight]{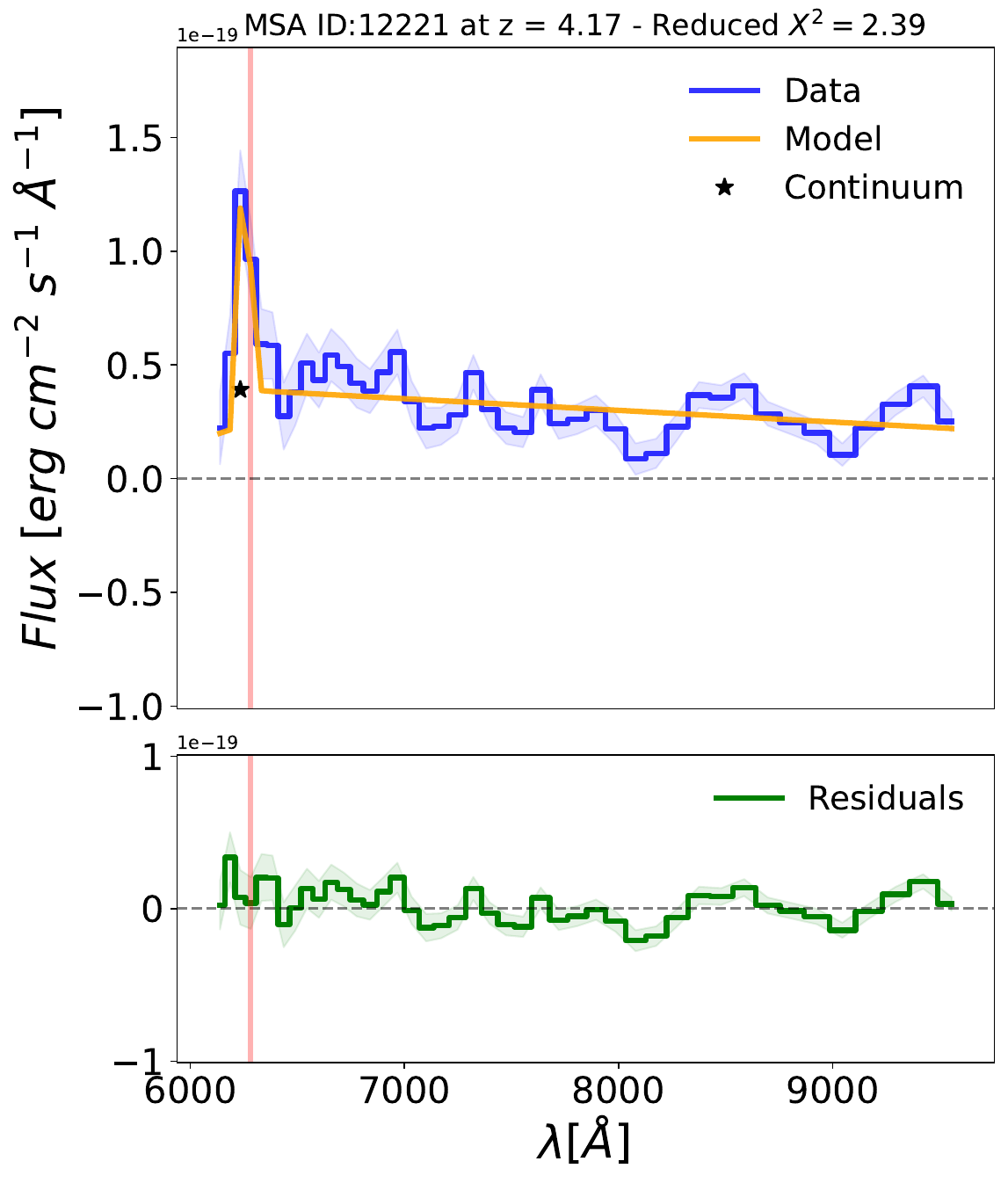}%

\caption{See the description of Fig.~\ref{fig:EW_fit}.}
\label{fig:EW_fit_new5}
\end{figure*}

\end{document}